\documentclass[12pt]{article}

\usepackage{bbm}
\usepackage{epsfig}
\usepackage{array}
\usepackage{float}
\usepackage{dsfont}
\usepackage{amstext}
\usepackage{rotating}
\usepackage{a4}
\usepackage{a4wide}
\usepackage{cite}

\def\ra{\rightarrow}
\def\be{\begin{equation}}
\def\ee{\end{equation}}
\def\gs{\mathrel{
   \rlap{\raise 0.511ex \hbox{$>$}}{\lower 0.511ex \hbox{$\sim$}}}}
\def\ls{\mathrel{
   \rlap{\raise 0.511ex \hbox{$<$}}{\lower 0.511ex \hbox{$\sim$}}}}

\newcommand{\ba}{\begin{array}{c}}
\newcommand{\baz}{\begin{array}{cc}}
\newcommand{\barrr}{\begin{array}{rrr}}
\newcommand{\bad}{\begin{array}{ccc}}
\newcommand{\bav}{\begin{array}{cccc}}
\newcommand{\baf}{\begin{array}{ccccc}}
\newcommand{\bea}{\begin{equation} \begin{array}{c}}
\newcommand{\eea}{ \end{array} \end{equation}}
\newcommand{\ea}{\end{array}}
\newcommand{\D}{\displaystyle}

\newcommand{\nue}{\nu_e}
\newcommand{\nuebar}{\overline{\nu}_e}
\newcommand{\numu}{\nu_\mu}
\newcommand{\nutau}{\nu_\tau}
\newcommand{\numubar}{\overline{\nu}_\mu}
\newcommand{\nutaubar}{\overline{\nu}_\tau}

\newcommand{\eps}{\epsilon}

\newcommand{\gsim}{\raise0.3ex\hbox{$\;>$\kern-0.75em\raise-1.1ex\hbox{$\sim\;$}}} 
\newcommand{\lsim}{\raise0.3ex\hbox{$\;<$\kern-0.75em\raise-1.1ex\hbox{$\sim\;$}}}

\begin{document}

\title{%\vspace{-2cm}
\hfill {\small arXiv: 0711.4517 [hep-ph]} 
\vskip 0.4cm
\large \bf
Flavor Ratios of Astrophysical Neutrinos: 
Implications for Precision Measurements}
\author{
Sandip Pakvasa$^a$\thanks{email: 
\tt pakvasa@phys.hawaii.edu}~\mbox{ 
},~~Werner Rodejohann$^b$\thanks{email: 
\tt werner.rodejohann@mpi-hd.mpg.de}~\mbox{ 
},~~Thomas J.~Weiler$^c$\thanks{email: 
\tt tom.weiler@vanderbilt.edu}
\\\\
{\normalsize \it$^a$Department of Physics and Astronomy,}\\
{\normalsize \it University of Hawaii, Honolulu, HI 96822, USA}\\ \\ 
{\normalsize \it$^b$Max--Planck--Institut f\"ur Kernphysik,}\\
{\normalsize \it  Postfach 103980, D--69029 Heidelberg, Germany} \\ \\ 
{\normalsize \it$^c$Department of Physics and Astronomy,} \\
{\normalsize \it Vanderbilt University, Nashville, TN 37235, USA}
}
\date{}
\maketitle
\thispagestyle{empty}
\vspace{-0.8cm}
\begin{abstract}
\noindent  
We discuss flavor-mixing probabilities and flavor ratios of 
high energy astrophysical neutrinos. In the first part of this paper, 
we expand the neutrino flavor-fluxes in terms of the small parameters 
$U_{e3}$ and $\pi/4 - \theta_{23}$, and show that there are universal 
first and second order corrections. The second order term 
can exceed the first order term, and so should be 
included in any analytic study. 
We also investigate the probabilities and ratios after a further expansion 
around the tribimaximal value of $\sin^2 \theta_{12} = 1/3$. 
In the second part of the paper, 
we discuss implications of deviations of initial flavor ratios 
from the usually assumed, idealized flavor compositions 
for pion, muon-damped, and neutron beam sources, viz., 
$(\nue : \numu : \nutau)=(1 : 2 : 0)$, $(0 : 1 : 0)$, 
and $(1 : 0 : 0)$, respectively. 
We show that even small deviations have significant 
consequences for the observed flavor ratios at Earth.
If initial flavor deviations are not taken into account in 
analyses, then false inferences for the values in the 
PMNS matrix elements (angles and phase) may result.

\end{abstract}

\newpage

\section{\label{sec:intro}Introduction}
There has been recently much discussion on the flavor mixing of high energy 
astrophysical neutrinos 
\cite{LP,fluxes,GR,Michael,Pasquale,xing2,WW,xing,KT,old,WR,MO,CP,kachneu,new}. 
Neutrino mixing modifies the initial flavor distribution of fluxes 
$\Phi_e^0 : \Phi_\mu^0 : \Phi_\tau^0$ in calculable ways.
In terrestrial neutrino telescopes such as 
IceCube \cite{ice} or KM3Net \cite{km3}, one can measure 
neutrino flavor ratios\cite{ratio1} and thereby 
obtain information on the neutrino parameters and/or the 
sources\footnote
{In principle, more exotic neutrino properties such as  
neutrino decay, Pseudo-Dirac structure, 
magnetic moments, interaction with dark energy, 
or breakdown of fundamental 
symmetries could also be studied 
\cite{decay,Pseudo,magn,HungPaes,CPT,MinSmi}.}.   
There are two essential ingredients to 
such analyses.  One is the initial flavor composition which depends on the 
nature of the production process at the source.  The other is 
the neutrino mixing scheme, in particular the values of the 
parameters governing neutrino mixing. The latter plays the role 
of altering the flavor mix from the original due to non-trivial 
lepton mixing. We will discuss in this paper 
precision issues for both these aspects. 

The current best-fit values as well as the 
allowed $1\sigma$ and $3\sigma$ ranges of the 
oscillation parameters are \cite{data}: 
\begin{eqnarray} \label{eq:data}
\sin^2 \theta_{12}
 &=& 0.32^{+0.02\,, \,0.08}_{-0.02\,, \,0.06} ~, \nonumber \\
\sin^2\theta_{23} &=& 0.45^{+0.07\,, \,0.20}_{-0.07\,, \,0.13} ~,\\
|U_{e3}|^2 &<& 0.02~(0.04)~.\nonumber
\end{eqnarray}
The CP phase is completely unknown and has a range between zero and $2\pi$.  
The current information on the mixing parameters \cite{APS} 
therefore suggests that $\theta_{13}$ and the deviation 
from maximal atmospheric neutrino mixing are small parameters and therefore 
to expand the formulae in terms of them. This is quite useful in order to 
obtain an analytical understanding of basically all phenomenological 
problems of interest. Furthermore, the deviation from the 
value $\sin^2 \theta_{12} = \frac 13$ is also small and thereby a third 
small expansion parameter is introduced. 
An idealized description for the leptonic mixing, 
or Pontecorvo-Maki-Nakagawa-Sakata (PMNS) matrix $U$, 
which is nevertheless perfectly compatible 
with all experimental information is tribimaximal mixing \cite{tbm} 
\be \label{eq:tbm}
U \simeq U_{\rm TBM} = \frac{1}{\sqrt{6}}
\left( 
\barrr
\D 2 & \D \sqrt{2}  & 0 \\[0.cm]
\D -1 & \D \sqrt{2}  & \D -\sqrt{3}\\[0.cm]
\D -1 & \D \sqrt{2}  & \D \sqrt{3}
\ea 
\right)~.
\ee
The transition probabilities of astrophysical neutrinos are 
functions of the elements of $U$, whose usual PDG parameterization is 
given in the Appendix. Expanding these probabilities 
in terms of the small parameters motivated by the observed 
oscillation phenomenology is one of the purposes of this paper. 
Previously, expansions in terms of $|U_{e3}|$ and 
$\pi/4-\theta_{23}$ have been discussed. 
In this case, we will show here first that in addition to the known 
universal first order correction \cite{xing,WR} 
there is a second order universal correction 
which can exceed the first order one and has to be 
included in analytical studies. Furthermore, by  
expanding the probabilities in terms of the deviation from 
$\sin^2 \theta_{12} = \frac 13$ we find extremely compact expressions.

In what concerns the initial flavor mix at the source, 
there are three simple and -- as we will make clear in 
this paper -- idealized possibilities: 
\begin{itemize}
\item The most conventional one is  neutrino emission from purely 
hadronic processes, such as $ p + p \rightarrow \pi^\mp 
\rightarrow \mu^\mp + \stackrel{(-)}{\nu_\mu} \rightarrow 
e^\mp +  \stackrel{(-)}{\nu_e} + \nu_\mu + \overline{\nu}_\mu$, 
in which the initial neutrino flavor mix is 
identical to the one in atmospheric neutrinos: 
\be \label{eq:120}
\Phi_e^0 : \Phi_\mu^0 : \Phi_\tau^0 = 1 : 2 : 0~.
\ee
If the dominant decay process is 
$p + \gamma \rightarrow \pi^+ + X$ then the flavor mix is 
still $1 : 2 : 0$, 
however the initial $\overline{\nu}_e$ flux is absent. 
This can in principle 
be checked by taking advantage of the Glashow resonance 
reaction $\nuebar+e^-\rightarrow W^-$~\cite{LP,GR1,GR}, 
which occurs at an incident $\nuebar$~energy of $6.3$ PeV.

\item The muons may lose energy 
so that the $\stackrel{(-)}{\nu_e}$ flux is depleted at the energies of 
interest. This can happen in a variety of ways 
\cite{010c,010mod,010b,010ba,010a,lipari,kachneu}; 
e.g., muons can lose energy in strong magnetic fields, or get absorbed in matter. 
In these cases the initial flavor composition is simply 
\be \label{eq:010}
\Phi_e^0 : \Phi_\mu^0 : \Phi_\tau^0 = 0 : 1 : 0~
\ee
without any electron or tau neutrinos. Specific models have been discussed.
In general, one expects that any 
pion source will have a transition to a muon-damped composition, 
with the transition energy depending on the source properties.
Qualitatively, muon damping occurs when the muon's energy-loss length 
$\sim\frac{E}{dE/dx}$ is shorter than its decay length 
$(E/m_\mu)\, c  \tau_0$,
i.e., when $dE/dx \gs c \tau_0/m_\mu$.

\item The third case is that of sources which emit dominantly neutrons. 
These neutrons originate 
in the photo-dissociation of heavy nuclei.
The decays of the neutrons give rise to an initial pure ``$\beta$-beam''  
of $\overline{\nu}_e$~\cite{100a}, i.e., 
\be \label{eq:100}
\Phi_e^0 : \Phi_\mu^0 : \Phi_\tau^0 = 1 : 0 : 0~,
\ee
with no $\nu_e$ or any muon or tau neutrinos. 
\end{itemize}

One other expected source of both pion-decay neutrinos and 
neutron-decay neutrinos,
in separated energy regions~\cite{Engel:2001hd}, is the GZK nucleonic reaction 
$p_{\rm CR} + \gamma_{\rm CMB} \rightarrow \Delta^+ \rightarrow n + \pi^+$. 
The subsequent pion and muon decays produce neutrinos with energies 
about 20 times (i.e., $m_\pi/m_N\times 1/4$) below 
$E_{\rm GZK}\sim 5\times 10^{19}$~eV,
while neutron decay produces $\nuebar$'s with energies about a thousand times 
(i.e., $\beta$-decay $Q$-value/$m_n$) below $E_{\rm GZK}$.

We will emphasize in the present work that the above three flavor mixes 
in Eqs.~(\ref{eq:120}, \ref{eq:010}, \ref{eq:100}) are idealizations. 
Realistically, one should expect deviations from 
these simple flux compositions. 
These deviations should be taken into account in analyses to avoid 
incorrect conclusions about the violation/conservation of CP, the 
octant of $\theta_{23}$, or the magnitude of $|U_{e3}|$.
We shall outline the need for care with several examples.

This paper is built up as follows. 
In Section~\ref{sec:formalism} we discuss an expansion of the mixing 
probabilities and flux ratios 
up to second order (i) in the small 
parameters $\sin^2 \theta_{23} - \frac 12$ and $|U_{e3}|$, and eventually,
(ii) in the small parameter $\sin^2 \theta_{12} - \frac 13$. 
In Section \ref{sec:uncertain} we assess the validity of 
the idealized and often used initial flavor compositions given in 
Eqs.~(\ref{eq:120}, \ref{eq:010}, \ref{eq:100}).
We illustrate the effects of impure neutrino mixes with 
various examples. We sum up and conclude in Section \ref{sec:concl}.

\section{\label{sec:formalism}Mixing Probabilities and Flux Ratios 
up to Second Order}
We discuss in this Section approximate formulae for the 
mixing probabilities and flux ratios, and
general properties of the relevant flux ratios.
For illustration and insight, we will expand formulas in terms of small parameters,
but we use the exact expressions for 
all plots of flux ratios presented in this paper.

An expansion up to second order 
in the small parameters\footnote{The importance of second 
order terms has also been mentioned in Ref.~\cite{MO}.} 
reveals universal correction terms~\cite{PRW}.
We will expand first in terms of the two parameters 
$\sin^2 \theta_{23} - \frac 12$ and $|U_{e3}|$,
related to breaking of the 
$\numu\leftrightarrow\nutau$~symmetry~\cite{mutausym}.
To reveal the dependence on these two parameters, 
it is sometimes useful to fix the solar neutrino mixing 
with the phenomenologically valid relation $\sin^2 \theta_{12} = \frac 13$.

\subsection{\label{sec:gen}Mixing Probabilities}

\subsubsection{Expansion in Terms of 
{\boldmath $|U_{e3}|$ and $\epsilon\equiv\frac{\pi}{4}-\theta_{23}$} }
The distances to most high energy neutrino sources are 
quite large compared to oscillation lengths 
$\lambda_{jk}=4\pi\,E/\Delta m^2_{jk}$, where 
$\Delta m^2_{jk} = m_j^2 - m_k^2$ is the mass-squared difference 
and $E$ the neutrino energy. 
Even for energies as high as the GZK-cutoff $\sim 5\times 10^{19}$~eV,
this is so.
Consequently, the terms involving the mass-squared 
differences in the oscillation probabilities are effectively averaged out; 
oscillation probabilities are reduced to mixing probabilities.
The $\nu_\alpha\leftrightarrow\nu_\beta$ mixing probabilities are 
\be \label{eq:Pab}
P_{\alpha \beta} = \sum\limits_i |U_{\alpha i}|^2 \, 
|U_{\beta i}|^2~.
\ee
Starting from an initial flux composition 
$\Phi_e^0 : \Phi_\mu^0 : \Phi_\tau^0$, 
the measurable neutrino flux at Earth is given by 
\be \label{eq:flux}
\Phi_\alpha = \sum\limits_\beta P_{\alpha \beta}\, \Phi_\beta^0\,.
\ee
If tribimaximal mixing (given in Eq.~(\ref{eq:tbm})) 
is assumed, then the flavor-propagation probabilities are simply 
\be \label{eq:Ptbm}
P_{\rm TBM}  = \frac{1}{18}
\left( 
\bad
10 & 4 & 4 \\
 4 & 7 & 7 \\
 4 & 7 & 7  
\ea
\right)~.
\ee 
The flavor ordering for both column and row indices of this symmetric matrix are 
$(e,\ \mu,\ \tau)$.

Neglecting the small parameters $\pi/4-\theta_{23}$ and $\theta_{13}$, 
the probabilities $P_{\alpha \beta}$ (and therefore the observable 
flux ratios) are functions solely of the solar neutrino mixing angle.
The exact expressions for the mixing probabilities (given in the Appendix) 
are fourth order polynomials in the $\sin \theta_{ij}$. 
It is therefore more useful to expand the formulae in terms of small 
parameters and truncate after the quadratic terms. 
We will first expand in terms of 
\be \label{eq:eps}
|U_{e3}|~\mbox{ and } \epsilon \equiv \frac{\pi}{4} - \theta_{23} = 
\frac 12 - \sin^2 \theta_{23} + {\cal O}(\eps^3)~.
\ee 
The explicit PDG parametrization of the PMNS matrix is given 
in the Appendix. Writing the result in matrix form yields: 
\bea \label{eq:res1a}
P \equiv \D 
\left( 
\bad 
P_{ee} & P_{e \mu} & P_{e \tau} \\
P_{e \mu} & P_{\mu \mu} & P_{\mu \tau} \\
P_{e \tau} & P_{\mu \tau} & P_{\tau \tau} 
\ea
\right) 
\simeq   
\left( 
\bad
1 - 2 \, c_{12}^2 \, s_{12}^2 &   c_{12}^2 \, s_{12}^2 
&  c_{12}^2 \, s_{12}^2 \\
 c_{12}^2 \, s_{12}^2 & \frac 12 \, (1 - c_{12}^2 \, s_{12}^2) & 
\frac 12 \, (1 - c_{12}^2 \, s_{12}^2)  \\
 c_{12}^2 \, s_{12}^2 & \frac 12 \, (1 - c_{12}^2 \, s_{12}^2) 
& \frac 12 \, (1 - c_{12}^2 \, s_{12}^2) 
\ea
\right)  \\[0.2cm] \D 
\\
- \frac 12 \, (1 - 2 \, c_{12}^2 \, s_{12}^2 )  \, |U_{e3}|^2  
\left( 
\barrr
4 & -2 & -2 \\ 
-2 & 1 & 1 \\ 
-2 & 1 & 1 
\ea
\right) 
+ \Delta 
\left( 
\barrr
0 & 1 & -1 \\ 
1 & -1 & 0 \\ 
-1 & 0 & 1
\ea
\right) 
+ 
\frac 12 \, \overline{\Delta}^2
\left( 
\barrr
0 & 0 & 0 \\ 
0 & 1 & -1 \\ 
0 & -1 & 1
\ea
\right)~,
\eea
As it must, every row and every column sums to 1. 
The correction linear in $|U_{e3}|$ and $\epsilon$ is 
given by \cite{xing,WR} 
\be \label{eq:Delta}
\Delta \equiv \frac 14 \, \sin 4 \theta_{12} \, 
\cos \delta ~ |U_{e3}| + \frac 12 \, \sin^2 2 \theta_{12} 
\, \epsilon~,
\ee
and the correction quadratic in $|U_{e3}|$ and $\epsilon$ is
\begin{eqnarray} \label{eq:Delta2} %\nonumber 
\overline{\Delta}^2 &\equiv& 
\sin^2 2 \theta_{12} \, \cos^2 \delta ~ |U_{e3}|^2 
+ 4 \, (1 - \cos^2 \theta_{12} \, \sin^2 \theta_{12}) \, \epsilon^2 %\\[0.2cm]
 - \sin 4 \theta_{12} \,  \cos \delta ~ |U_{e3}| \, \epsilon \\[0.2cm]
&=& \left(\sin 2 \theta_{12} \cos  \delta ~ |U_{e3}| - 
\epsilon \,\cos 2\theta_{12}  \right)^2   + 3\,\epsilon^2 \nonumber ~.
\end{eqnarray}
The latter is a new result, as is the observation that 
the universal second order correction $\overline{\Delta}^2$ is 
positive semidefinite.
With the current $1\sigma$ and ($3\sigma$) ranges of the 
oscillation parameters, one finds the following 
ranges for $\Delta$ and $\overline{\Delta}^2$:
\be \label{eq:ranges}
(-0.104)\ -0.043 \le \Delta\le 0.069\ (0.117)\,,
\quad {\rm and}\quad
\overline{\Delta}^2 \le 0.061\ (0.179)~.
\ee
It is seen that the second order correction 
can exceed the first order correction. Consequently, the first 
order correction to the flux ratios alone is not sufficient to accurately 
describe the phenomenology.
The second order correction needs inclusion in analytical studies. 
The reason for the large second order term are the sizable numerical 
coefficients, especially the one in front of $\epsilon^2$ 
(we have checked that the higher order terms have smaller coefficients).  

Explicitly, the individual mixing probabilities are 
\begin{eqnarray} \label{eq:probs}
%\bad 
P_{ee} & \simeq & \nonumber 
(1 - 2 \, c_{12}^2 \, s_{12}^2) (1 - 2 \, |U_{e3}|^2) ~,\\[0.2cm] 
P_{e \mu} & \simeq  & \nonumber 
c_{12}^2 \, s_{12}^2 + \Delta + (1 - 2 \, c_{12}^2 \, s_{12}^2) \, |U_{e3}|^2 
~,\\[0.2cm]
P_{e \tau} & \simeq  & \nonumber 
c_{12}^2 \, s_{12}^2 - \Delta + (1 - 2 \, c_{12}^2 \, s_{12}^2) \, |U_{e3}|^2
~,\\[0.2cm]
P_{\mu \mu} & \simeq & %\nonumber 
\frac 12  \left(1 - c_{12}^2 \, s_{12}^2\right) - \Delta + \frac 12 \, 
\overline{\Delta}^2 - \frac 12 \, (1 - 2 \, c_{12}^2 \, s_{12}^2) 
\, |U_{e3}|^2~,\\[0.2cm]
P_{\mu \tau} & \simeq &  \nonumber 
\frac 12  \left(1 - c_{12}^2 \, s_{12}^2\right) - \frac 12 \, 
\overline{\Delta}^2 - \frac 12 \, (1 - 2 \, c_{12}^2 \, s_{12}^2) 
\, |U_{e3}|^2~,\\[0.2cm] 
P_{\tau \tau} & \simeq & \nonumber 
 \frac 12  \left(1 - c_{12}^2 \, s_{12}^2\right) + \Delta + \frac 12 \, 
\overline{\Delta}^2 - \frac 12 \, (1 - 2 \, c_{12}^2 \, s_{12}^2) 
\, |U_{e3}|^2~.
\end{eqnarray}
This may be compared with the same in the case of exact 
tribimaximal mixing, given in Eq.~(\ref{eq:Ptbm}). 
The explicit terms depending on $|U_{e3}|^2$ 
are small, because the maximal value of $(1 - 2 \, c_{12}^2 \, s_{12}^2) 
\, |U_{e3}|^2$ is 0.011 (0.031) at $1\sigma$ (3$\sigma$), 
which is well below the maximal values of $|\Delta|$ and 
$\overline{\Delta}^2$.
In addition, for the initial flavor 
composition of $\Phi_e^0 : \Phi_\mu^0 : \Phi_\tau^0 = 1 : 2 : 0$,
the terms with explicit $|U_{e3}|^2$ cancel in the fluxes. 
Consequently, the  first and second 
order corrections $\Delta$ and $\overline{\Delta}^2$
attain a universal status.
$P_{ee}$ and $P_{\mu \tau}$ receive only quadratic corrections, 
where the correction to $P_{ee}$ depends only\footnote
{This remains true 
for $P_{ee}$ even when oscillations do not average out.} 
on $\theta_{13}$ and 
not on $\delta$ or $\theta_{23}$. 
Note that the universal second order term 
shows up only in $P_{\mu\mu}$, $P_{\mu\tau}$ and $P_{\tau\tau}$, 
i.e., in the $\mu$--$\tau$ sector. This will be of interest later 
when we discuss neutron beam sources, in which these probabilities,  
and therefore $\overline{\Delta}^2$, do not show up in the flux ratios. 

Inserting the tribimaximal value $\sin^2 \theta_{12} = \frac 13$ in the expressions 
for $\Delta$ and $\overline{\Delta}^2$ gives a more illustrative 
result.  We define the resulting universal corrections as 
\bea \D  \label{eq:Deltbm}
\Delta_{\rm TBM} = \frac 19 
\left( 
\sqrt{2} \, \cos \delta \, |U_{e3}| + 4 \, \epsilon
\right)~,\\[0.2cm] \D 
\overline{\Delta}^2_{\rm TBM} = \frac 49 
\left( 
2 \cos^2 \delta \, |U_{e3}|^2 + 7 \, \epsilon^2 - \sqrt{2} 
\, \cos \delta \, |U_{e3}| \, \epsilon
\right)~.
\eea
If for instance $\theta_{13} = 0$ and $|\epsilon|$ is larger than $1/7$,  
then $\overline{\Delta}_{\rm TBM}^2$ exceeds $|\Delta_{\rm TBM}|$. 
Note that $\Delta = \Delta_{\rm TBM}$ plus quadratic terms 
and that $ \overline{\Delta}^2 = \overline{\Delta}^2_{\rm TBM}$ plus 
cubic terms. 
Hence, the ranges of $\Delta_{\rm TBM}$ and 
$\overline{\Delta}_{\rm TBM}^2$ are changed little from the ranges 
of $\Delta$ and $\overline{\Delta}^2$ given in Eqs.~(\ref{eq:ranges}).\\

It is also interesting to consider the allowed range of 
the second order correction $\overline{\Delta}^2$ in the case of a vanishing 
first order correction $\Delta$. We find that, at $1\sigma$ and ($3\sigma$),
\be \label{eq:Del0Del2}
\mbox{ if } \Delta = 0\,,\ {\rm then}\ 
\overline{\Delta}^2 \le 0.029\ (0.093)\,.
\ee
In particular if the oscillation parameters lie outside their current 
$1\sigma$ ranges, then the mixing probabilities and flux ratios can deviate 
up to ten percent from their tribimaximal values even if the 
first order correction vanishes.

\begin{figure}[ht]
\begin{tabular}{lr}
\includegraphics[width=7cm,height=5cm]{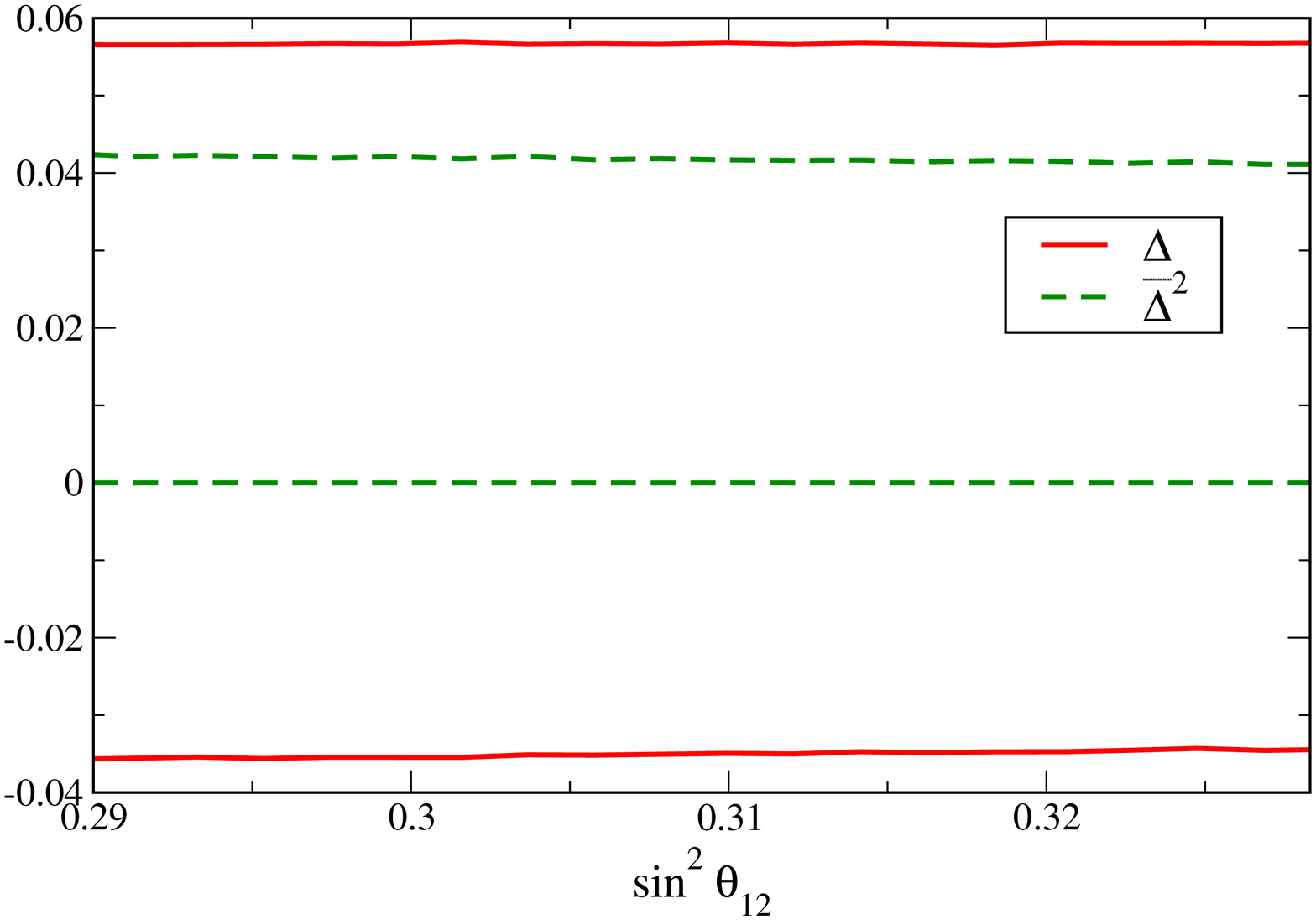} &
\includegraphics[width=7cm,height=5cm]{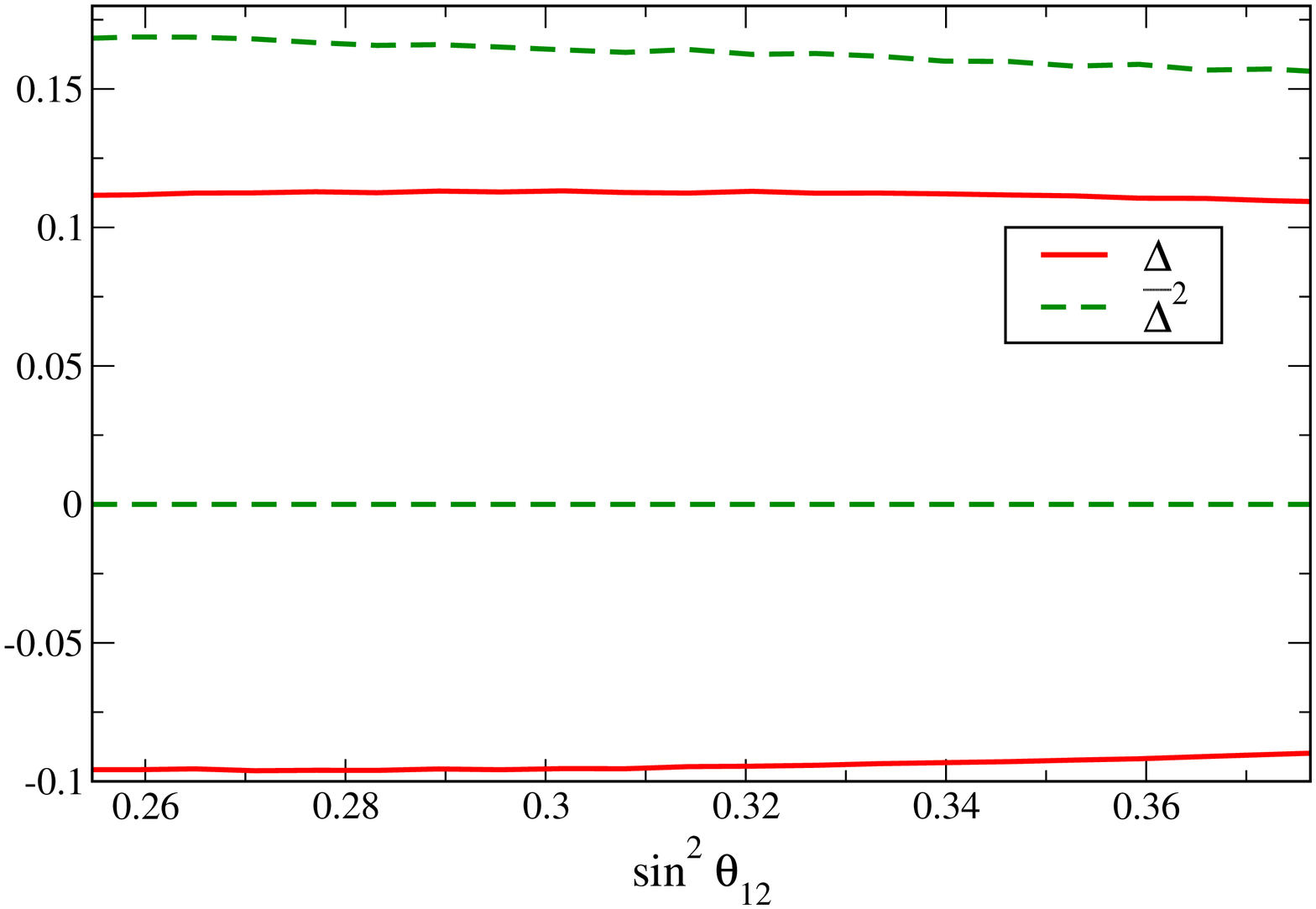}\\
\includegraphics[width=7cm,height=5cm]{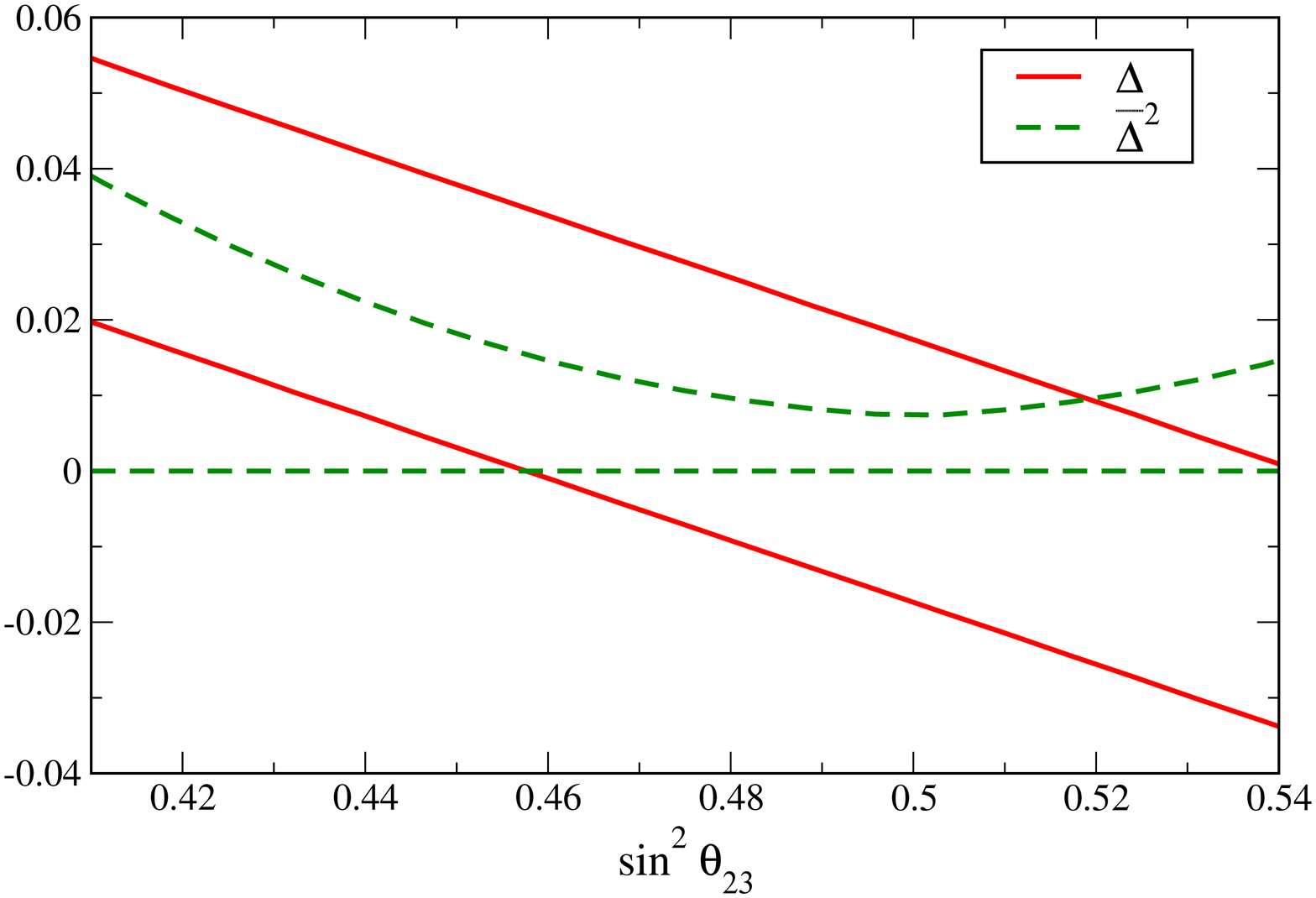} &
\includegraphics[width=7cm,height=5cm]{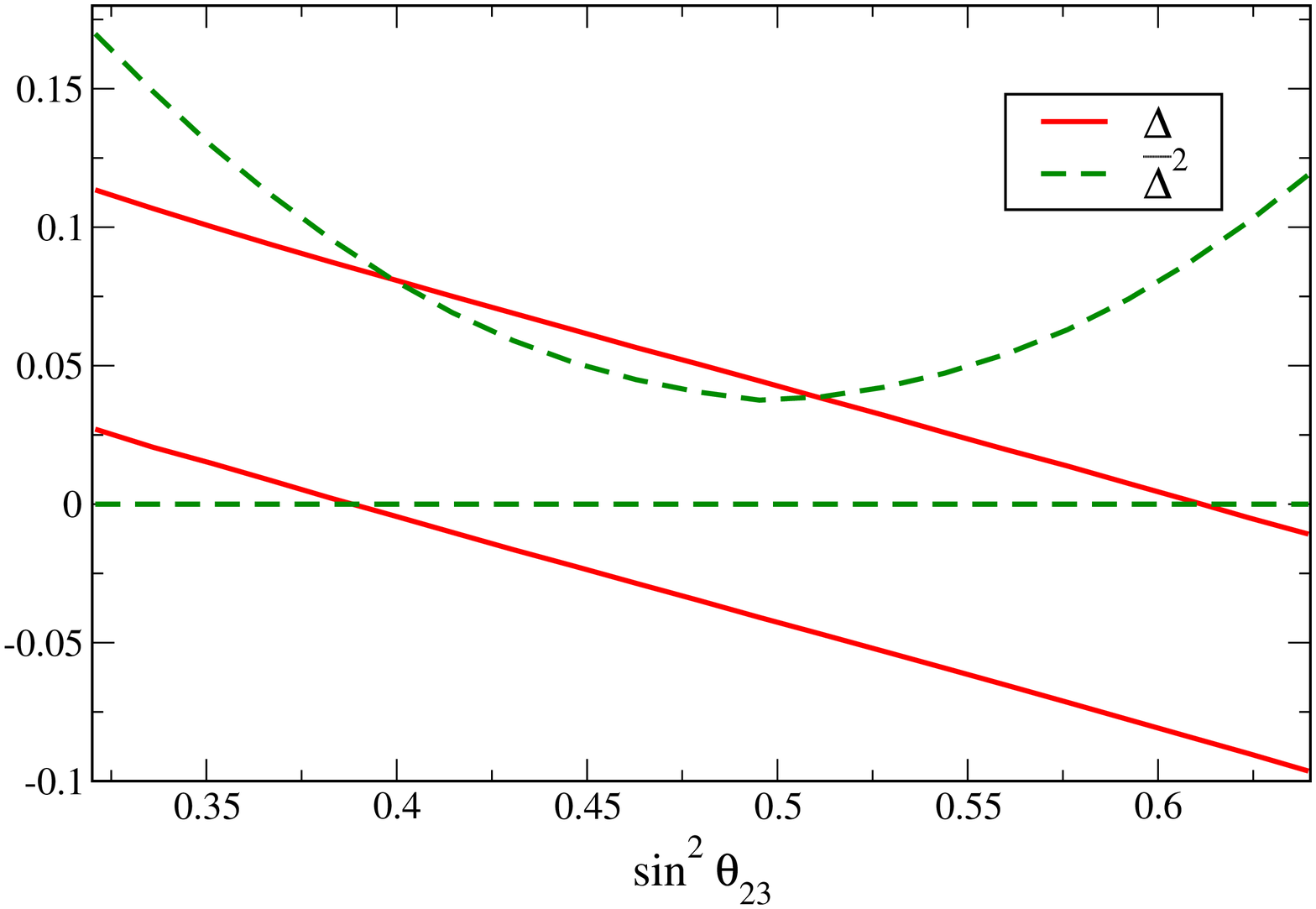} \\
\includegraphics[width=7cm,height=5cm]{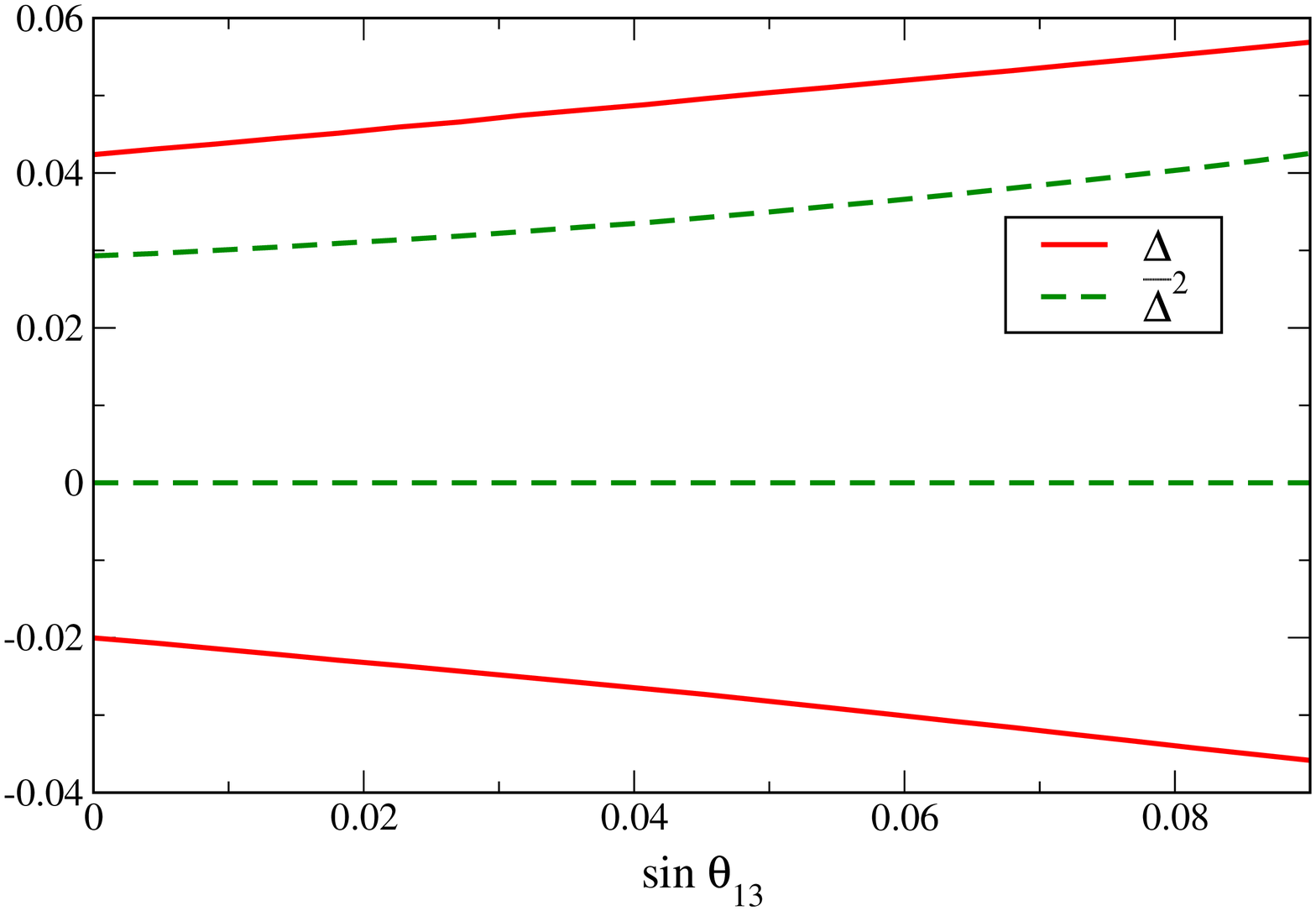} &
\includegraphics[width=7cm,height=5cm]{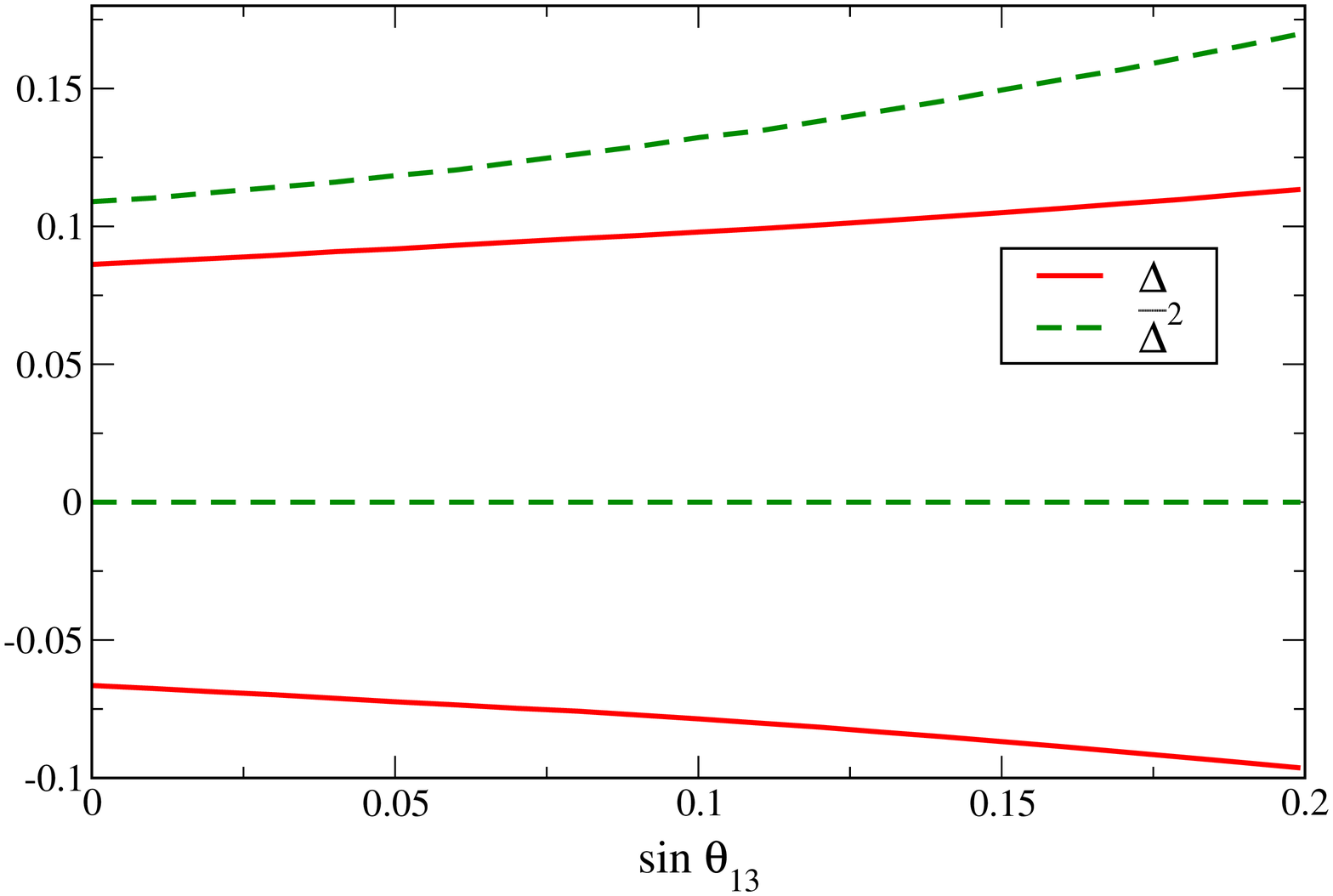} \\
\includegraphics[width=7cm,height=5cm]{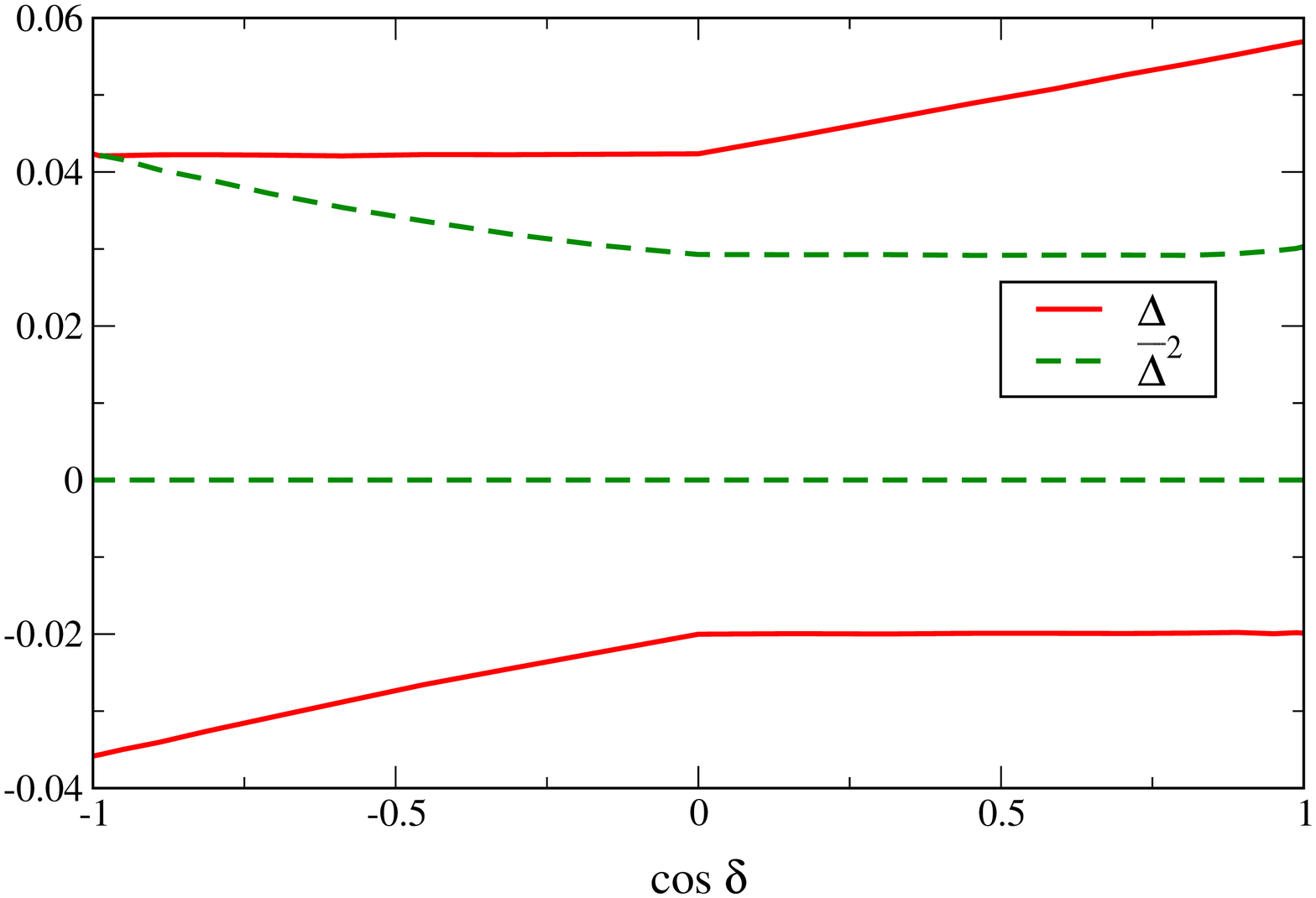} &
\includegraphics[width=7cm,height=5cm]{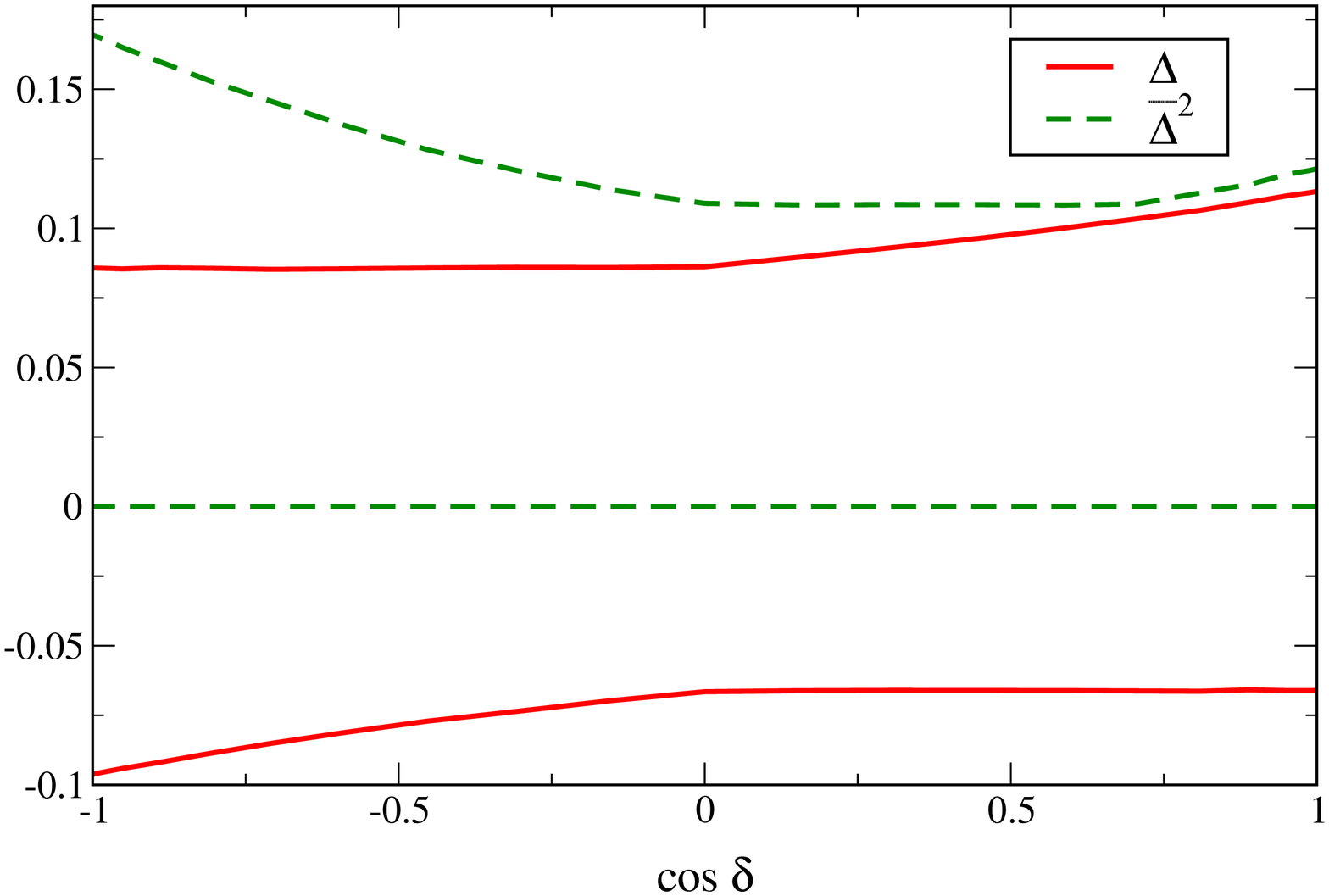}
\end{tabular}
\caption{\label{fig:Deltamima}The minimal and maximal allowed values of the 
universal first and second order parameters $\Delta$ and 
$\overline{\Delta}^2$ as a function of the neutrino mixing parameters. 
The observables not specified in the horizontal axis were varied over 
their currently allowed $1\sigma$ (left) 
and $3\sigma$ (right) ranges.}
\end{figure}

Note that in the definition of $\Delta$ in Eq.~(\ref{eq:Delta}), 
the factor in front of $\frac 12 - \sin^2 \theta_{23}$ is larger 
and has a smaller range 
than the factor in front of $|U_{e3}| \, \cos \delta$. 
To be precise, for the 
allowed $3\sigma$ range of solar neutrino mixing, 
$\frac 14  \sin 4 \theta_{12}$ ranges from 0.12 to 0.21, whereas 
$\frac 12  \sin^2 2 \theta_{12} $ ranges from 0.38 to 0.48. Consequently, 
the sensitivity to deviations from maximal 
atmospheric neutrino mixing 
is better than the sensitivity to deviations from $|U_{e3}|=0$.
The latter is also mitigated by 
the dependence on the unknown CP phase $\delta$. 
Comparing $\overline{\Delta}^2$ from Eq.~(\ref{eq:Delta2}) with 
the first order parameter $\Delta$, we note that 
the factor in front of $\epsilon^2$ in $\overline{\Delta}^2$ 
is larger than the one in front of 
$\epsilon$ in $\Delta$. Also in $\overline{\Delta}^2$,
the factor in front of $\eps^2$ has a smaller range 
than the one in front of $|U_{e3}|^2$: 
$\sin^2 2 \theta_{12}$ lies between 
0.75 and 0.64, whereas $4 \, (1 - \cos^2 \theta_{12} \, 
\sin^2 \theta_{12})$ 
ranges from 3.06 to 3.25, when the allowed $3\sigma$ values of 
$\theta_{12}$ are inserted. Again, sensitivity to 
deviations from maximal atmospheric neutrino mixing is favored over 
sensitivity to $|U_{e3}|$. 
In Fig.~\ref{fig:Deltamima} we show the minimal and maximal allowed values of 
$\Delta$ and $\overline{\Delta}^2$ as a function of the neutrino mixing 
parameters. There is almost no dependence on $\theta_{12}$. 
The strongest dependence is on $\theta_{23}$. 
The statistically weak preference for $\sin^2\theta_{23} = 0.45$ would 
mean for $\theta_{13}=0$ that $\Delta \simeq 0.02$ is 
positive and $\overline{\Delta}^2 \simeq 0.008$. 

\begin{figure}[ht]
\begin{tabular}{lr}
\epsfig{file=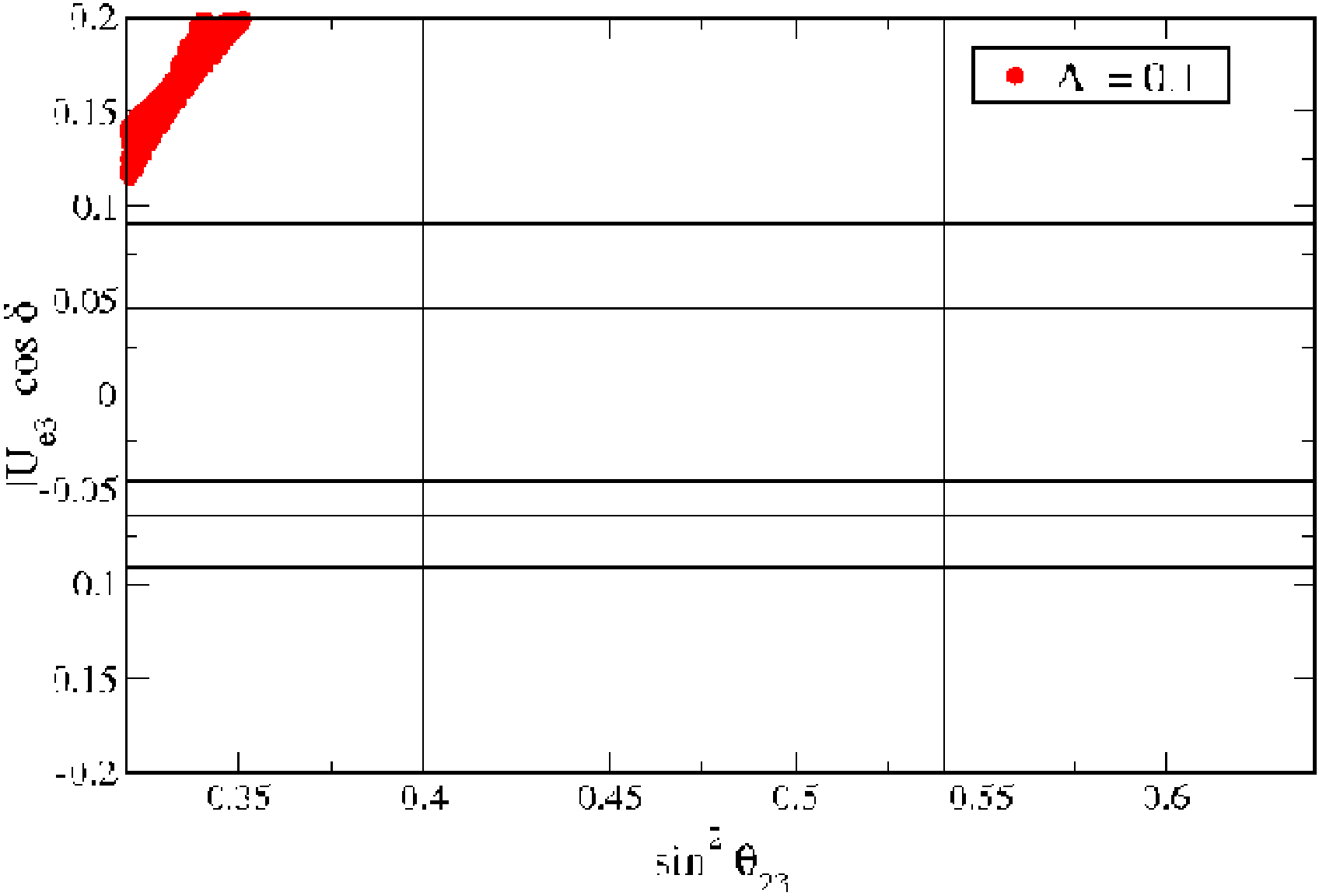,width=7cm,height=5cm} & 
\epsfig{file=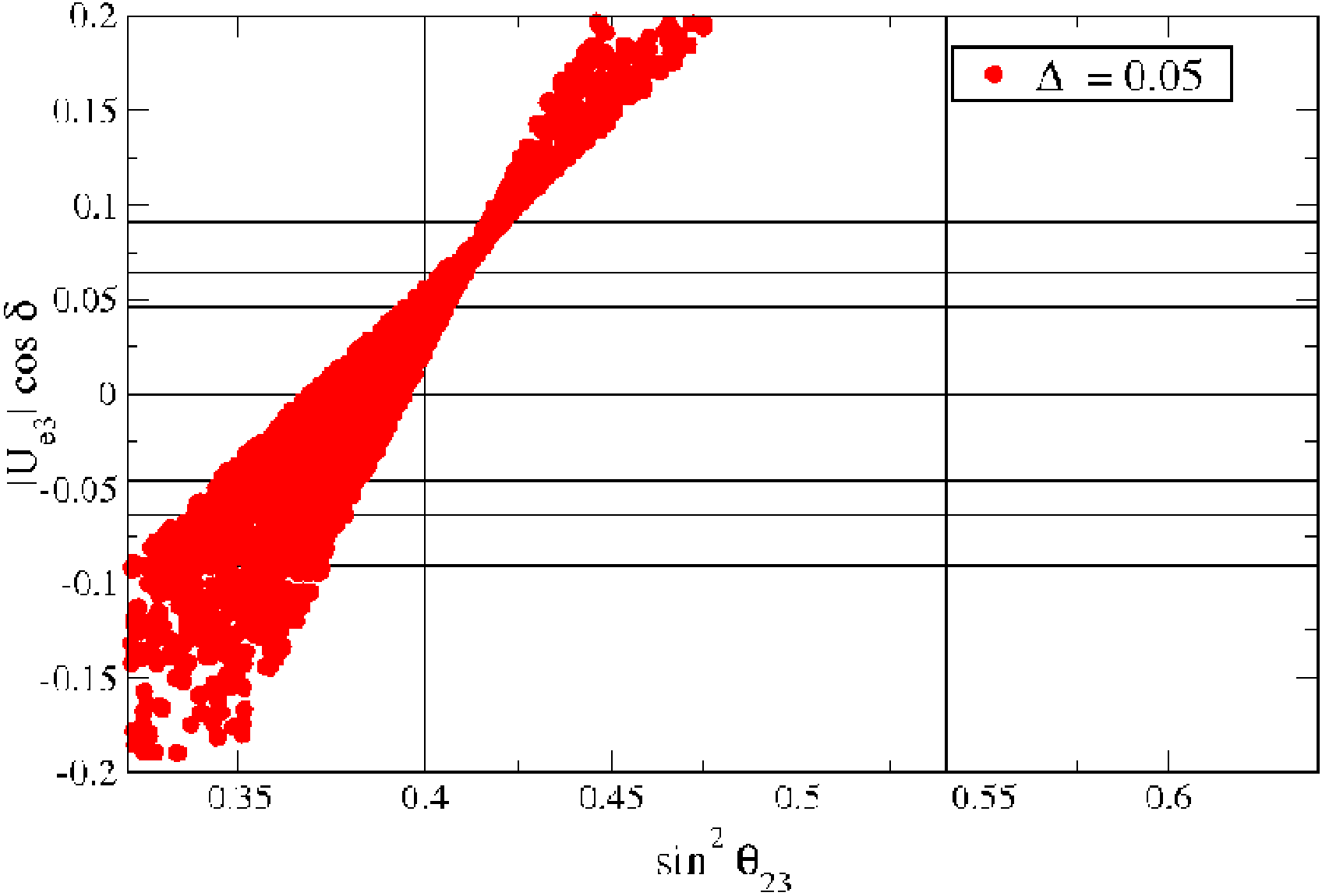,width=7cm,height=5cm} \\ 
\epsfig{file=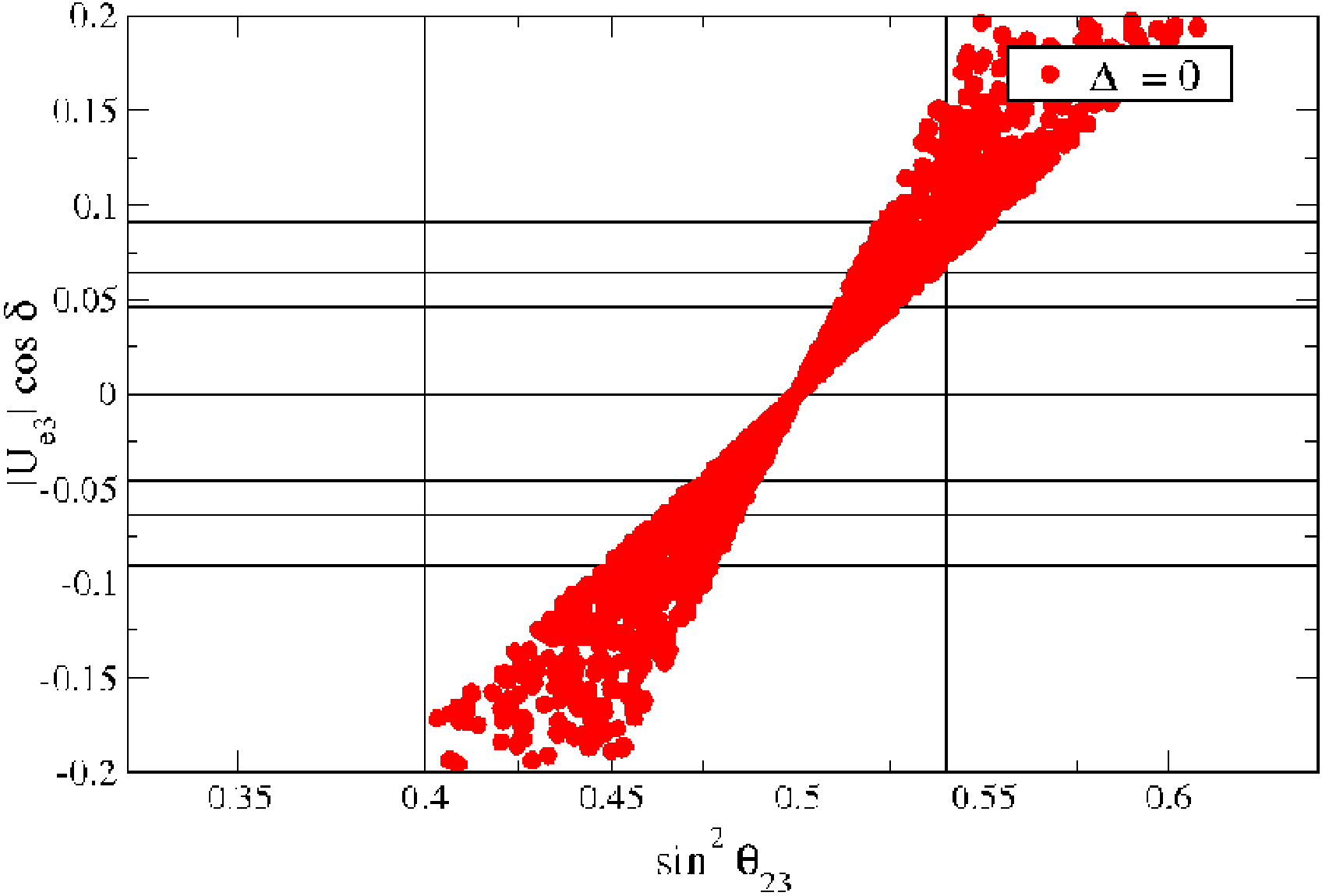,width=7cm,height=5cm} & 
\epsfig{file=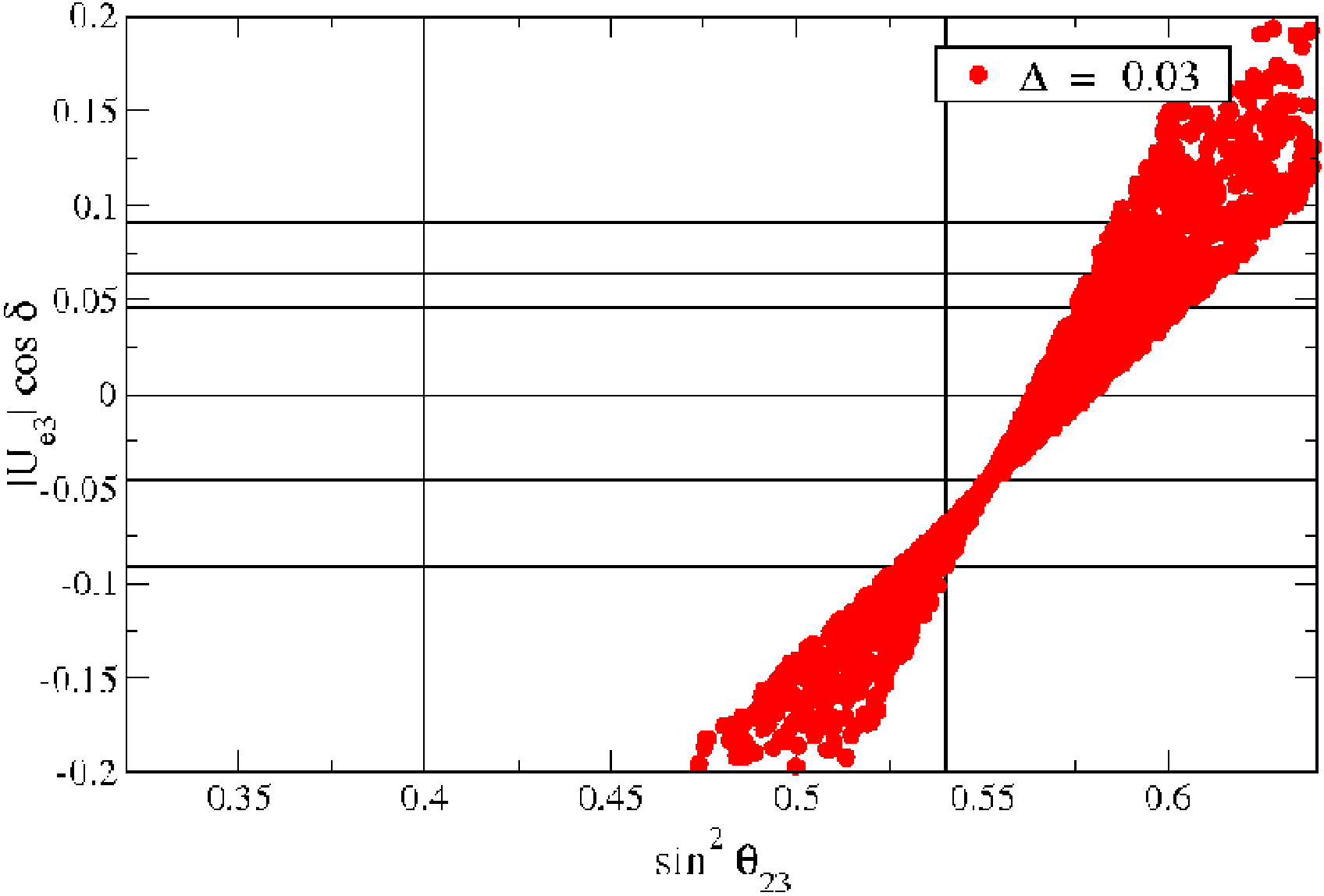,width=7cm,height=5cm} \\  
\epsfig{file=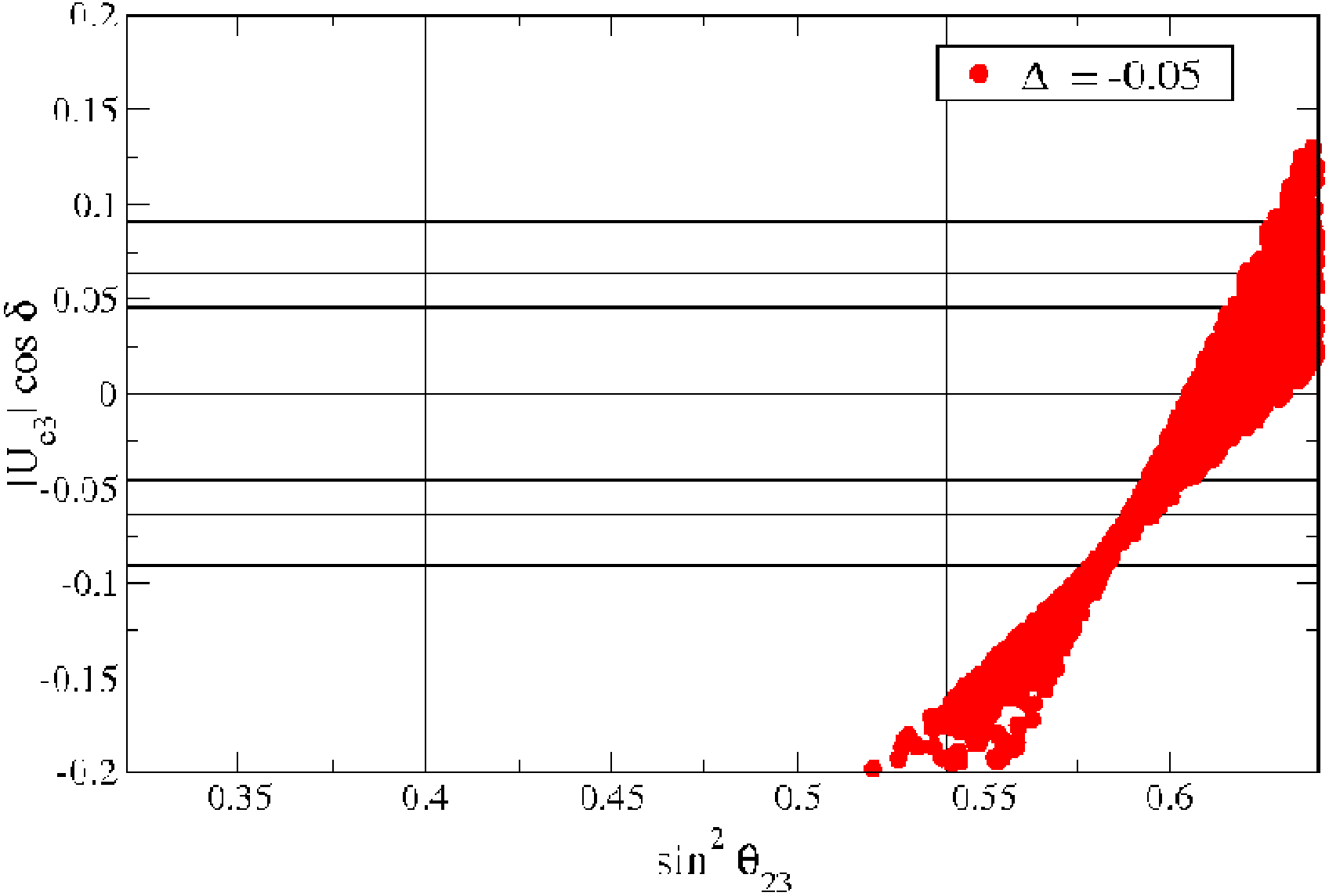,width=7cm,height=5cm}  &  
\epsfig{file=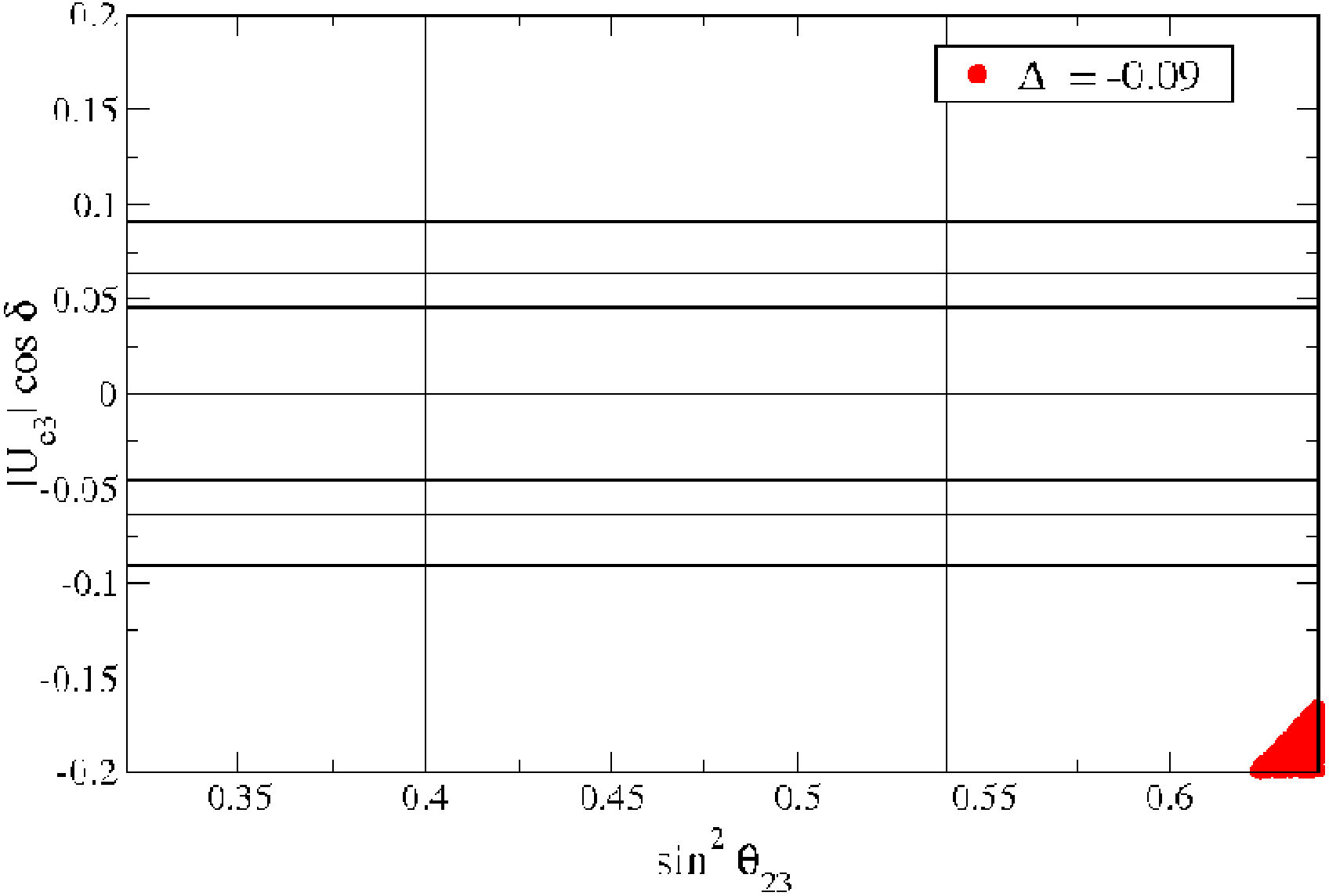,width=7cm,height=5cm} \\
\end{tabular}
\caption{\label{fig:Delta}Distribution of $|U_{e3}| \, \cos \delta$ 
against $\sin^2 \theta_{23}$ if $\Delta$ 
takes certain indicated values. 
Indicated also is the allowed $1\sigma$ range of $\theta_{23}$ and 
of $|U_{e3}| \, \cos \delta$ for (from top to bottom above zero) $\delta=0$, 
$\delta = \pi/4$ and $\delta = \pi/3$. The value $\delta = \pi/2$ means 
$|U_{e3}| \, \cos \delta = 0$.}
\end{figure}

\begin{figure}[ht]
\begin{tabular}{lr}
\epsfig{file=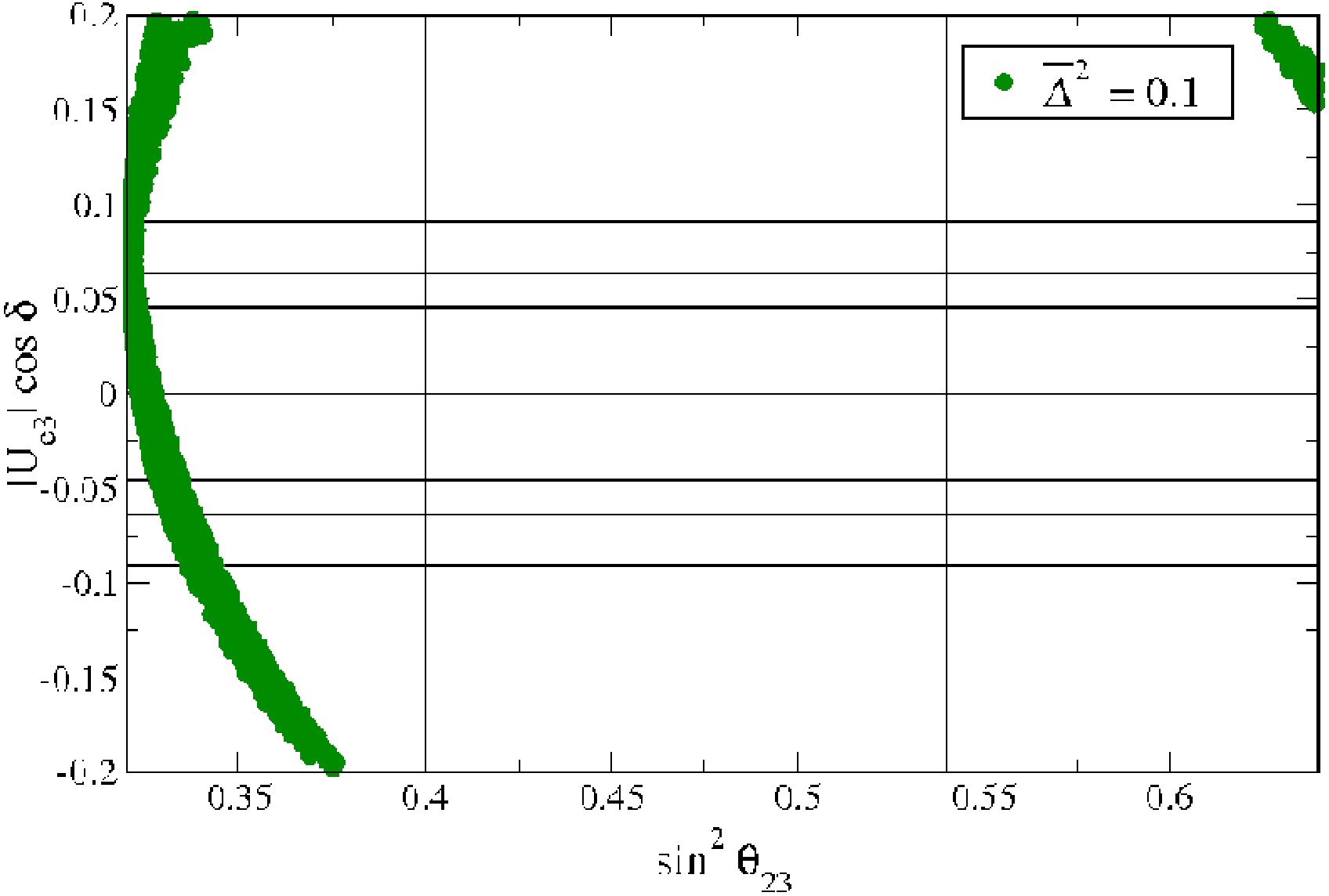,width=7cm,height=5cm} & 
\epsfig{file=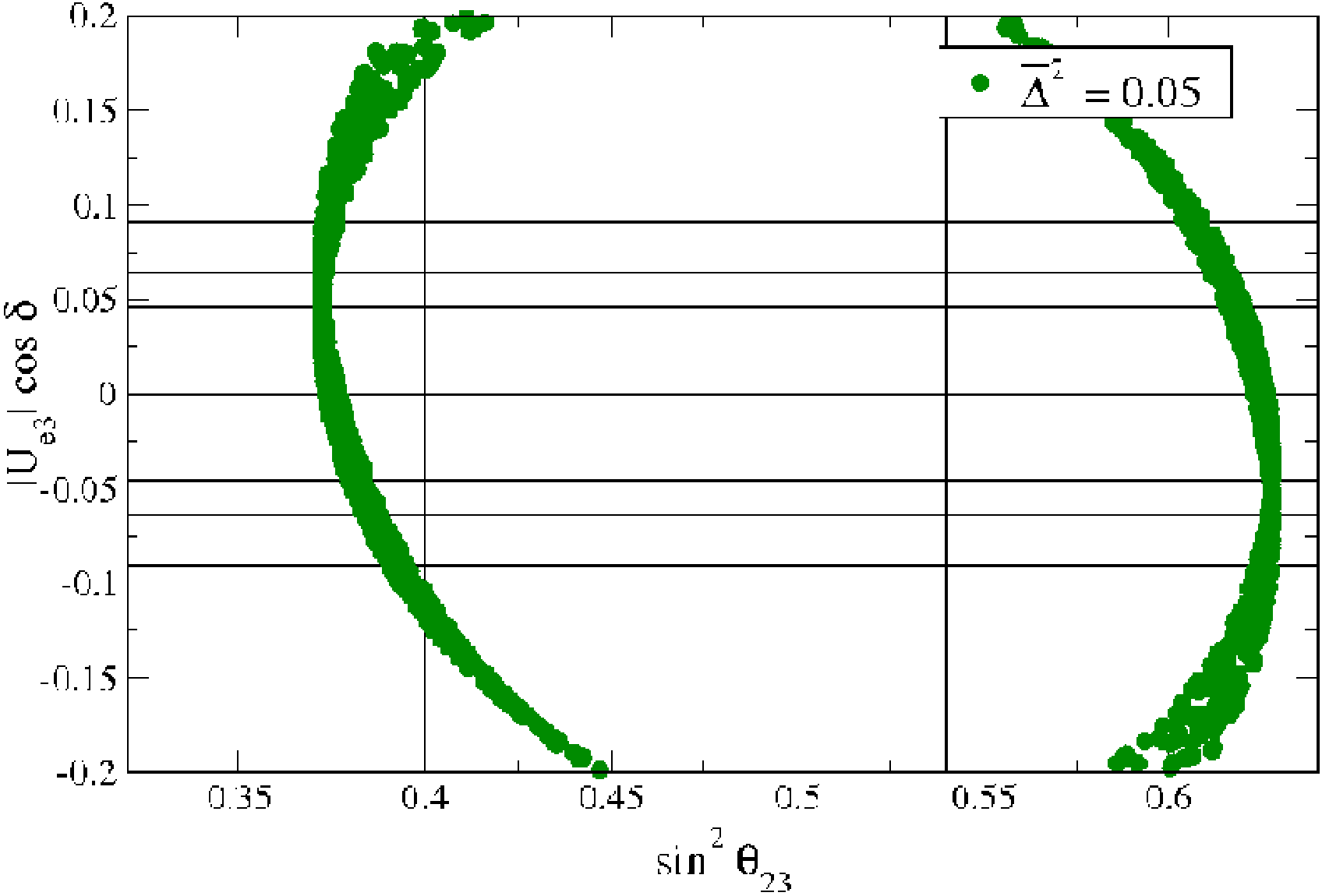,width=7cm,height=5cm} \\
\epsfig{file=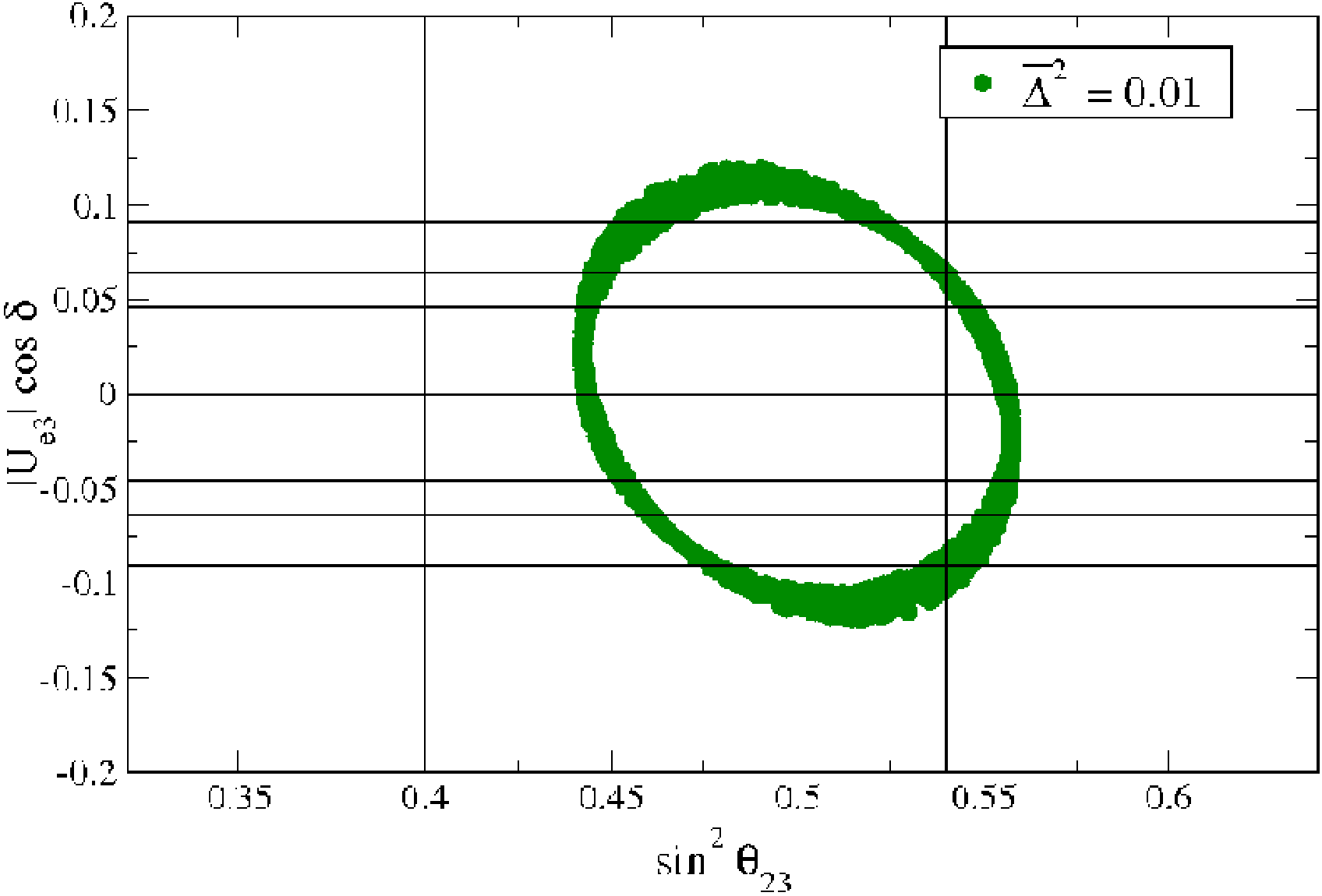,width=7cm,height=5cm} & 
\epsfig{file=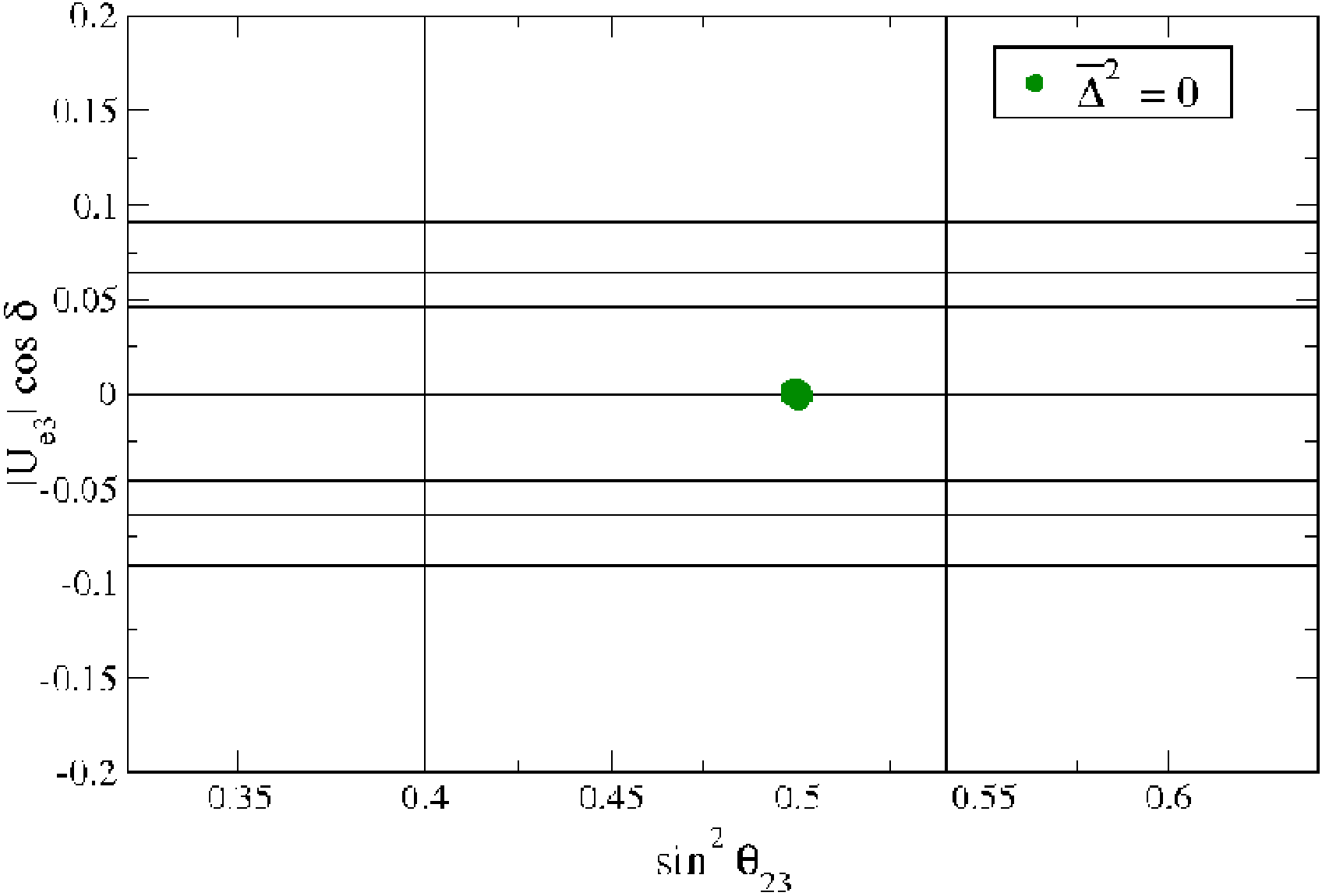,width=7cm,height=5cm} \\  
\end{tabular}
\caption{\label{fig:Delta2}Distribution of $|U_{e3}| \, \cos \delta$ 
against $\sin^2 \theta_{23}$ if $\overline{\Delta}^2$ 
takes certain indicated values. 
Indicated also is the allowed $1\sigma$ range of $\theta_{23}$ and 
of $|U_{e3}| \, \cos \delta$ for (from top to bottom above zero) $\delta=0$, 
$\delta = \pi/4$ and $\delta = \pi/3$. The value $\delta = \pi/2$ means 
$|U_{e3}| \, \cos \delta = 0$.}
\end{figure}

We show in Fig.~\ref{fig:Delta} the distribution of $|U_{e3}| \, \cos \delta$ 
against $\sin^2 \theta_{23}$ for several values of $\Delta$, where we 
allow all mixing angles to vary in their allowed $3\sigma$ ranges. 
We also indicate the $1\sigma$ ranges for $\sin^2 \theta_{23}$ 
and for $|U_{e3}| \, \cos \delta$ if $\delta = 0, \pi/4$ and $\pi/3$. 
Fig.~\ref{fig:Delta2} shows the same for $\overline{\Delta}^2$. 
Only if $\epsilon = |U_{e3}| \, \cos \delta = 0$ can $\overline{\Delta}^2$ 
vanish exactly.
The first order correction $\Delta$ may also vanish, if~\cite{xing,WR} 
\be \label{eq:Delta0}
|U_{e3}| \, \cos \delta = \left(\sin^2 \theta_{23} - \frac 12  \right)
\tan 2 \theta_{12} %\simeq 2\sqrt{2} 
%\left(\sin^2 \theta_{23} - \frac 12  \right)\,;
~.
\ee
The tribimaximal value of $\tan 2 \theta_{12}$ is $2\sqrt{2}$.

\subsubsection{Expansion in terms of 
{\boldmath $|U_{e3}|$, $\eps\equiv\frac{\pi}{4} - \theta_{23}$, 
and $\epsilon' \equiv \sin^2 \theta_{12} - \frac 13$} }
We can simplify the expressions even more when we introduce a third small 
parameter, taking advantage of the closeness of $\sin^2 \theta_{12}$ to 
$\frac 13$. Let us define\footnote
{A related expansion can be found 
in Ref.~\cite{PRW}.}
\be \label{eq:eps1}
\epsilon' \equiv \arcsin \sqrt{\frac 13} - \theta_{12} = 
\frac{3}{2\sqrt{2}}  \left(\frac 13 - \sin^2 \theta_{12} \right) 
+ {\cal O}(\eps'^2)~.
\ee
The best-fit value of $\sin^2 \theta_{12} = 0.32$ corresponds 
to $\epsilon' = 0.0142$. From the previous Subsection it is not 
difficult to obtain the elements of the flavor-propagation matrix $P$. 
They are~\cite{PRW}
\be \label{eq:probs2}
\bad \D 
P_{e e} = \frac{1}{18} \left(10 + 4 \, A \right) ~,& \D 
P_{e \mu} = \frac{1}{18} \left(4 - 2\, A + B \right) ~,& \D 
P_{e \tau} = \frac{1}{18} \left(4 - 2 \, A - B \right) ~,\D \\[0.3cm]
\D P_{\mu \mu} = \frac{1}{18} \left(7 + A - B + C \right) ~,&\D  
P_{\mu \tau} = \frac{1}{18} \left(7 + A - C \right) ~,& \D 
P_{\tau \tau} = \frac{1}{18} \left(7 + A + B + C \right)~,\D 
\ea
\ee
where 
\begin{eqnarray} \label{eq:ABC}\nonumber 
A &=& 2\sqrt{2} \, \epsilon' + 7 \, \epsilon'^2 
- 5 \, |U_{e3}|^2 ~,\\
B &=& 2\left(\sqrt{2} \, |U_{e3}| \, \cos \delta + 4 \, \epsilon 
- 4 \sqrt{2} \, \epsilon \, \epsilon' 
+ 7 \, \epsilon' \, |U_{e3}| \, \cos \delta \right) ~,\\ 
C &=& 4 \left( 
2 \, |U_{e3}|^2 \, \cos^2 \delta + 7 \, \epsilon^2 - \sqrt{2} \, 
\epsilon \, |U_{e3}| \, \cos \delta \nonumber 
\right) ~.
\end{eqnarray}
Note that $C = 9 \, \overline{\Delta}^2_{\rm TBM}$, the latter being 
previously defined in Eq.~(\ref{eq:Deltbm}). 
If $\theta_{23}$ is maximal and $|U_{e3}| \cos \delta$ vanishes, then 
$B = C = 0$ and only $A \neq 0$, unless in addition 
$\sin^2 \theta_{12} = \frac 13$. 
Just as with $\overline{\Delta}^2$, the correction factor $C$ is positive 
semi-definite and appears exclusively in the $\mu$--$\tau$ sector. 
The probability $P_{ee}$ receives contributions only from $A$, whereas 
$P_{\mu\tau}$ is not corrected by $B$. Deviations from 
$\sin^2 \theta_{12} = \frac 13$ show up mainly in $A$. 

At $1\sigma$ and ($3\sigma$), we obtain the ranges of 
these ($A$, $B$, $C$)~parameters.  The ranges are 
\bea
(-0.418)\ -0.126 \le A \le 0.117\ (0.284)\,,\\[0.2cm]
(-1.956)\ -0.778 \le B  \le 1.255\ (2.177)\,,\\[0.2cm]
 0 \le C \le 0.465\ (1.356)\,.
\eea
The tribimaximal values for these parameters are zero.

Both expansions, one in terms of ($\Delta$, $\overline{\Delta}^2$) 
and the other in terms of ($A$, $B$, $C$), are useful. 
To disentangle the dependence on $\theta_{12}$ from the dependences 
on $|U_{e3}|$ and $\theta_{23} - \pi/4$, the expansion defined 
by ($\Delta$, $\overline{\Delta}^2$) may be more helpful; 
the ($\Delta$, $\overline{\Delta}^2$) expansion has
the full dependence on $\theta_{12}$ included in it. 
On the other hand, the range of $\epsilon'$ is smaller than the range 
of $\epsilon$ and $|U_{e3}|$. Therefore, to a good approximation one may set 
$\sin^2 \theta_{12}$ to $\frac 13$. Below, we concentrate on 
the dependences on $|U_{e3}|$ and $\eps\equiv\pi/4-\theta_{23}$.

\subsection{\label{sec:fluxes} Flavor Ratios}
With the help of the probabilities 
in Eq.~(\ref{eq:probs}) and a given initial flavor composition 
$\Phi_e^0 : \Phi_\mu^0 : \Phi_\tau^0$ it is easy to obtain approximate 
formulae for flux compositions or ratios. We will focus here 
on the ratio of muon neutrinos to the total flux $\Phi_{\rm tot}$,
and on 
the ratio of $\nu_e$ to $\nu_\tau$ \cite{ratio1,ratio2}: 
\bea \label{eq:ratios} \D
T \equiv \frac{\Phi_\mu}{\Phi_{\rm tot}} 
~~\mbox{ and }~~
\D R \equiv \frac{\Phi_e}{\Phi_\tau} 
~.
\eea
Muon neutrinos with energies $\gsim 10^2$~GeV 
can be identified via muons emerging from the shower. 
Electromagnetic showers from $\nu_e$ charged current 
reactions may be identifiable.
At energies $\gsim 10^6$~GeV, $\nu_\tau$ 
can be identified by double-bang or lollipop signatures. Sometimes the ratio 
$\Phi_\mu/(\Phi_e + \Phi_\tau)$ is discussed in the literature. It 
is trivially related to our ratio $T$, being equal to $T/(1 - T)$.

In the 6.3~PeV energy region, the $\nuebar$ flux becomes easy to measure,
due to the enhanced rate of the ``Glashow resonance''~\cite{LP,GR,GR1}.  
The reaction is $\nuebar+e^-\rightarrow W^-$.
Thus we also define the ratio 
\be
Q \equiv \frac{\Phi_{\overline e}}{\Phi_{\rm tot}} ~,
\ee
where $\Phi_{\overline e}$ is the $\nuebar$ flux.

In the rest of this paper, we will focus mostly on the $T$ and $R$ ratios. 
With the formulas we have given in Section~\ref{sec:gen}, 
all other flux ratios are easy to obtain.

\subsubsection{\label{sec:120}Pion Sources}
Pion-sources present a initial neutrino flux ratios  
$\Phi_e^0 : \Phi_\mu^0 : \Phi_\tau^0 = 1 : 2 : 0$. 
It holds in this case that at Earth, 
$\Phi_e = \frac 13 \, \Phi_{\rm tot} \, (P_{ee} + 2 \, P_{e \mu})$, 
$\Phi_\mu = \frac 13 \, \Phi_{\rm tot} \, (P_{e\mu} + 2 \, P_{\mu \mu})$, 
and 
$\Phi_\tau = \frac 13 \, \Phi_{\rm tot} \, (P_{e\tau} + 2 \, P_{\mu \tau})$, 
These simplify to 
\be \label{eq:flux120} 
\mbox{pion sources: }\quad 
\left(\Phi_e : \Phi_\mu : \Phi_\tau \right)= 
\left(1 + 2 \, \Delta : 1 - \Delta + \overline{\Delta}^2 
: 1 - \Delta - \overline{\Delta}^2\right)~.
\ee
Here we have used the ($\Delta$, $\overline{\Delta}^2$) expansion.  
The single (and small) terms containing only $|U_{e3}|^2$, which are 
present besides $\Delta$ and $\overline{\Delta}^2$ in Eq.~(\ref{eq:probs}),  
drop out of these expressions. 
If one uses the ($A$, $B$, $C$) expansion, which includes $\epsilon'$ as well, 
then one finds that $A$ drops out of the ratios, to leave: 
\be \label{eq:flux120a} 
\mbox{pion sources: }\quad 
\left( \Phi_e : \Phi_\mu : \Phi_\tau \right)= 
\left( 1 + \frac{B}{9} : 1 - \frac{B}{18} + \frac{C}{9 }
: 1 - \frac{B}{18} - \frac{C}{9} \right) ~.
\ee
These relations illustrate that 
deviations from the ``canonical'' $1 : 1 : 1$ result 
are of order ($\Delta$, $\overline{\Delta}^2$), or 
($\frac{1}{10}\,(B,\ C)$) and can therefore exceed 10\%.  
Another result from these formulae is that the ratio 
\be \label{eq:mutau}
\frac{\Phi_\mu}{\Phi_\tau} \simeq 1 + 2 \,  \overline{\Delta}^2 = 
1 + \frac{2}{9 } \, C
\ee 
is always larger than or equal to one. We checked that this is also 
true for the full expression. Hence, there cannot be more $\nu_\tau$ 
than $\nu_\mu$ at Earth if the initial flux composition is $1 : 2 : 0$. 

Reorganizing the results in Eq.~(\ref{eq:flux120}), we get the 
ratios of our interest.
They are\footnote{The expression for $R$ does not exactly follow from 
Eq.~(\ref{eq:probs}), but is obtained by evaluating and 
expanding the full fraction. The difference is however negligible.}
\bea \label{eq:Tp} \D
T 
\simeq \frac 13 \, (1 - \Delta + \overline{\Delta}^2)
~~\mbox{ and }~~
\D R 
\simeq 1 + 3 \, \Delta + \overline{\Delta}^2 + 3 \, \Delta^2~.
\eea
Numerically, using the full expressions, $T$ lies between 0.32 and 0.39, while 
$R$ ranges from 0.82 to 1.48. Therefore, deviations of more than 15\% for $T$, 
and almost 50\% for $R$ can be expected.

The ratio of electron neutrinos to all 
other neutrino flavors is interesting in that it receives no 
quadratic corrections. The ratio is 
\be\label{eq:etot}
\frac{\Phi_e}{\Phi_{\rm tot}} \simeq \frac 13 \left( 
1 + 2 \, \Delta \right) \,.
\ee 
The tribimaximal values of the ratios $T$, $R$, and $\Phi_e/\Phi_{\rm tot}$ 
are clearly $\frac 13$, 1, and $\frac 13$, respectively. 
 
\subsubsection{\label{sec:010} Muon-Damped Sources}
The fluxes at Earth for muon-damped sources are found to be 
\bea 
\mbox{muon-damped sources:}\quad 
\left(\Phi_e : \Phi_\mu : \Phi_\tau  \right)= 
\left(P_{e \mu} :  P_{\mu \mu} : P_{\mu \tau}  \right)
\stackrel{\rm TBM}{=}(4 : 7 : 7) \\[0.2cm]
\longrightarrow \left(4 - 2 \, A + B : 7 + A - B + C : 7 + A - C \right)
~,
\eea 
where we first insert the tribimaximal mixing values, and then show 
the result with corrections. 
From these, we get the ratios of interest:
\begin{eqnarray} \label{eq:Tm} \D
&T& 
= P_{\mu\mu} 
\simeq 
\frac{1}{18} \left(7 + A - B + C \right) 
\simeq \frac{7}{18} - \Delta 
+ \frac 12 \, \overline{\Delta}^2 
~,\\[0.2cm]~~
\D \hspace{-0cm} &R& 
= \frac{P_{e\mu}}{P_{\mu\tau}} 
\simeq \nonumber
\frac{4 - 2\, A + B}{7 + A - C} 
\simeq \frac 17 
\left(4 + 18 \, \Delta + \frac{36}{7} \, 
\overline{\Delta}^2  
\right) 
~,
\end{eqnarray} 
where we have inserted on the far right-hand sides, the value 
$\sin^2 \theta_{12} = \frac 13$. 
The quantities $\Delta$ and $\overline{\Delta}^2$ are therefore 
the quantities $\Delta_{\rm TBM}$ and $\overline{\Delta}^2_{\rm TBM}$
defined earlier in Eq.~(\ref{eq:Deltbm}).
However, as mentioned below Eq.~(\ref{eq:Deltbm}), the differences are small. 
For exact tribimaximal mixing we 
have $T = \frac{7}{18}$ and $R = \frac 47$. 
Refs.~\cite{Michael} and \cite{CP} have proposed to use these sources to 
probe $\theta_{23}$ and the CP phase, respectively.

\subsubsection{\label{sec:100} Neutron Beam Sources}
Neutron beam sources have an initial $1 : 0 : 0$ flavor mix. 
We find for the flavor decomposition at Earth, 
\bea 
\mbox{neutron beam:}\quad 
\left(\Phi_e : \Phi_\mu : \Phi_\tau  \right) 
= \left(P_{e e} :  P_{e \mu} : P_{e \tau}  \right)
\stackrel{\rm TBM}{=}(5 : 2 : 2)\\[0.2cm]
\longrightarrow \left(10 + 4 \, A : 4 - 2 \, A + B : 4 - 2 \, A - B  \right)
~.
\eea 
Again, we give results for exact tribimaximal mixing, 
followed by the corrected expression. 
From these, we get the flavor ratios of interest:
\bea \label{eq:Tn}
T = P_{e \mu} \simeq \D \frac{1}{18} \left(4 - 2\, A + B\right) 
\simeq c_{12}^2 \, s_{12}^2 + \Delta \simeq \frac 29 + \Delta 
~,\\[0.2cm]
\D R = \frac{P_{ee}}{P_{e\tau}} \simeq 
\frac{10 + 4 \, A}{4 - 2\, A - B} 
\simeq 
\frac{1 - 2 \, c_{12}^2 \, s_{12}^2 }{c_{12}^2 \, s_{12}^2 } \, 
\left( 1 + \frac{\Delta}{c_{12}^2 \, s_{12}^2 } \right) 
\simeq \frac 52 \left(1 + \frac 92 \, \Delta\right)
~.
\eea
Tribimaximal values are $\frac 29$ and 
$\frac 52$, respectively. 
Note that $C$, or alternatively $\overline{\Delta}^2$, which show up only 
in the $\mu$--$\tau$ sector, do not appear in neutron beam formulae. 
The ratio of $\overline{\nu}_e$ to the total flux, initially unity, is simply 
$Q = P_{ee} \stackrel{\rm TBM}{=} \frac 59 \rightarrow 
(10 + 4 \, A)/18  $.

\subsubsection{\label{sec:sum}Summarizing the Flavor Ratios}
\begin{figure}[ht]
\begin{center}
\epsfig{file=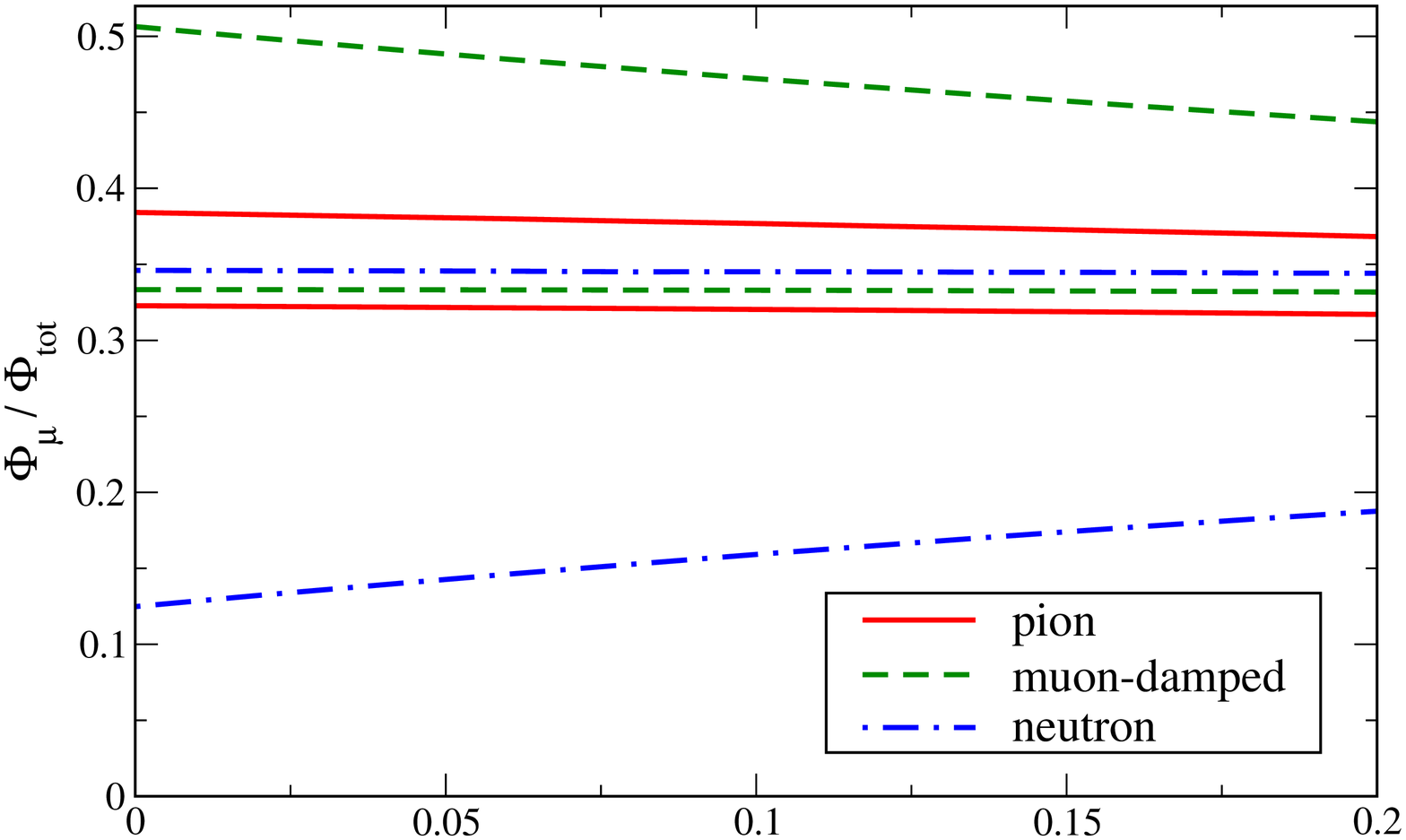,width=12cm,height=7.58cm} 
\end{center}
\caption{\label{fig:zetaT}The minimal and maximal values of the 
ratio $T$ of muon neutrinos to the total flux, obtained by varying oscillation 
parameters within their $3\sigma$ ranges.
The horizontal axis labels deviations 
($\zeta$ or $\eta$) from the idealized flux compositions,
parameterized as  
$1 : 2 \, (1 - \zeta) : 0$ for pion sources, $\eta : 1 : 0$ for muon-damped 
sources, and $1 : \eta : 0$ for neutron beam sources. 
The very left side of the plot is therefore the allowed range for  
a pure flux composition.}
\end{figure}

We summarize the situation for the flux ratios $T$ and $R$. 
Tables \ref{tab:pion}, \ref{tab:muon} and \ref{tab:neutron} show their  
ranges for the currently allowed $3\sigma$ ranges of the 
oscillation parameters. 
In the case of exact tribimaximal mixing, $T$ is $1/3$, 
$7/18 \simeq 0.39$, and $2/9 \simeq 0.22$ for pion, muon-damped 
and neutron sources, respectively. If future neutrino oscillation 
experiments show that deviations from tribimaximal are small, 
then measurements of $T$ with $\simeq 10\%$ precision would  
distinguish between pion and muon-damped sources.
However, a low value of $T\sim 2/9$ would clearly indicate neutron 
sources\footnote
{Alternatively, the same statements apply to the 
ratio $\Phi_\mu/(\Phi_e + \Phi_\tau)$, 
which is $1/2$, $7/11 \simeq 0.64$ and $2/7 \simeq 0.29$.}. 
On the other hand, currently allowed nonzero values 
of $|U_{e3}|$, $\epsilon$ or $\epsilon'$ lead to a possible overlap of the 
ratio $T$ for all sources. This is shown in Fig.~\ref{fig:zetaT}, where 
we display the allowed ranges of $T$ for the $3\sigma$ ranges of the 
oscillation parameters. 
The range at the very left side of the plot is what is relevant to this 
discussion (the relevance of the rest of the plot concerns flux uncertainties,
which we introduce in the next Section). 

\begin{figure}[ht]
\begin{center}
\epsfig{file=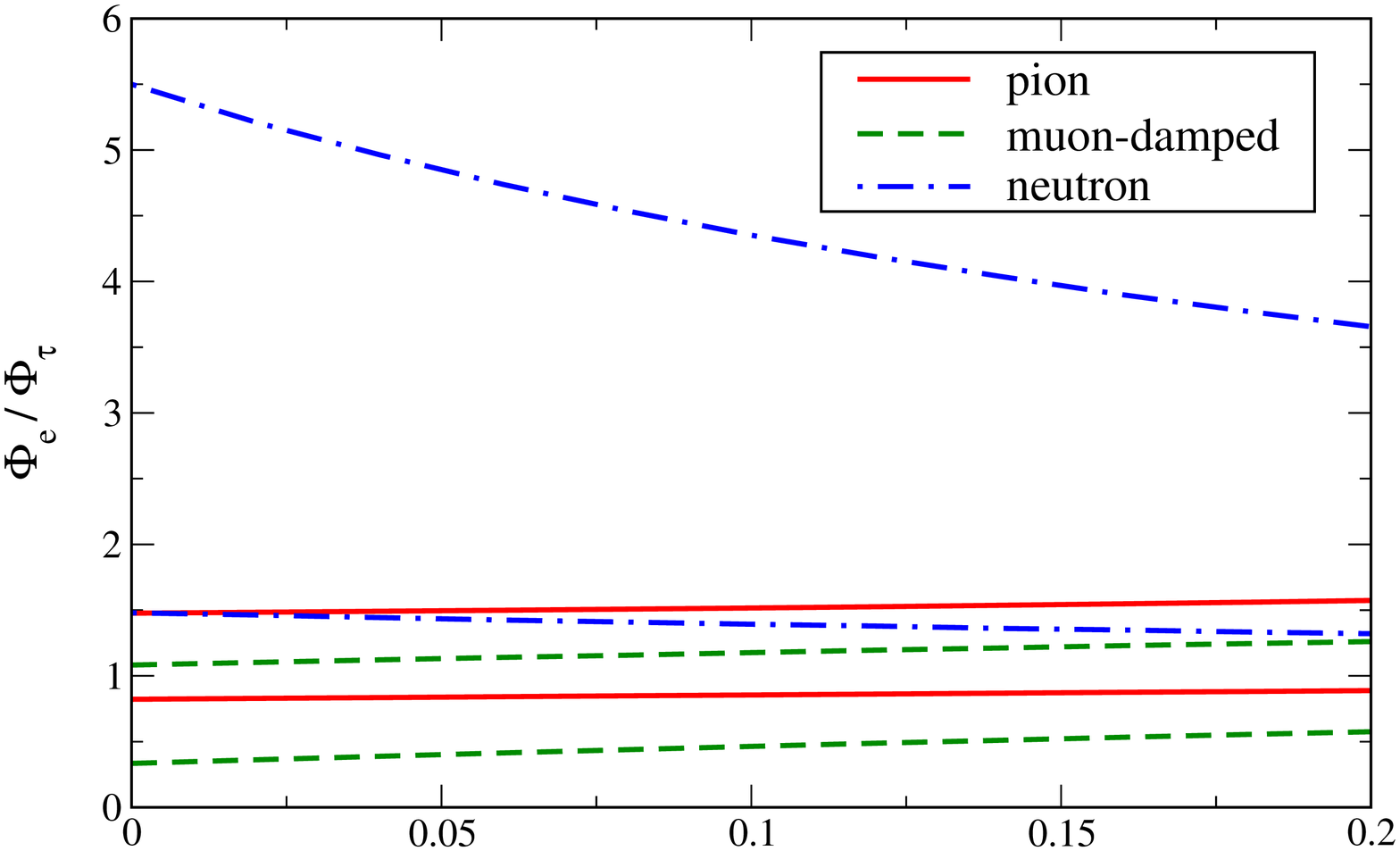,width=12cm,height=8.139cm} 
\end{center}
\caption{\label{fig:zetaR}Same as previous Figure for the ratio $R$ of 
electron to tau neutrinos.}
\end{figure}

$R$, the ratio of electron to tau neutrinos, is for tribimaximal mixing 
$1$, $4/7$ and $5/2$, respectively, for the pion, muon-damped, 
and neutron sources. 
Hence, much less precision would be required in 
order to distinguish the different source types.  
A large value of $R$ would indicate a neutron source. 
However, the correction terms 
due to nonzero $|U_{e3}|$, $\epsilon$ and $\epsilon'$ are seen to 
have rather large pre-factors, so that again overlap can be expected. 
Results with these uncertainties are shown in Fig.~\ref{fig:zetaR}.  

\begin{figure}[ht]
\begin{center}
\epsfig{file=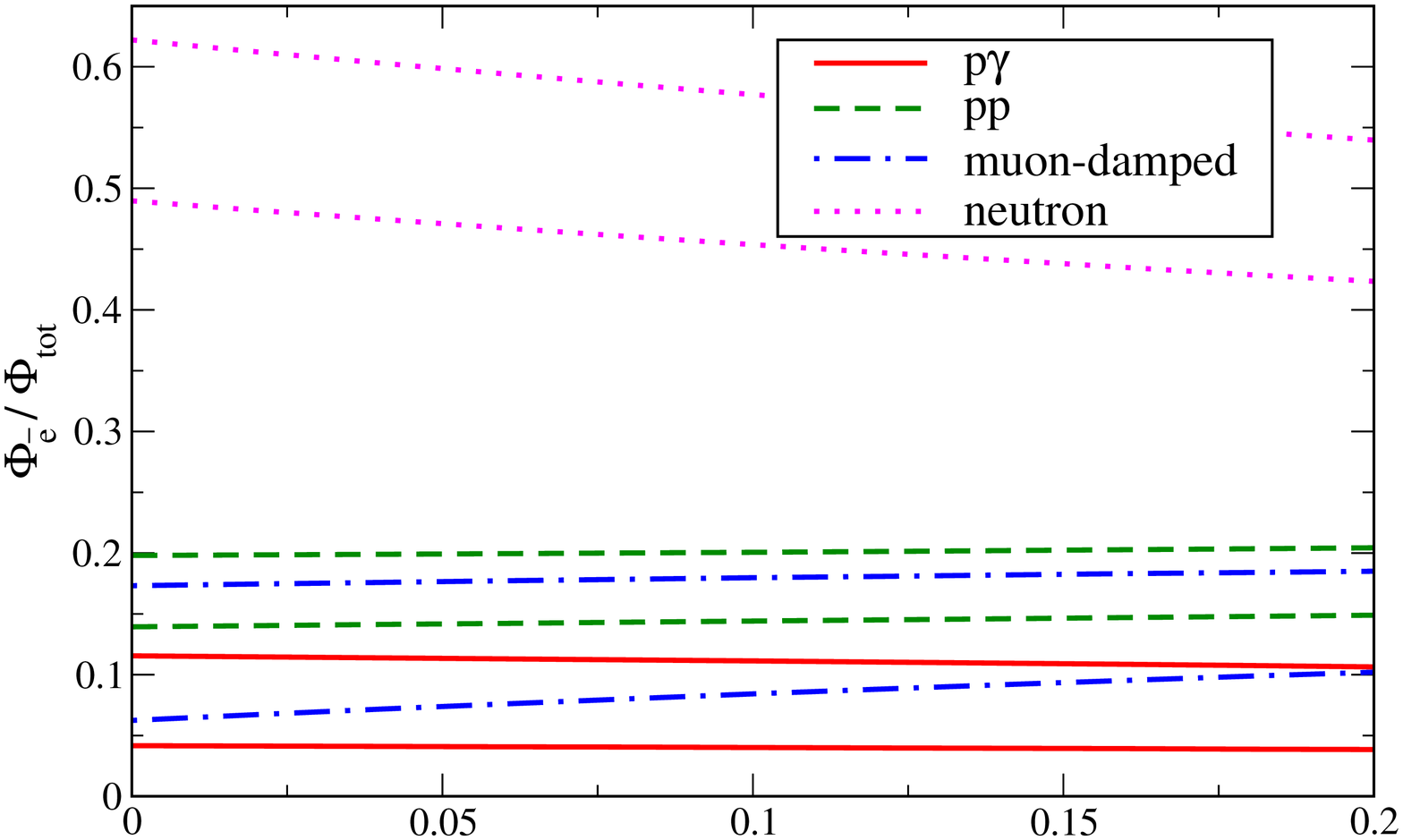,width=12cm,height=8.139cm} 
\end{center}
\caption{\label{fig:zetaQ}Same as previous Figure for the ratio $Q$ of 
$\overline{\nu}_e$ to all neutrinos.}
\end{figure}

We show in Fig.~\ref{fig:zetaQ} the ratio 
$Q$ of $\overline{\nu}_e$ to the total flux. 
Recall that this ratio is important for the $\nuebar+e^-\rightarrow W^-$ reaction
in the $\sim 6.3$~PeV energy bin. 
With the $Q$ observable, the neutron source (``cosmic $\beta$-beam'')
is clearly differentiated from the other source types.
An interesting proposal~\cite{GR1} is that a measurement of $Q$ as the ratio  
of the resonant $\nuebar + e^-\rightarrow W^-$ 6.3~PeV bin 
to any off-resonance bin can differentiate between two 
nuances of the pion source, namely $pp$ and $p\gamma$ beam dumps.
The latter has no $\nuebar$'s at production. 
Fig.~\ref{fig:zetaQ}, in which neutrino luminosities are 
assumed equal for all source types, 
reveals that $pp$ and $p\gamma$ are in principle distinguishable.
In fact, it has been argued that the neutrino luminosity from $pp$ 
is larger than that from $p\gamma$ by a factor $\sim 2.4$ 
(see \cite{GR1} and references therein).
This implies a better statistical determination of $Q$ with the $pp$ origin,
and an additional signal for discriminating $pp$ and $p\gamma$.
On the other hand, the muon-damped source also scales with this 2.4~factor,
potentially confusing the discrimination between $pp$ and $p\gamma$ sources.
Also, it has been noted that the $\gamma\gamma\rightarrow\mu^+\mu^-$ reaction may
provide some $\nuebar$'s in a hot $p\gamma$ environment~\cite{Razzaque:2005ds}.

We note that for pion sources the ratios depend very little on 
$\theta_{12}$. For the other sources, the dependence is stronger, 
as (unlike for pion sources) in the limit of 
$\epsilon = U_{e3} = 0$ the ratios depend on 
$\theta_{12}$. In general, however, for equal deviations 
from the tribimaximal values, the dependence on $\theta_{12}$ 
is weaker than the dependences on $\theta_{23}$ and $|U_{e3}|\,\cos\delta$. 
Nevertheless, we note that schematically one may write 
\bea  \label{eq:scheme}
\mbox{Ratio}(\mbox{pion}) = 
c_0 + c_1 \, \Delta + c_2 \, \overline{\Delta}^2~~,~\\[0.2cm]
\mbox{Ratio}(\mbox{muon-damped}) 
= f_0(\theta_{12}) + f_1(\theta_{12}) \, \Delta 
+ f_2(\theta_{12}) \, \overline{\Delta}^2~~,~\\[0.2cm]
\mbox{Ratio}(\mbox{neutron}) 
= g_0(\theta_{12}) + g_1(\theta_{12}) \, \Delta ~.
\eea 
The zeroth order expression for pion sources 
does not depend on $\theta_{12}$. In addition, typically we have 
that the magnitude of $c_0/c_{1,2}$ is larger than the magnitudes of 
$f_0(\theta_{12})/f_{1,2}(\theta_{12})$ and 
$g_0(\theta_{12})/g_{1,2}(\theta_{12})$. 
For instance, for the ratio $T$, we find from the above 
that $|c_0/c_1| = |c_0/c_2| = 1$, whereas 
$|f_0/f_1| = |f_0/f_2| = \frac 12 \, (1 - c_{12}^2 \, s_{12}^2) 
\simeq 7/18$ and $|g_0/g_1| = c_{12}^2 \, s_{12}^2 \simeq 2/9$. 
Therefore, the ratios for muon-damped and neutron sources are 
significantly more sensitive to $\Delta$ (and 
$\overline{\Delta}^2$), and thus on deviations from vanishing $U_{e3}$ 
and maximal $\theta_{23}$. 
This simple fact is the reason that 
flux ratios of muon-damped and neutron beam sources are better suited to 
probe such deviations. Noting further that $\Delta$ and 
$\overline{\Delta}^2$ depend strongly on $\theta_{23}$, we can expect 
that neutrino telescopes will be most sensitive to 
the $\theta_{23}$ parameter.

\section{\label{sec:uncertain} Uncertainties in Initial Flavor Composition}
Up to this point we have obtained expressions for the flux ratios in the 
cases of exact initial flavor composition. However, as indicated 
in the Introduction, one expects on general grounds 
some deviations from the idealized $1 : 2 : 0$ (pion source), 
$0 : 1 : 0$ (muon-damped source), or $1 : 0 : 0$ (neutron source) 
flavor ratios. The implications of non-idealized source ratios, 
on the extraction of the neutrino mixing parameters, 
have not been previously studied in detail. 
In this Section, we will first estimate the 
amount of ``impurity'' in the initial flux compositions.
After that, we will discuss some 
examples leading to incorrect inferences if care is not taken.

\subsection{\label{sec:estimate} Estimating Realistic Flavor-Flux Compositions} 
When the source of neutrinos is a hadronic beam dump 
which produces pions and consequently neutrinos from 
$\pi \ra \mu \, \nu_\mu \ra e \, \nu_\mu \, \nu_\mu \, \nu_e$,
the pion decay chain, naive counting gives a flux ratio for 
$\nu_e$ to $\nu_\mu$ of $1 : 2$ and no $\nutau$.
However, the ``wrong-helicity'' muon polarization from pion decay 
makes the $\nu_\mu$ from its decay 
softer, thus reducing the effective $\nu_\mu$ count. This effect depends
on the injection spectrum. 
For a canonical spectral index of $\alpha = 2$, 
the predicted initial flavor ratio is $1 : 1.86$ 
(see the detailed discussion in Ref.~\cite{lipari}). 
In addition, the production and decay of kaons also 
produces neutrinos. The kaon decay modes which produce charged pions in the 
final state (e.g., $K_S \ra 2 \pi$, $K^\pm \ra 2 \pi$, $K_L\ra 3\pi$) 
give rise to the same ratio of $1 : 1.86$. 

We now extend the analysis of Ref.~\cite{lipari} to include 
leptonic and semi-leptonic $K$ decays, 
and production and decay of heavy flavors. 
The appreciable $K^\pm \ra \mu^\pm + \stackrel{(-)}{\nu_\mu}$ 
chain and the 
three-body semi-leptonic decay modes of $K^\pm$ and especially $K_L$ 
also produce neutrinos.
The $K^\pm \ra \mu^\pm + \stackrel{(-)}{\nu_\mu}$ chain 
has a $\nu_e : \nu_\mu$ ratio of $1 : 2.8$ and the 
$K_L$ semi-leptonic decays give 
a ratio of $1 : 0.75$.  We use a $K/\pi$ ratio of $0.15$ in our 
estimates, although the dependence on this ratio is rather weak. 
When all the modes are added to the pion  
chain neutrinos, and all the branching ratios and 
neutrino energy spectra are taken into account (following the methods 
described in \cite{lipari}), the final flavor mix 
ends up being surprisingly close to the value $1 : 1.86$, which was 
obtained by including only the pion chain. 
Finally, at high energies, production of the heavy flavors  
$c$ and $b$ is expected. 
The semi-leptonic decay of these heavy flavors gives rise  
to so-called ``prompt'' neutrinos, characterized by the 
flux ratio for $\nu_e : \nu_\mu$ of $1 : 1$.
There is also a small flux 
of $\nu_\tau$ from the decay mode $D_s \ra \tau \, \nu_\tau$, which has a 
branching ratio of about $6\%$. Bottom quark decays also produce 
some $\nu_\tau$, but $b$-production is down compared to 
$c$-production by a factor of about $30$. 

After taking all cross-sections, decaying species, branching  
ratios, and neutrino energy distributions into 
account, our final estimate of the corrected flavor mix is 
\be \label{eq:est} 
\Phi_e^0 : \Phi_\mu^0 : \Phi_\tau^0 = 1.00 : 1.852 : 0.001 ~.
\ee 

Remarkably, the $\nu_e : \nu_\mu$ ratio remains very close to the original ratio 
$1 : 1.86$ due to just pion decay. 
Notice that the $\nu_\tau$ content is a negligible $0.1\%$ at most. 
The main source of uncertainty in our estimate is the spectral index at 
injection, $\alpha$. 
Fig.~3 of Ref.~\cite{lipari} shows that the ratio $1 : 1.86$ is modified 
by a few percent to  
$1 : 1.9$  if $\alpha = 1.7$, and to  $1 : 1.8$ if $\alpha = 2.3$.

We quantify the deviations with the single parameter $\zeta$:
\be \label{eq:flux_zeta}
\Phi_e^0 : \Phi_\mu^0 : \Phi_\tau^0 = 1 : 2 \, (1 - \zeta) : 0  ~,
\ee 
and we can expect $\zeta$ to be $\sim 0.1$.
The ratio $1 : 1.86$ for electron to muon neutrinos 
corresponds to $\zeta = 0.07$. 

Damped muon sources result when muon energy loss
mechanisms are operative. 
The effects of muon energy loss for the flavor mix
have been discussed by a number of authors 
\cite{010c,010mod,010b,010ba,010a,lipari,kachneu} for a number 
of sources (such as gamma ray bursts),
and models (such as Waxman-Bahcall type).
From Refs.~\cite{010b,lipari,kachneu} one can draw 
two general conclusions: first, the onset of muon-damping 
happens abruptly in energy. 
In Ref.~\cite{kachneu}, the onset is within a factor of 
2 to 3 of $10^8$ GeV. 
Second, the $\nu_e$ flux may be severely suppressed, but it 
never goes to zero.
It never falls below 4\% in Ref.~\cite{kachneu}, and never below 
 2\% in Ref.~\cite{lipari}. 
Consequently, we parameterize the 
initial muon-damped sources as
$(\eta : 1 : 0)$, where $\eta$ is at least a few percent. 

Finally, for the neutron sources a pion pollution of 
order $10\%$ has been estimated in Ref.~\cite{Michael}.  
So for the muon-damped and neutron sources, we introduce the single 
parameter $\eta$ 
and use for the initial flux ratios 
\be \label{eq:eta}
\Phi_e^0 : \Phi_\mu^0 : \Phi_\tau^0 
= \eta : 1 : 0\quad (\mbox{muon-damped})\,,
~~~\mbox{ and }~~ 1 : \eta : 0\quad ({\rm neutron})\,.
\ee

For typical values of $\zeta$ and $\eta$,
we show in Tables \ref{tab:pion}, \ref{tab:muon} and \ref{tab:neutron} 
the ranges of the flux ratios under consideration. 
The oscillation parameters 
are varied in their currently allowed $3\sigma$ ranges. 
Results for pion sources are the most stable.  
We also give in the Tables the 
values of the flux ratios if the neutrino parameters are fixed to their 
tribimaximal values.
One concludes that impure initial flux compositions should be taken 
into account when discussing the prospects of inferring neutrino 
parameters with flux ratio measurements.

It is worth mentioning that another possible source of deviations 
from idealized flux ratios is new physics.
For example, $3+2$ 
sterile neutrino scenarios \cite{3+2,GoRo} (still allowed even  
after the MiniBooNE results~\cite{mini}), 
can cause deviations of the flux ratios of order $10\%$, thereby 
interfering with the program of inferring deviations 
from three-flavor tribimaximal 
mixing \cite{sterile,GoRo}.

We note that a different parameterization for the initial flavor ratios 
is given in Ref.~\cite{xing2}. There the unit normalization of the sum 
of the flavor ratios is emphasized by introducing two polar 
angles for the vector on the unit sphere, 
$\Phi_e : \Phi_\mu : \Phi_\tau = \sin^2 \xi \, \cos^2 \zeta : 
\cos^2 \xi \, \cos^2 \zeta : \sin^2 \zeta$. 
Although this parameterization (the angular $\zeta$-parameter of 
Ref.~\cite{xing2} is not to be confused with our small $\zeta$-parameter 
defined in Eq.~(\ref{eq:flux_zeta})) and ours are equivalent, ours 
does have the advantage that the introduced parameters in 
Eqs.~(\ref{eq:flux_zeta}) and (\ref{eq:eta}) are small, 
thereby allowing the perturbative expansions 
we present in the next subsections.
Also, the focuses in Ref.~\cite{xing2} and in our work are different.
The former emphasizes determination of the initial 
flux composition, i.e., determining $\xi$ and $\zeta$, when the 
neutrino mixing parameters are known with sufficient precision. 
In our work we include the influence of uncertainties from 
both the initial fluxes and the neutrino mixing parameters upon the 
experimental program,
with particular attention paid to the 
consequences for extraction of precise neutrino parameters.

\subsection{\label{sec:effects120} Parameterized Pion Sources}
Justifiably neglecting the $\nu_\tau$ contribution, the  
initial flux composition from pion sources is 
$\Phi_e^0 : \Phi_\mu^0 : \Phi_\tau^0 = 1 : 2 \, (1 -\zeta):0$.  
In the limit of $\zeta = 0$, we recover 
the ratios of Section~\ref{sec:fluxes}. 
The detectable fluxes behave according to 
\begin{eqnarray} \label{eq:120zeta}
\Phi_e &\propto & \nonumber 
P_{ee} + 2 \, (1 - \zeta) \, P_{e \mu} = 
1 - 2 \, \zeta \, c_{12}^2 \, s_{12}^2 
+ 2 \, \Delta \, (1 - \zeta) - 2 \, \zeta \, 
(1 - 2 \, c_{12}^2 \, s_{12}^2 ) \, |U_{e3}|^2 ~,\\[0.2cm]\nonumber 
\Phi_\mu &\propto& 
P_{e\mu} + 2 \, (1 - \zeta) \, P_{\mu \mu} = 
1 - \Delta - \zeta \, (1 - c_{12}^2 \, s_{12}^2) + \overline{\Delta}^2 + 
2 \, \zeta \, \Delta +  \zeta \, 
(1 - 2 \, c_{12}^2 \, s_{12}^2 ) \, |U_{e3}|^2~,\\[0.2cm]
\Phi_\tau &\propto & \nonumber
P_{e \tau} + 2 \, (1 - \zeta) \, P_{\mu \tau} = 
1 - \Delta - \zeta \, (1 - c_{12}^2 \, s_{12}^2) - \overline{\Delta}^2 
+  \zeta \, (1 - 2 \, c_{12}^2 \, s_{12}^2 ) \, |U_{e3}|^2~,
\end{eqnarray}
plus cubic terms in the small parameters. 
With the alternate ($A$, $B$, $C$) expansion, 
we have 
\begin{eqnarray} \label{eq:120zeta2}
\Phi_e &\propto & \nonumber 
1 + \frac{B}{9} - 2 \, \zeta \, (4 - 2 \, A + B) ~,\\[0.2cm]
\Phi_\mu &\propto& 
1 - \frac{B}{18} + \frac{C}{9 }
- 2 \, \zeta \, (7 + A - B + C)   ~,\\[0.2cm]
\Phi_\tau &\propto & \nonumber
1 - \frac{B}{18} - \frac{C}{9} 
- 2 \, \zeta \, (7 + A - C)  ~.
\end{eqnarray}
We show in Fig.~\ref{fig:zetaT} the allowed range of the ratio $T$ for 
pion sources as a function of $\zeta$, 
while Fig.~\ref{fig:zetaR} shows the same for $R$, and Fig.~\ref{fig:zetaQ} 
the same for $Q$. 
The largest effect is seen in the ratio $Q$, 
in the case of a $p\gamma$ neutrino source.  

Impure sources will still lead to deviations from 
``pure'' values of the flux ratios, 
even if neutrino mixing is exactly tribimaximal:
\be
T_{\rm TBM} = \frac{9 - 7 \, \zeta}{27 - 18 \, \zeta }
\simeq \frac 13 \left( 1 - \frac{\zeta}{9} \right)~\mbox{ and }~
R_{\rm TBM}  = \frac{9 - 4 \, \zeta}{9 - 7 \, \zeta }
 \simeq  1 + \frac \zeta3~.
\ee
The deviation is seen to be stronger in $R_{\rm TBM}$. 

\begin{figure}[ht]
\begin{center}
\epsfig{file=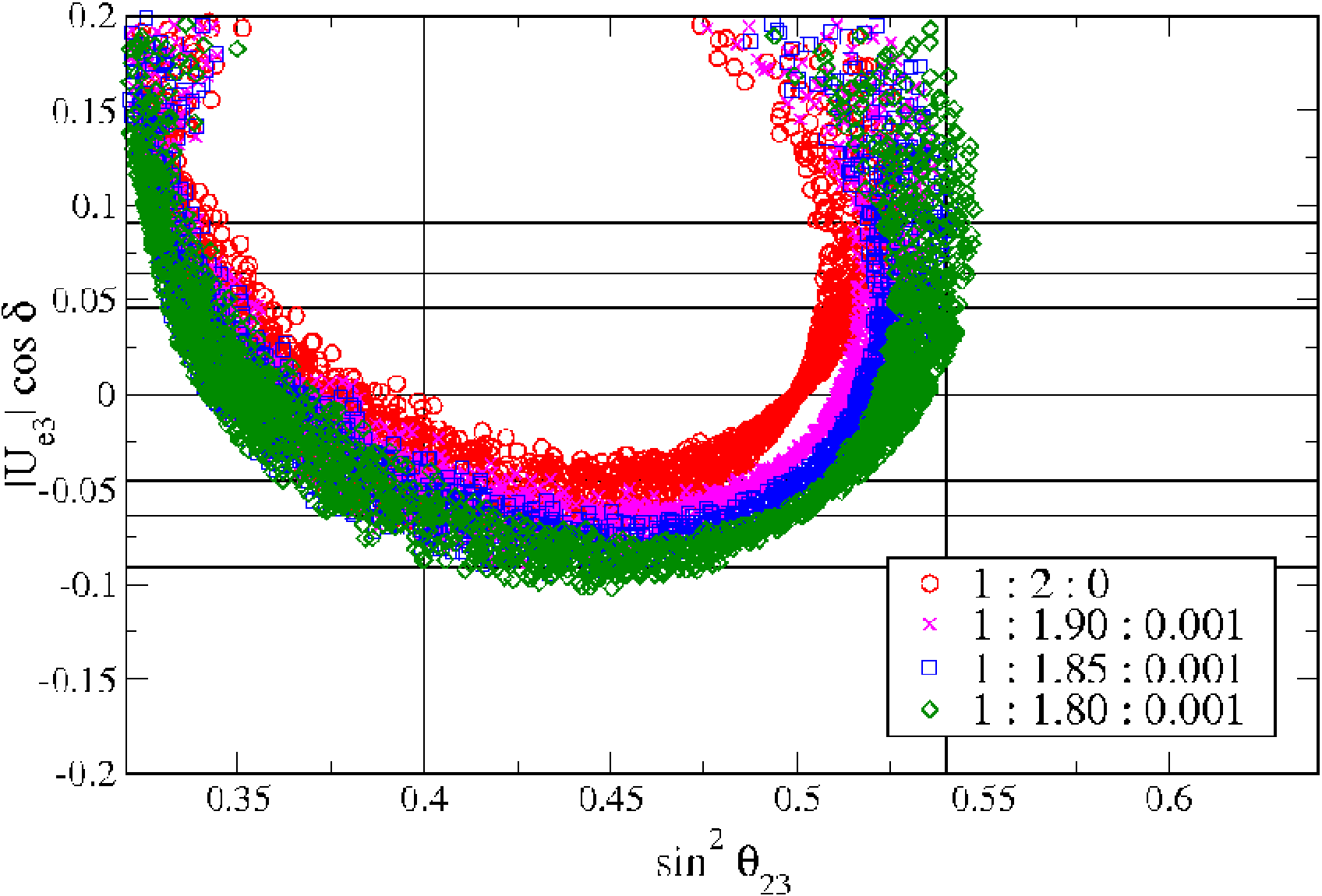,width=8cm,height=7.139cm} 
\epsfig{file=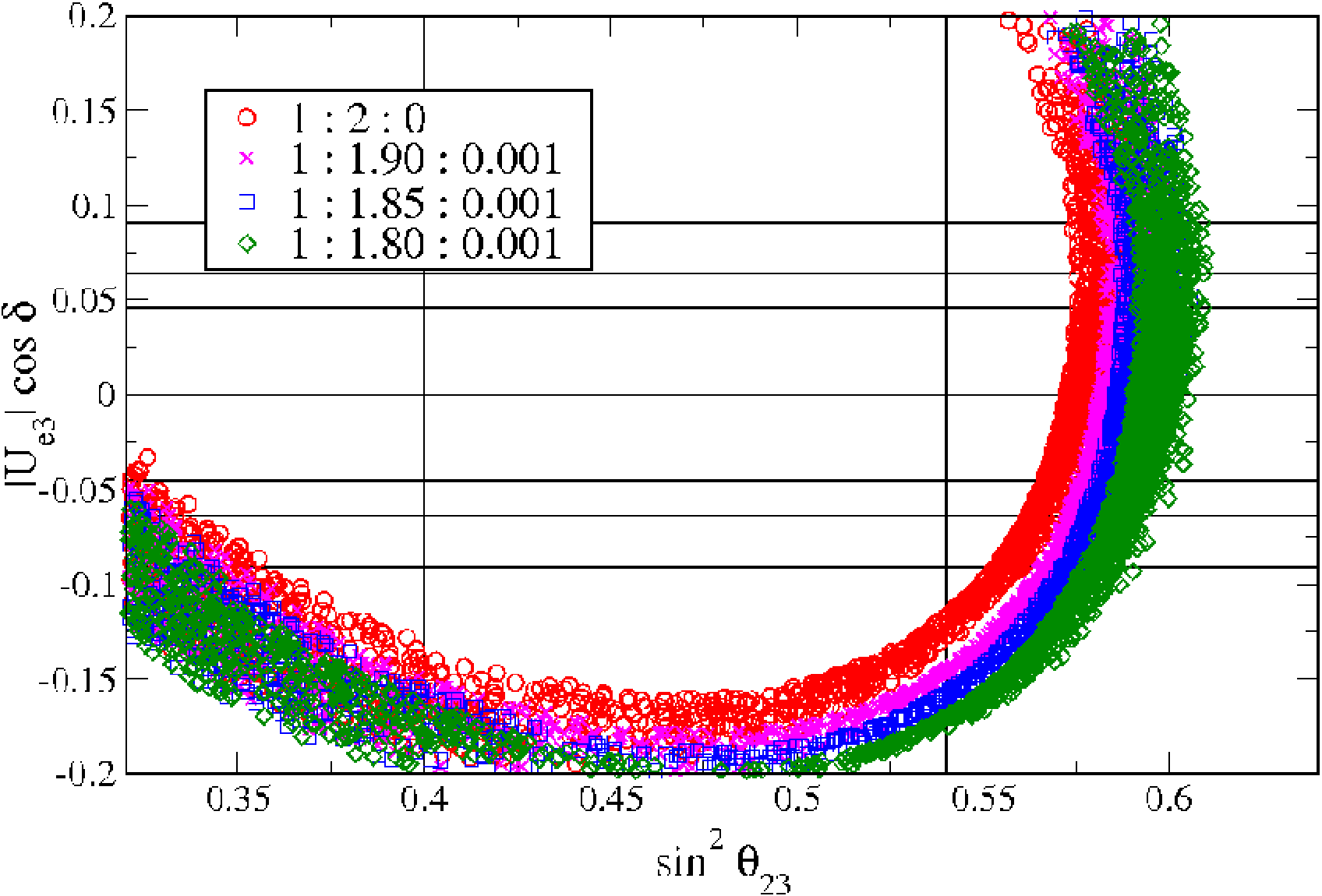,width=8cm,height=7.139cm}
\end{center}
\caption{\label{fig:mutot_zeta}Distribution of $|U_{e3}| \, \cos \delta$ 
against $\sin^2 \theta_{23}$ if the flux ratio $\Phi_\mu/\Phi_{\rm tot}$ 
is measured to be $\frac 13$ (left) and 0.35 (right), for different 
initial flavor compositions. The red 
circles are for pure $1 : 2 : 0$, the magenta crosses are for 
$1 : 1.90 : 0.001$, the  blue squares are for $1 : 1.85 : 0.001$ and 
the green diamonds are for $1 : 1.80 : 0.001$. 
Indicated also is the allowed $1\sigma$ range of $\theta_{23}$ and 
of $|U_{e3}| \, \cos \delta$ for (from top to bottom above zero) $\delta=0$, 
$\delta = \pi/4$ and $\delta = \pi/3$. The value $\delta = \pi/2$ means 
$|U_{e3}| \, \cos \delta = 0$.}
\end{figure}

\begin{figure}[ht]
\begin{center}
\epsfig{file=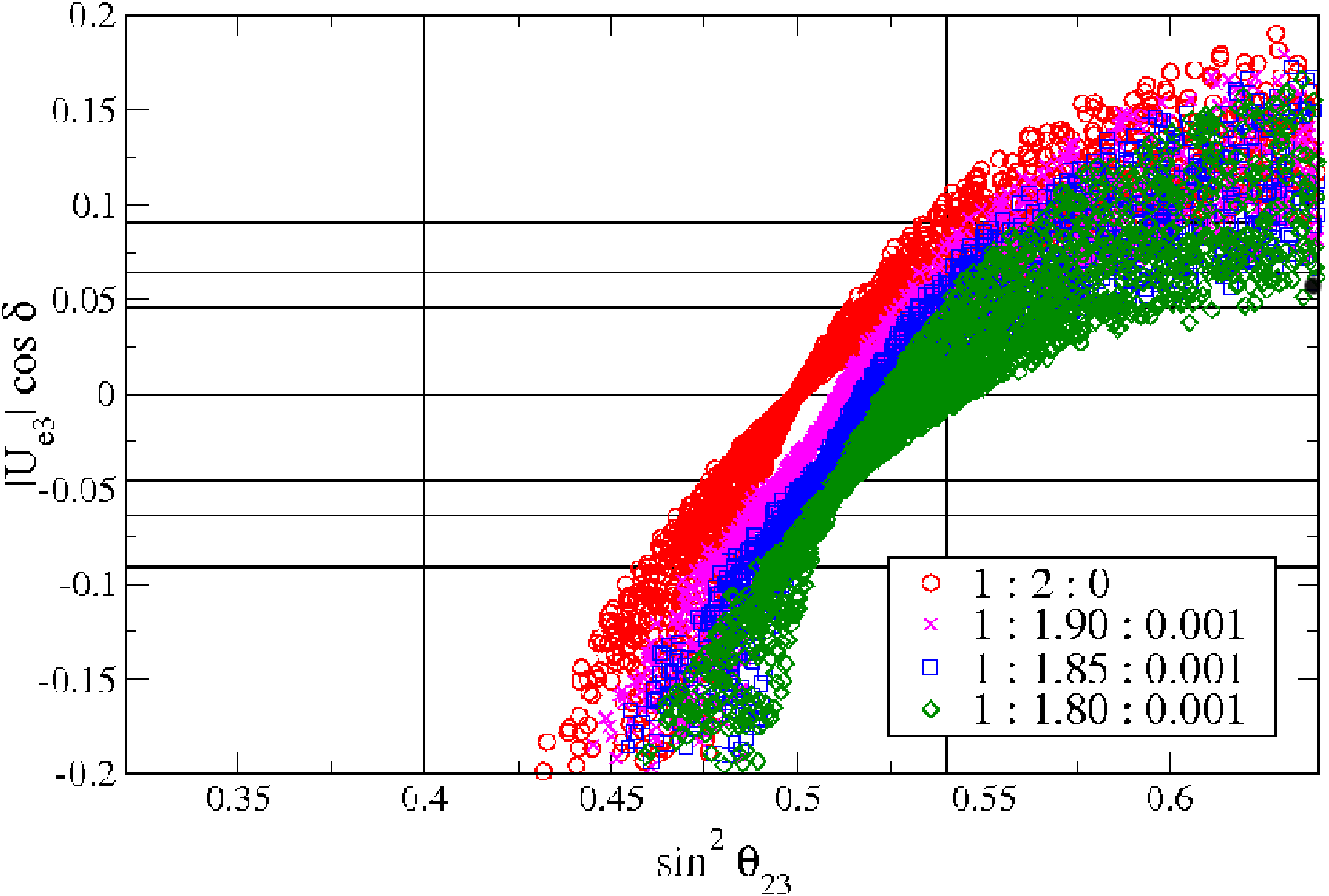,width=8cm,height=7.139cm} 
\epsfig{file=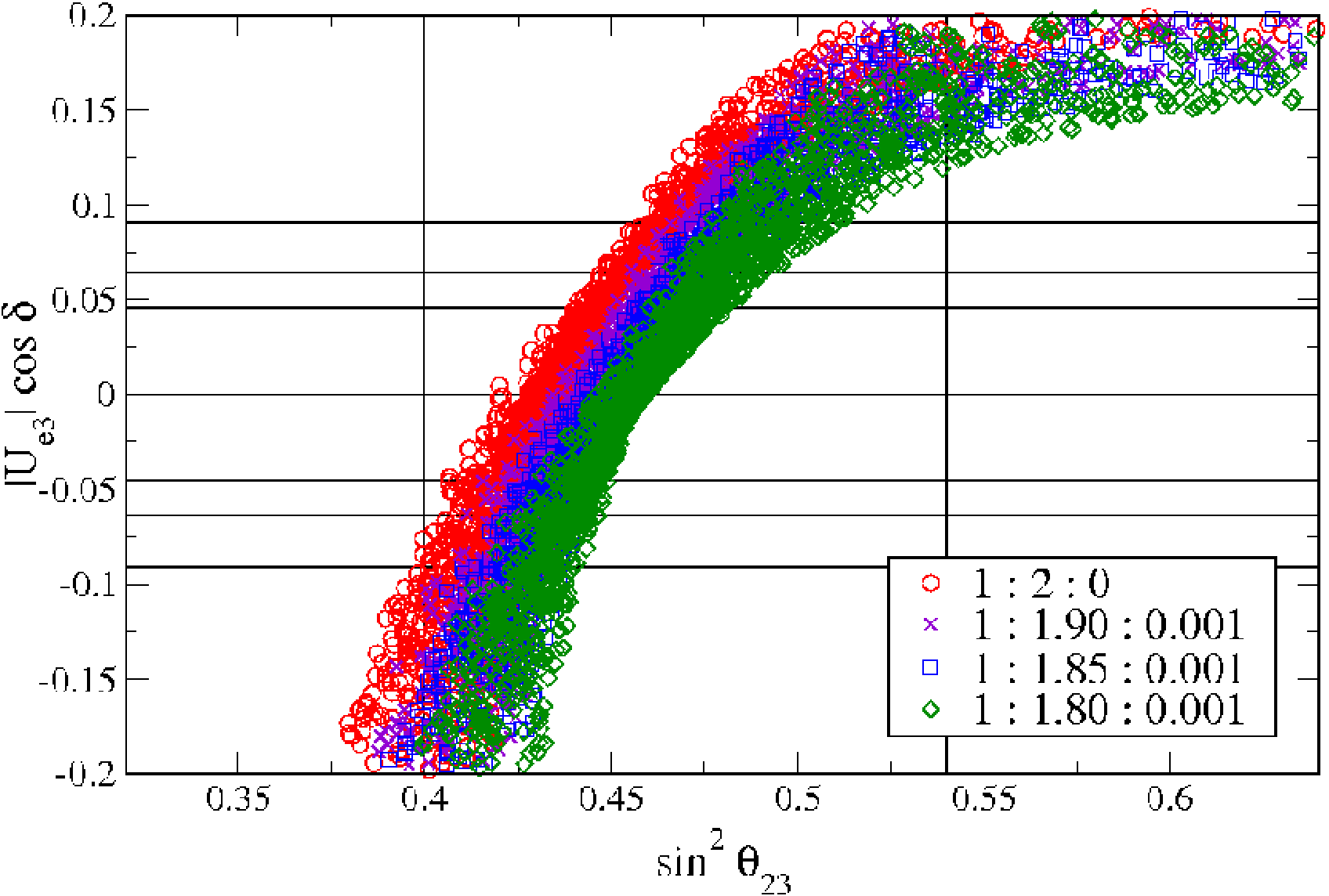,width=8cm,height=7.139cm} 
\end{center}
\caption{\label{fig:etau_zeta}Same as previous Figure for the flux ratio 
$\Phi_e/\Phi_{\tau}$ assumed to be 1 (left) and 1.1 (right). }
\end{figure}

Possible nonzero $\theta_{13}$ and non-maximal $\theta_{23}$ may be 
compensated by the ``impurity factor'' $\zeta$. 
The flux ratio $\Phi_e /\Phi_{\rm tot}$ illustrates this ``confusion''. 
From Eq.~(\ref{eq:etot}) we know that for $\zeta = 0$ the ratio is 
given by $\frac 13 (1 + 2 \, \Delta)$. From Eq.~(\ref{eq:120zeta}), 
it is (neglecting the very small single terms depending on $|U_{e3}|^2$) 
easily obtained that a nonzero $\Delta$ 
is compensated by an initial flux uncertainty 
if the following relation holds: 
\be
\zeta = -\frac{3 \, \Delta}{1 - 3 \, c_{12}^2 \, s_{12}^2}
\simeq-9\,\Delta\,.
\ee
This value of uncertainty returns the ratio to the 
value $\frac 13$, even though $\Delta \neq 0$. 
The ratio of muon and tau neutrino fluxes provides another example.
Compensation occurs if 
\be
\zeta = - \frac{\Delta}{\overline{\Delta}^2} \,.
\ee
In Fig.~\ref{fig:mutot_zeta} we show the distribution of 
$|U_{e3}| \cos \delta$ against $\sin^2 \theta_{23}$ if the flux ratio 
$T$ is measured to be $\frac 13 $ and 0.35. We choose a pure source with 
$1 : 2 : 0$ and three different impure sources motivated 
by our estimates in Section \ref{sec:estimate}.  
Fig.~\ref{fig:etau_zeta} shows the same for $R = 1$ and $R = 1.1$. 
It is obvious that the covered area in parameter space changes 
considerably when the flux composition is varied. 
Table \ref{tab:pion} gives the numerical values of the ranges. 
Though the upper and lower limits are barely modified by varying 
$\zeta$, there are easily situations in which nonzero $\zeta$ is dramatic. 
For example, if $T$ was measured to 
be $\frac 13$ and it was known that 
$\sin^2 \theta_{23} = \frac 12$, then one would infer for 
an ideal pion source 
that $|U_{e3}| \, \cos \delta = 0$. If in addition $|U_{e3}|$ was known to be 
nonzero, then $\delta$ would be $\pi/2$ and CP-violation would be inferred.
However, if instead, the true value is $\zeta = 0.1$, 
then $|U_{e3}| \, \cos \delta \simeq 0.06$ would hold, and if $|U_{e3}|$ 
was known to be 0.06 then CP may or may not be broken. 

The simple examples given here show that care has to be taken when 
conclusions about neutrino mixing parameters are drawn from flux 
ratio measurements.

\subsection{\label{sec:effects010} Parameterized Muon-Damped Sources}
Here we investigate the impact of the non-idealized initial muon-damped source ratios 
$\Phi_e^0 : \Phi_\mu^0 : \Phi_\tau^0 = \eta : 1 : 0$. 
Figs.~\ref{fig:zetaT}, \ref{fig:zetaR} and \ref{fig:zetaQ} show the 
minimal and maximal values of the ratios $T$, $R$ and $Q$. 
In general, the 
dependence on initial flavor deviations is larger here than it was 
with pion sources. 
The fluxes behave according to 
\begin{eqnarray} \nonumber 
\Phi_e & \propto & P_{e\mu} + \eta \, P_{ee} = 
\frac{1}{18} 
\left( 
4 - 2 \, A + B 
+ 2 \, \eta \, (5 + 2 \, A)  
\right) 
~,\\[0.2cm] 
\Phi_\mu & \propto & P_{\mu\mu} + \eta \, P_{e\mu} = 
\frac{1}{18} 
\left( 
7 + A - B + C  
+ \eta \, (4 - 2 \, A + B) 
\right)
~,\\[0.2cm] 
\Phi_\tau & \propto & P_{\mu\tau} + \eta \, P_{e\tau} = \frac{1}{18} 
\left( 
7 + A -  C 
+ \eta \, (4 - 2 \, A - B) 
\right)
~. \nonumber
\end{eqnarray}   
Table~\ref{tab:muon} shows the numerical effect of $\eta \neq 0$ for 
neutrino fluxes from 
muon-damped sources.  Comparing with Table~\ref{tab:pion} for pion sources,
one sees that the effect of source uncertainty 
is larger for muon-damped sources compared to pion sources. 

If neutrinos mix tribimaximally, then 
\be
T_{\rm TBM} = \frac{1}{18} \frac{7 + 4 \, \eta}{1 + \eta}
\simeq \frac{1}{18} \left( 7 - 3 \, \eta \right)~\mbox{ and }~ 
R_{\rm TBM} =  \frac{4 + 10 \, \eta}{7 + 4 \, \eta}
\simeq \frac 17 \left( 4 + \frac{54}{7} \, \eta \right)~. 
\ee
As with pion sources, the parameter dependence is stronger for $R_{\rm TBM}$ 
than $T_{\rm TBM}$.
 
\begin{figure}[ht]
\begin{center}
\epsfig{file=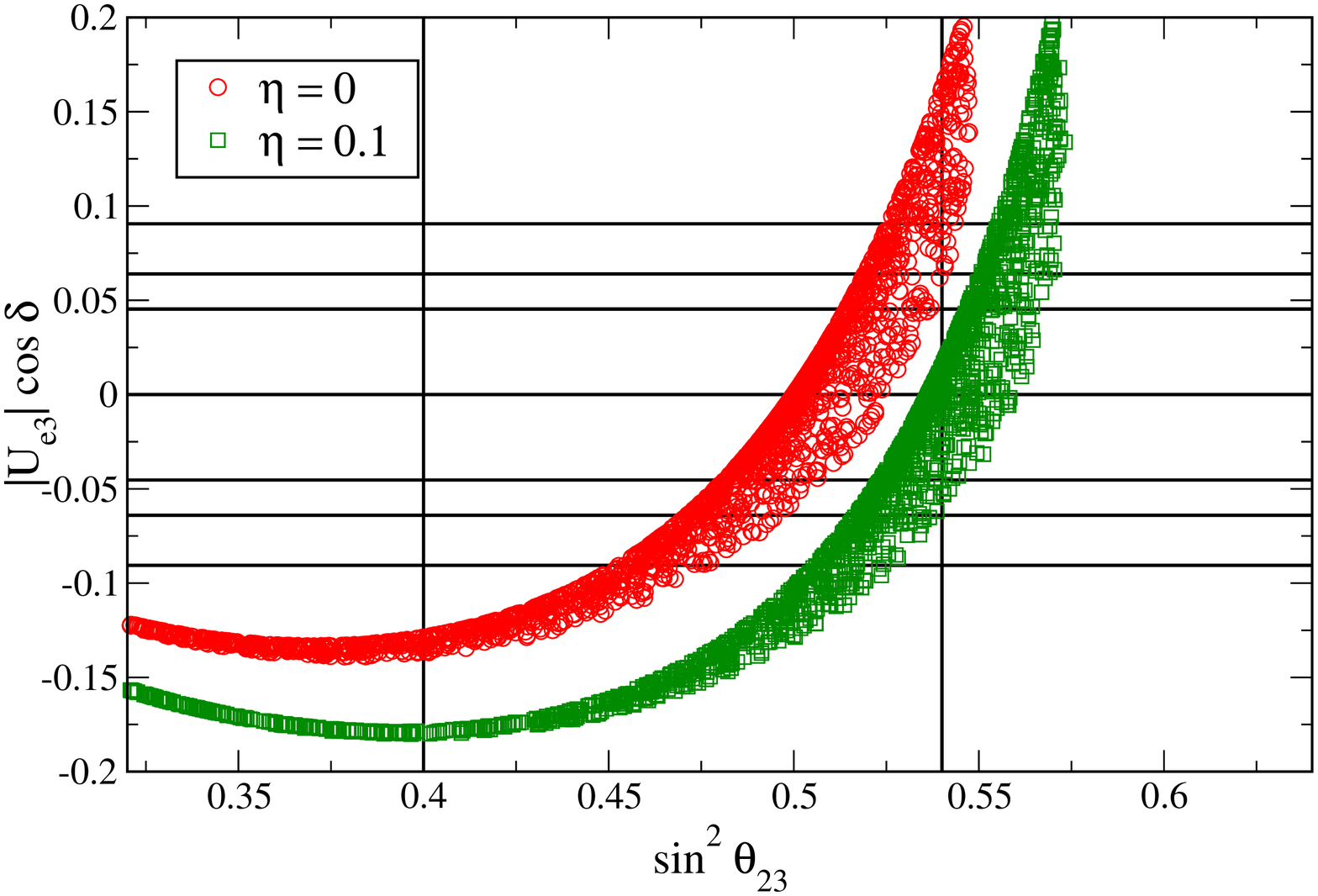,width=8cm,height=7.139cm} 
\epsfig{file=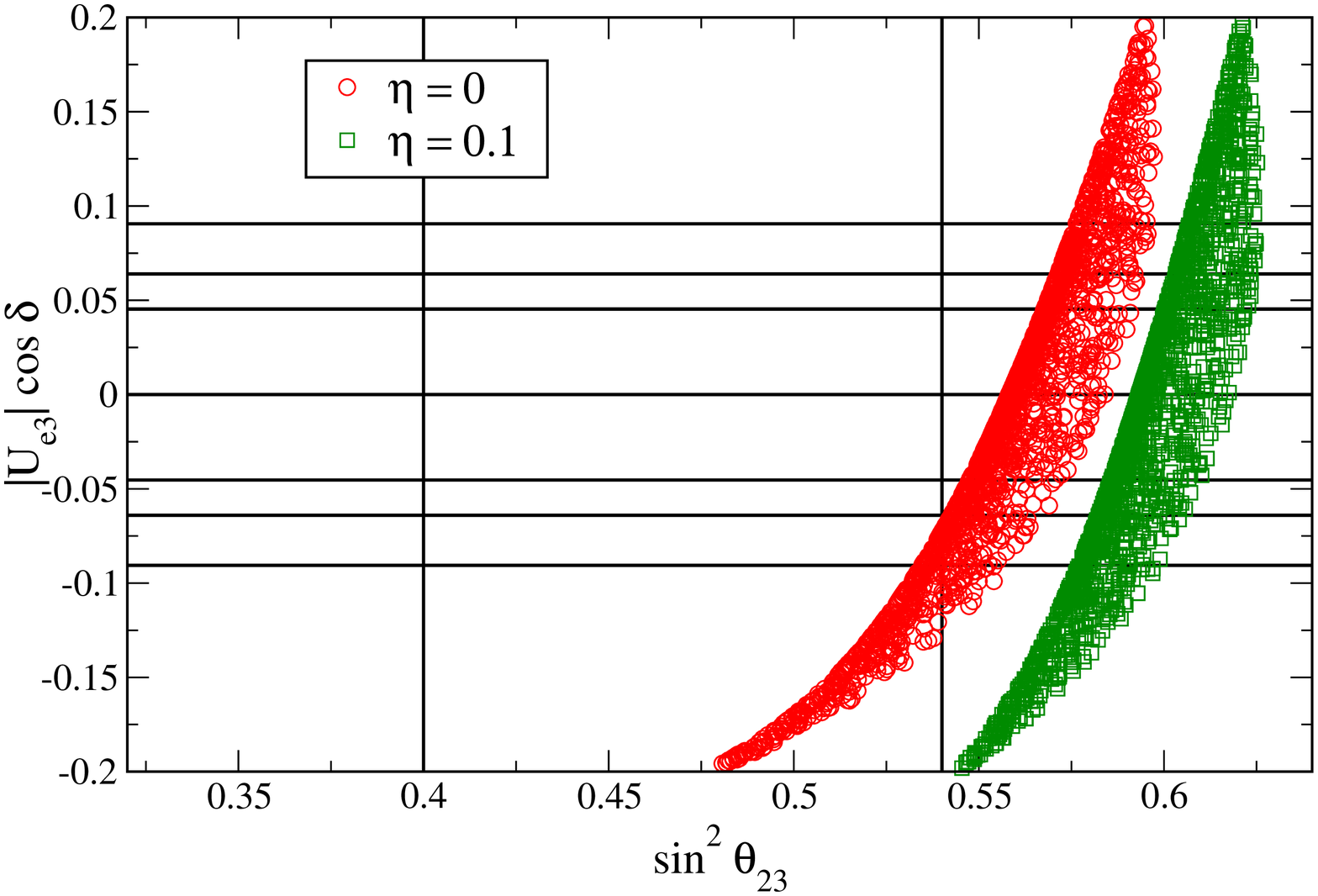,width=8cm,height=7.139cm} 
\end{center}
\caption{\label{fig:Tmu}Distribution of $|U_{e3}| \, \cos \delta$ 
against $\sin^2 \theta_{23}$ if the flux ratio $\Phi_\mu/\Phi_{\rm tot}$ 
is measured to be $7/18$ (left) and 0.42 (right);
the initial flavor mix is $\eta : 1 : 0$,
and $\sin^2 \theta_{12} = \frac 13$ is assumed.
}
\end{figure}

\begin{figure}[ht]
\begin{center}
\epsfig{file=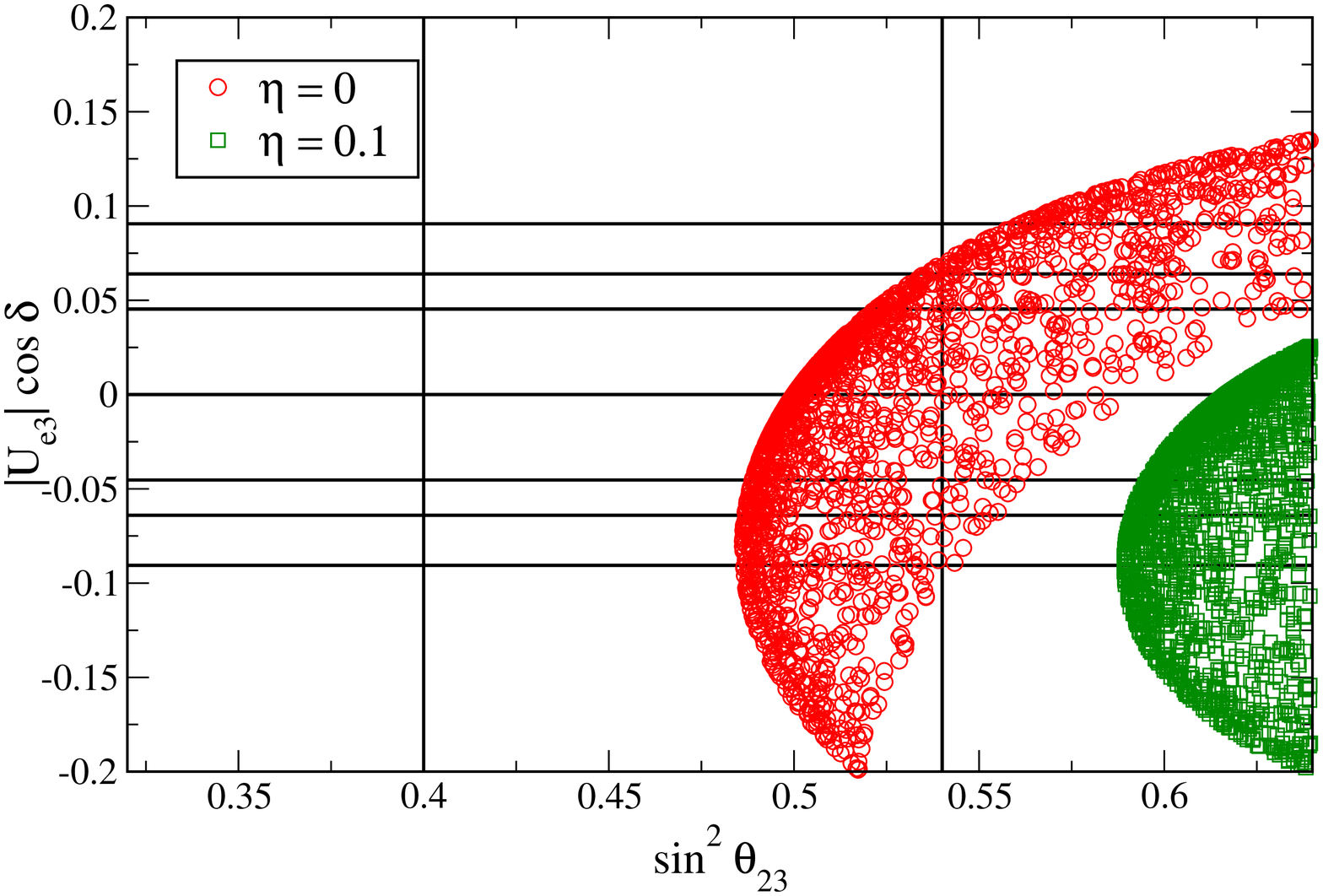,width=8cm,height=7.139cm} 
\epsfig{file=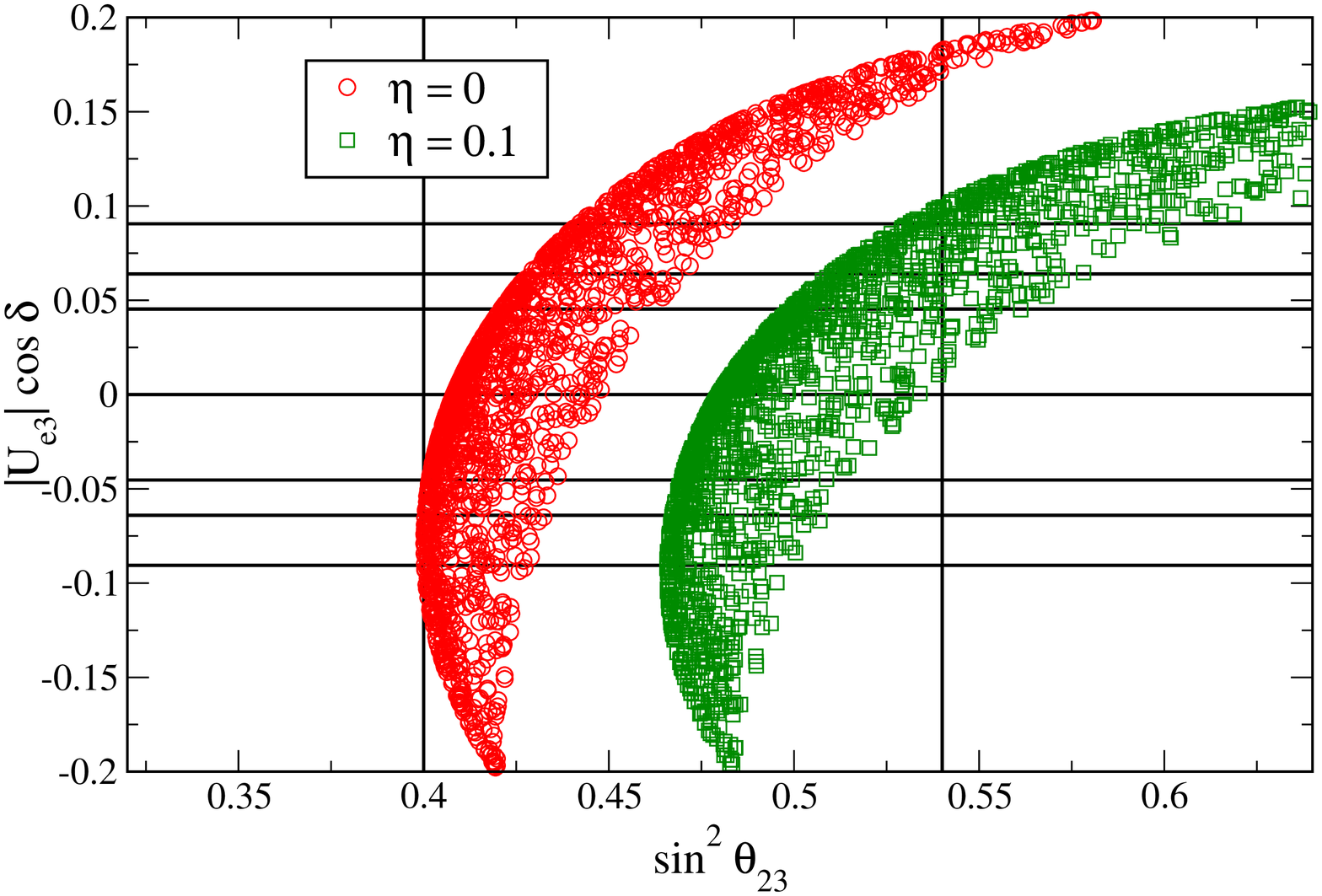,width=8cm,height=7.139cm}
\end{center}
\caption{\label{fig:Rmu}Same as previous Figure for the ratio of electron   
to tau neutrinos measured to be $7/18$ (left) and 0.7 (right).}
\end{figure}

Fig.~\ref{fig:Tmu} displays  
the distribution of $|U_{e3}| \cos \delta$ against $\sin^2 \theta_{23}$, 
taking $T$ to be $7/18$ and 0.42, and for simplicity, fixing 
$\sin^2 \theta_{12} = \frac 13$. 
For the two chosen values of $\eta=0$ and 0.1, 
the allowed areas do not meet, which shows again 
that the sensitivity on impure initial fluxes is stronger 
for muon-damped than for pion sources. 
Moreover, the dependence on the actual value of the ratio $T$ is weaker than 
the dependence on $\zeta$. 
In Fig.~\ref{fig:Rmu} we show correlated dependences, 
fixing the ratio $R$ to be $4/7$ and 0.7.

\begin{figure}[ht]
\begin{center}
\epsfig{file=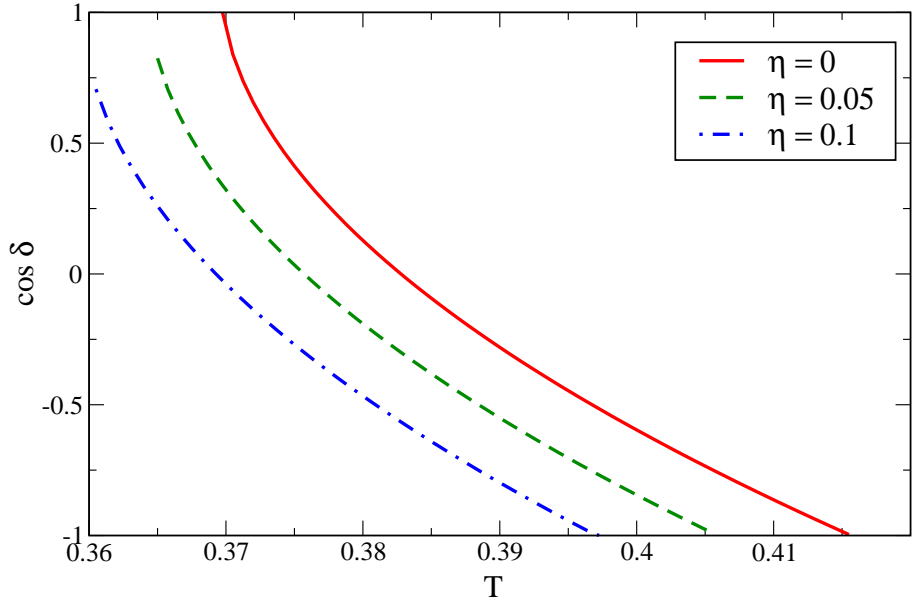,width=12cm,height=8.139cm} 
\end{center}
\caption{\label{fig:CPmu}Extracted value of $\cos \delta$ from an exact 
measurement of the ratio $T$ of muon neutrinos to the total flux for an initial 
flux composition $\eta : 1 : 0$. The oscillation parameters are fixed to 
$\theta_{23} = \pi/4$, $\sin^2 \theta_{12} = \frac 13$,  
and $|U_{e3}| = 0.15$.}
\end{figure}

Let us discuss a hypothetical but illustrative example.
If all neutrino parameters but $\delta$ were known exactly, and if 
$T$ were measured without any uncertainty, then the value of $\cos \delta$ 
can be extracted \cite{Michael,CP}. 
In Fig.~\ref{fig:CPmu} we display this possibility, taking optimistic  
values of the other parameters. We assume a large value of $\theta_{13}$ 
and maximal $\theta_{23}$ in order to maximize the dependence on 
$\cos \delta$. Suppose now that $T = 0.397$ were measured. Assuming 
that $\eta = 0$, i.e., a very pure muon-damped source, one would 
conclude that $\cos\delta = -0.5$, thereby inferring  leptonic 
CP-violation.  However, if in reality $\eta = 0.1$,
then the same $T_\mu = 0.397$ value would instead mean that $\cos \delta = -1$ 
and CP is conserved.

\begin{figure}[ht]
\begin{center}
\epsfig{file=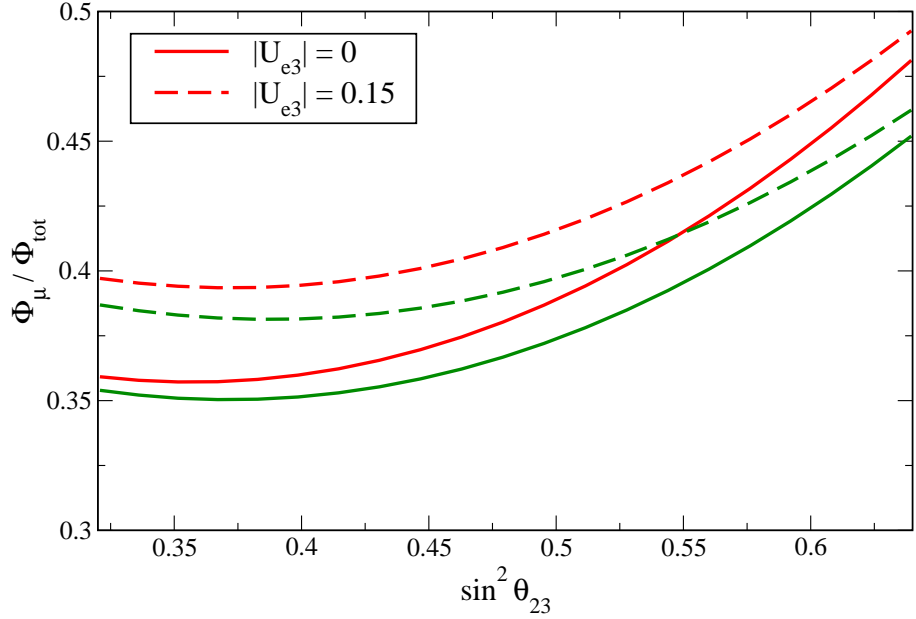,width=12cm,height=8.139cm} 
\end{center}
\caption{\label{fig:muzetat13}Dependence on $\sin^2 \theta_{23}$ 
of the ratio of muon neutrinos 
to the total flux. The red lines are for $0 : 1 : 0$, while the green 
lines below are for $0.1 : 1 : 0$. We have 
chosen $\delta = \pi$ and $\sin^2 \theta_{12} = \frac 13$. }
\end{figure}

Let us present another example using Fig.~\ref{fig:muzetat13}.
In Fig.~\ref{fig:muzetat13} the dependence 
of $T$ on $\sin^2 \theta_{23}$~\cite{Pasquale} is shown, 
for two values of $|U_{e3}|$ and for $\eta = 0$ and 0.1. 
The nonzero $\eta$ introduces a few percent uncertainty in the flux ratio, 
especially for $\sin^2 \theta_{23} \ge 0.5$. In general, the ratio decreases. 
The curves for different $|U_{e3}|$ 
do not meet when $\eta = 0$; however, for $\eta = 0.1$ they cross each other,
thereby destroying the possibility to disentangle the two chosen values 
of $|U_{e3}|$.

\subsection{\label{sec:effects100} Parameterized Neutron Beam Source}
Finally we come to the impact of an initial $\beta$-beam source  
which may not be pure, i.e., 
$\Phi_e^0 : \Phi_\mu^0 : \Phi_\tau^0 = 1 : \eta : 0$. 
In this case the flux ratios at Earth are given by
\begin{eqnarray} \nonumber 
\Phi_e & \propto & P_{e e} + \eta \, P_{e \mu} = 
\frac{1}{18} 
\left( 
10 + 4 \, A   
+ \eta \, (4 -  2 \, A + B)  
\right) 
~,\\[0.2cm] 
\Phi_\mu & \propto & P_{e\mu} + \eta \, P_{\mu\mu} = 
\frac{1}{18} 
\left( 
4 - 2 \, A + B    
+ \eta \, (7 + A - B + C) 
\right)
~,\\[0.2cm] 
\Phi_\tau & \propto & P_{e \tau} + \eta \, P_{ \mu\tau} = \frac{1}{18} 
\left( 
4 - 2 \, A - B  
+ \eta \, (7 + A - C) 
\right)
~. \nonumber
\end{eqnarray}

\begin{figure}[ht]
\begin{center}
\epsfig{file=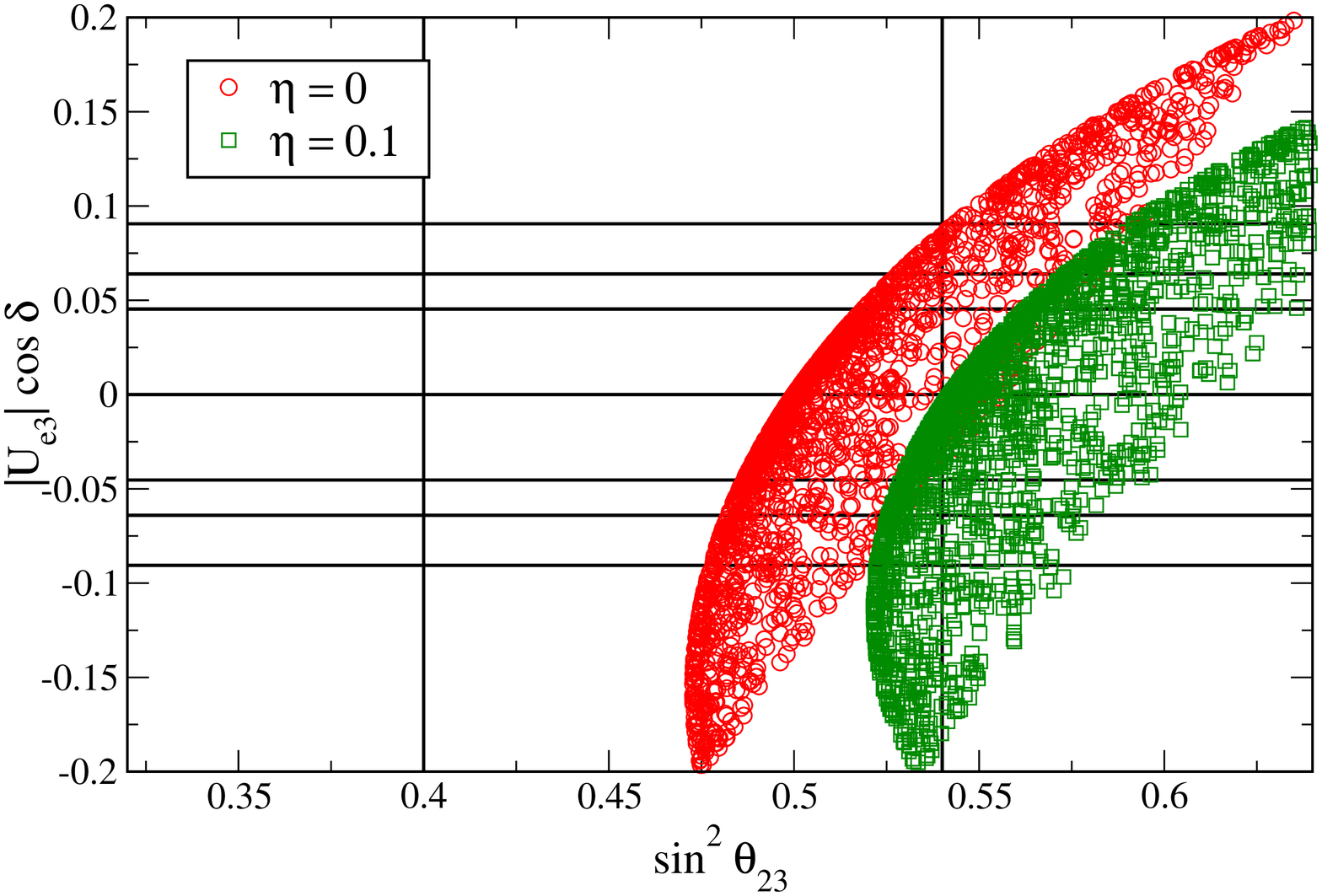,width=8cm,height=7.139cm} 
\epsfig{file=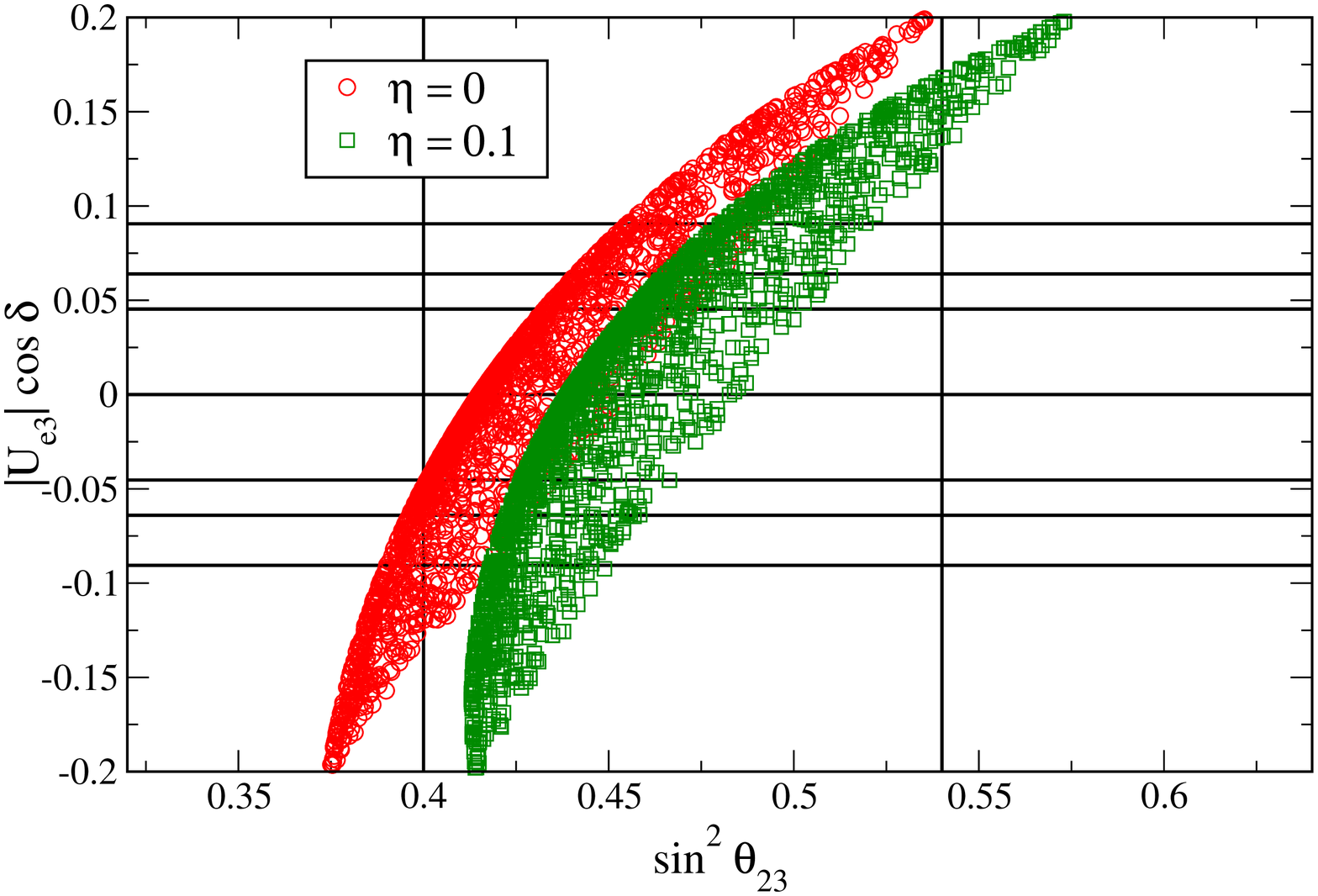,width=8cm,height=7.139cm} 
\end{center}
\caption{\label{fig:Tneut}
Distribution of $|U_{e3}| \, \cos \delta$ 
against $\sin^2 \theta_{23}$ if the flux ratio $\Phi_\mu/\Phi_{\rm tot}$ 
is measured to be $2/9$ (left) and 0.26 (right), 
for an initial flavor mix of $1 : \eta : 0$.}
\end{figure}

\begin{figure}[ht]
\begin{center}
\epsfig{file=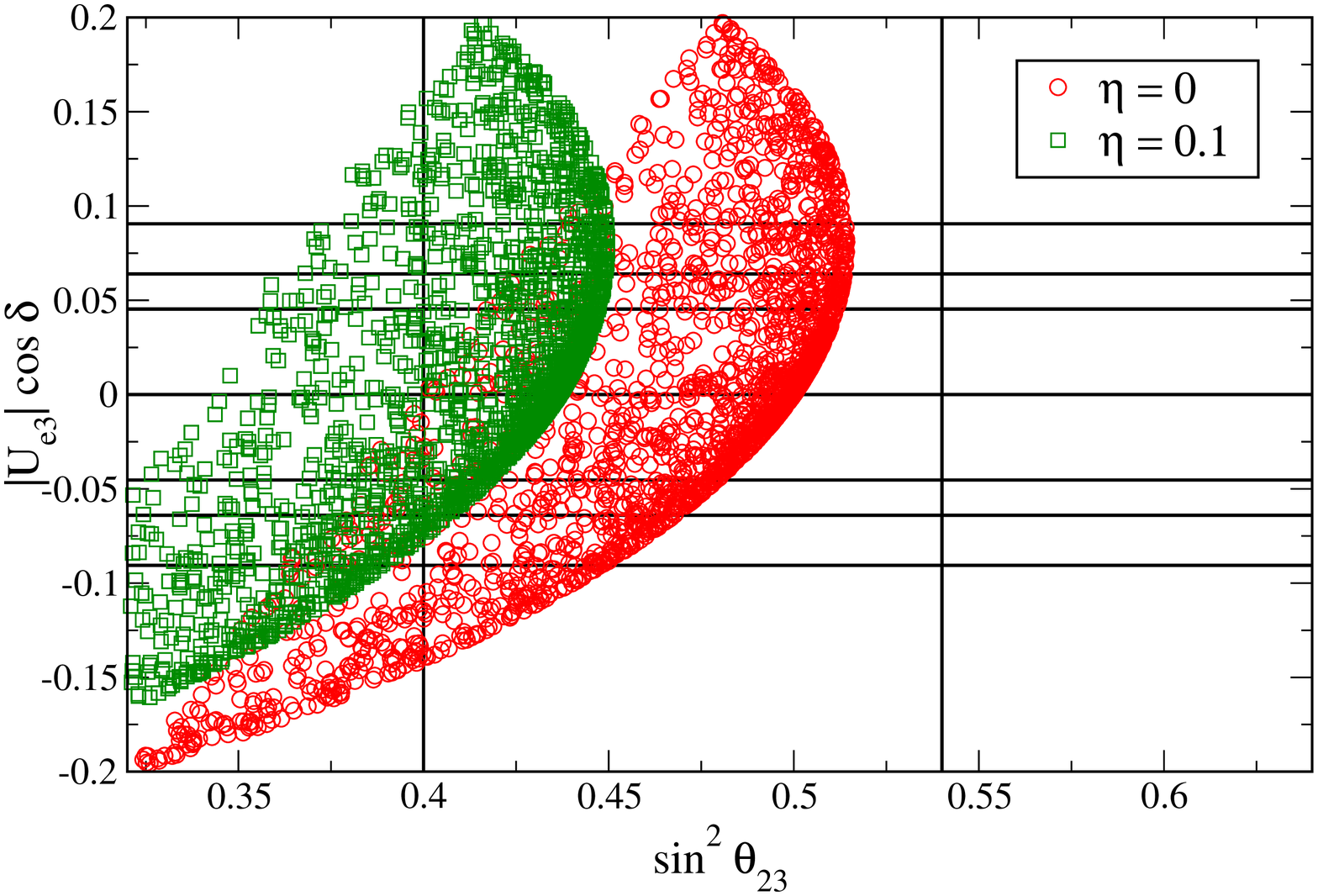,width=8cm,height=7.139cm}
\epsfig{file=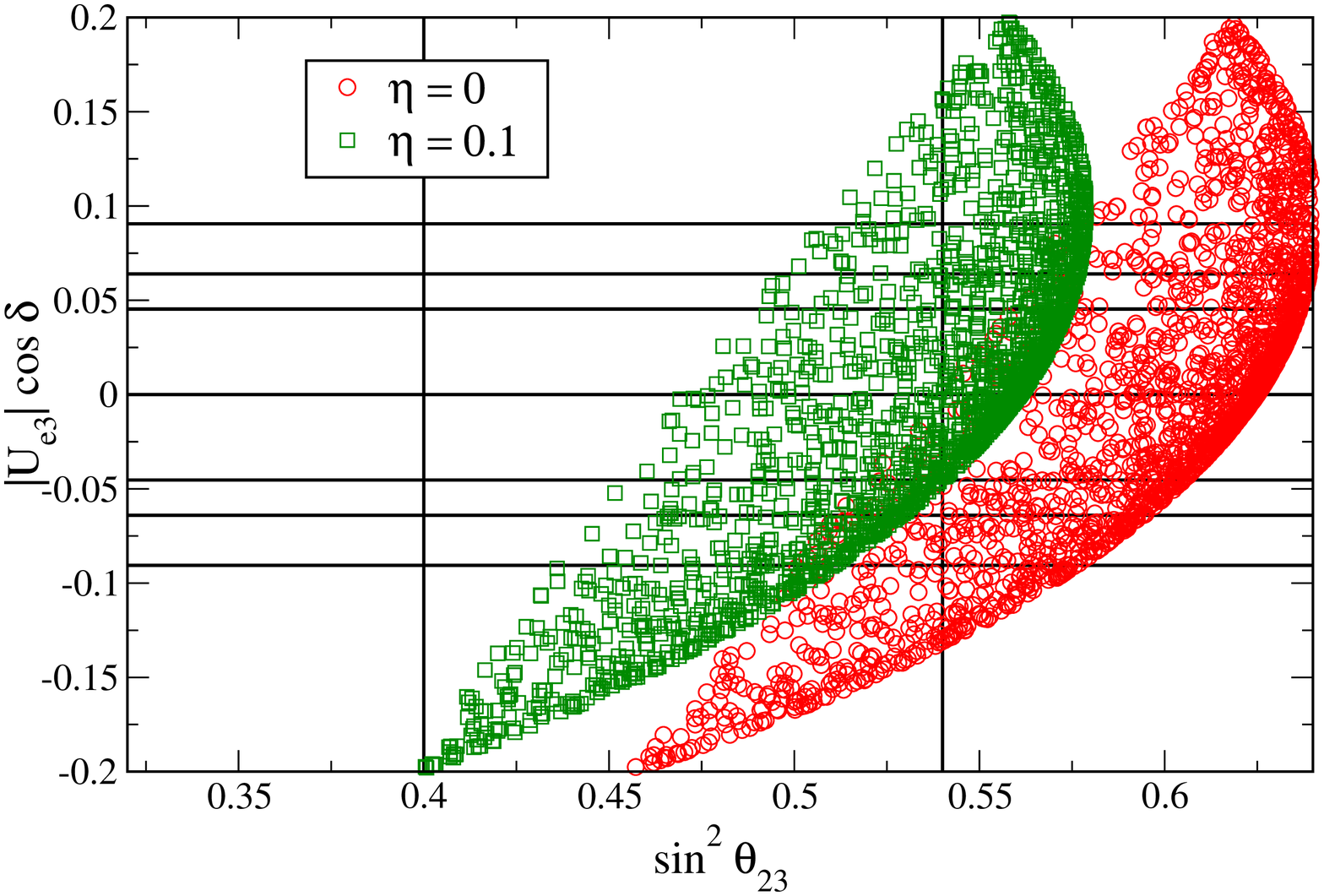,width=8cm,height=7.139cm} 
\end{center}
\caption{\label{fig:Rneut}Same as previous Figure for the ratio of electron   
to tau neutrinos measured to be 5/2 (left) and 2 (right).}
\end{figure}

The tribimaximal values for oscillation parameters leads to  
\be
T_{\rm TBM} = \frac{1}{18} \frac{4 + 7 \, \eta}{1 + \eta}
\simeq \frac{1}{9} \left( 2 + \frac 32 \, \eta \right)~\mbox{ and }~ 
R_{\rm TBM} =  \frac{10 + 4 \, \eta}{4 + 7 \, \eta}
\simeq \frac 52 \left( 1 - \frac{27}{20} \, \eta \right)~. 
\ee
More generally, we show in Figs.~\ref{fig:zetaT}, \ref{fig:zetaR} 
and \ref{fig:zetaQ} the 
minimal and maximal values of the ratios $T$, $R$ and $Q$ 
(flux composition 
$\nue :\nuebar :\numu :\numubar :\nutau :\nutaubar =0:1: \eta/2 : \eta/2 :0:0$) 
that result when the oscillation 
parameters are allowed to vary over their 3$\sigma$ ranges. 
As was the case with muon-damped sources, deviations from pure flux 
compositions have 
more impact on neutron sources than on pion sources.
And again, the dependence on the actual value of the ratio $T$ or $R$ 
is weaker than the dependence on $\zeta$. 
Table \ref{tab:neutron} confirms these remarks. 
In Figs.~\ref{fig:Tneut} and \ref{fig:Rneut} the distributions of 
$|U_{e3}| \cos \delta$ against $\sin^2 \theta_{23}$ for two 
characteristic values of $T$~(2/9 and 0.26) and $R$~(5/2 and 2) are given.

Recall that $\Delta$ but not $\overline{\Delta}^2$ appears in the 
flux ratios for neutron sources. 
This makes it possible to give a simple formula for the special case where 
the effect of nonzero $\Delta$ is exactly compensated by a nonzero $\eta$. 
If 
\be
\eta = 2 \, \frac{\Delta + (1 - 2 \, c_{12}^2 \, s_{12}^2 ) 
\, |U_{e3}|^2}{1 - 3 \, c_{12}^2 \, s_{12}^2} \simeq 
6 \left( \Delta + \frac 59 \, |U_{e3}|^2 \right)~,
\ee
then the tribimaximal value $T=c^2_{12}\,s^2_{12}$ necessarily results. 

\begin{figure}[ht]
\begin{center}
\epsfig{file=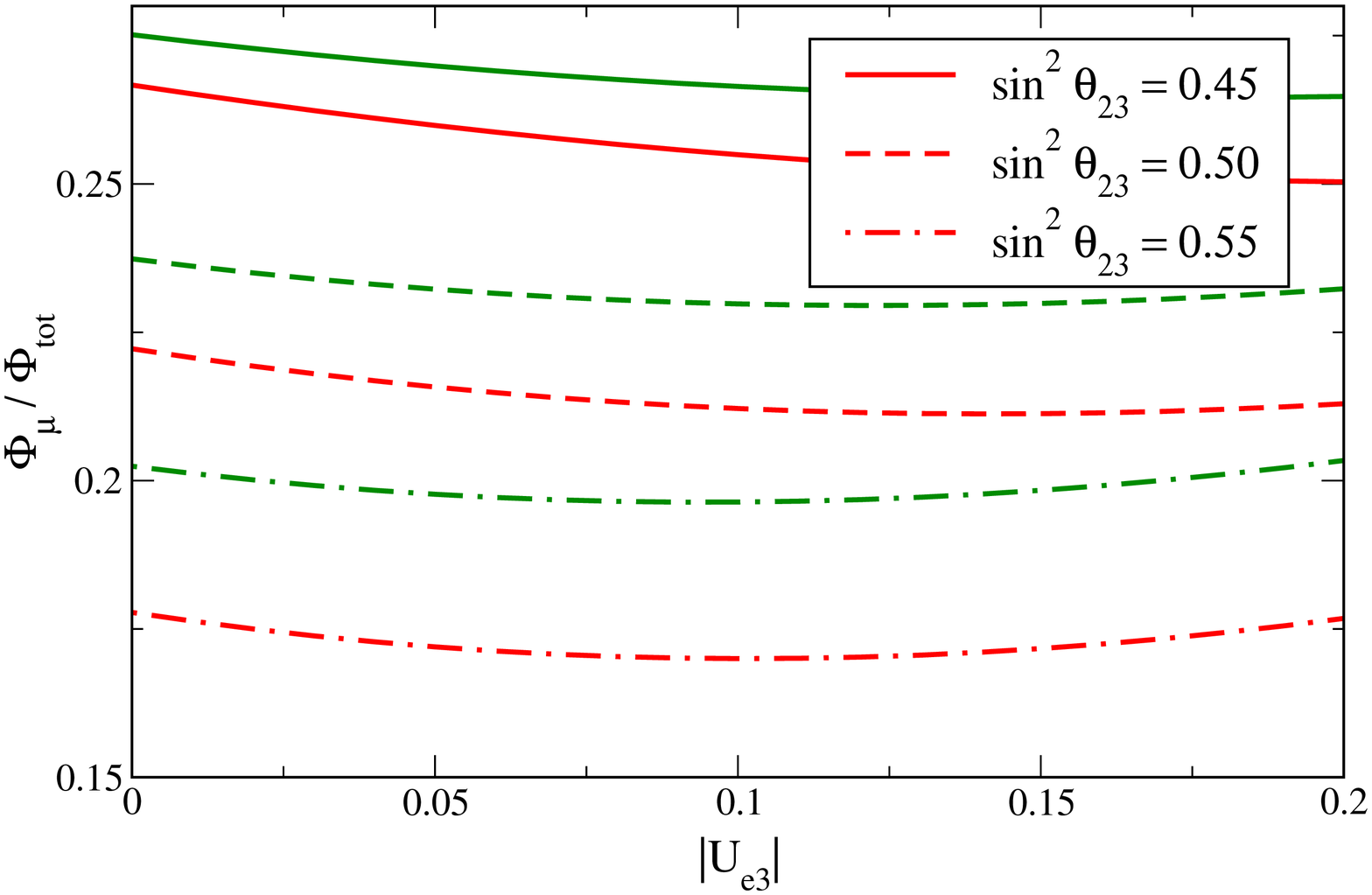,width=12cm,height=8.139cm} 
\end{center}
\caption{\label{fig:Nzetat23}Dependence on $\theta_{13}$ 
of the ratio of muon neutrinos 
to the total flux. The red lines are for $1 : 0 : 0$, while the green 
lines above are for $1 : 0.1 : 0$. We have 
chosen $\delta = \pi$ and $\sin^2 \theta_{12} = \frac 13$. }
\end{figure}

We give in Fig.~\ref{fig:Nzetat23} an example of the dependence of 
$T$ on $|U_{e3}|$ and $\theta_{23}$. 
It can be seen that impure initial flux compositions can 
influence statements on the octant of $\theta_{23}$. 
For instance, measuring $T = 0.2$ would rule out 
$\sin^2 \theta_{23} > 0.5$ only if $\eta = 0$. 
However, if $\eta=0.1$, then $T = 0.2$ 
is compatible with $\sin^2 \theta_{23} = 0.55$, and the octant of 
$\theta_{23}$ is different from the one inferred if $\eta = 0$.

\section{\label{sec:concl}Summary and Conclusions}

We have considered in this paper neutrino mixing and 
flux ratios of astrophysical neutrinos. We first have expanded the 
expressions in terms of small parameters 
$\epsilon = \pi/4 - \theta_{23}$ and $|U_{e3}|$,
while leaving $\theta_{12}$ free.  
The small parameters $\eps$ and $\Re \{U_{e3}\}$ measure the symmetry breaking
of $\numu \leftrightarrow \nutau$ interchange symmetry. 
With this expansion, 
we showed that the first and second order corrections 
which characterize the deviations from $\mu$--$\tau$ symmetry, 
$\Delta$ and $\overline \Delta^2$, appear universally.  
The universal corrections $\Delta$ and $\overline \Delta^2$ 
are given in Eqs.~(\ref{eq:Delta}, \ref{eq:Delta2}).
Each can take values as large as 0.1.
Compact results for the mixing probabilities, in terms of 
$\Delta$ and $\overline \Delta^2$, 
are shown in Eqs.~(\ref{eq:res1a}, \ref{eq:probs}). 

The second order term $\overline \Delta^2$ appears only in 
the $\mu$--$\tau$ sector (therefore 
it is not relevant for flux ratios from neutron beam sources) 
and is positive semidefinite. Because it 
can exceed the first order term, it is necessary to include it 
in analytical considerations.  It 
vanishes only for $\epsilon = |U_{e3}| \, \cos \delta = 0$, whereas 
the first order term can vanish also for nonzero values of $\eps$ and $|U_{e3}|$. 

In general, if the initial flavor mix is exactly $1 : 2 : 0$, then 
neutrino mixing transforms these ratios to 
$(\Phi_e : \Phi_\mu : \Phi_\tau) = 
(1 + 2 \, \Delta) : (1 - \Delta + \overline{\Delta}^2) 
: (1 - \Delta - \overline{\Delta}^2)$. Hence, there are always more 
muon than tau neutrinos upon arrival at Earth. The ratio of muon neutrinos 
to all neutrinos can deviate by more than 15\% from the tribimaximal 
value $\frac 13$, while the ratio of electron to tau neutrinos 
can deviate by up to 50\% from the 
tribimaximal value 1. 

As the solar neutrino mixing parameter $\sin^2 \theta_{12}$ 
is close to the tribimaximal value $\frac 13$, 
we next included $\epsilon' = \arcsin \sqrt{ \frac 13} - \theta_{12}$ 
in our set of expansion parameters. 
With this expansion set,
three universal corrections $A$, $B$ and $C$, defined in~\cite{PRW} and 
reproduced here in Eq.~(\ref{eq:ABC}), 
characterize the deviations from tribimaximal mixing. 
Very concise expressions for the mixing probabilities result~\cite{PRW}, 
as seen in Eqs.~(\ref{eq:probs2}).

In the second part of this paper, 
we investigated the purity of initial neutrino-flavor ratios expected from three
types of astrophysical sources. 
The initial flavor ratios commonly considered in the literature are 
(i) pion sources (with a complete pion decay chain),
having initial ratios $1 : 2 : 0$; 
(ii) muon-damped sources (initial pions but an incomplete decay chain),
having initial ratios $0 : 1 : 0$;
and 
(iii) neutron beam sources (so-called ``cosmic $\beta$-beams''),
with initial neutrino flavor ratios $1 : 0 : 0$. 
The idealized flavor ratios of all three source types 
are subject to small but important corrections. 
We investigated the 
effects of realistic corrections on the extraction of neutrino parameters from 
measurements of flux ratios. 
We found that the muon-damped and neutron beam sources 
are more sensitive to initial flavor deviations than is the pion source.
In addition, the muon-damped and neutron beam sources 
are also more sensitive to deviations of oscillation parameters $\theta_{23}$ and 
$U_{e3}$ from $\pi/4$ and zero, respectively. 
These sensitivities can be easily seen by considering the ratio 
$T$ of muon neutrinos to the total flux.
For $\sin^2 \theta_{12} = \frac 13$, this ratio reads 
\[ 
T = \frac{\Phi_\mu}{\Phi_{\rm tot}}\simeq \left\{ 
\bad 
\frac 13 \left( 1 - \Delta + \overline{\Delta}^2 - \frac 19 \, \zeta \right)\,,
& \mbox{pion source} & (1 : 2 \, (1 - \zeta) : 0)~,  \\[0.2cm]
\frac{7}{18} - \Delta + \frac 12 \, \overline{\Delta}^2 - 
\frac 16 \, \eta\,, & \mbox{muon-damped source} & (\eta : 1 : 0)~, \\[0.2cm]
\frac 29 + \Delta + \frac 16 \, \eta\,, & 
\mbox{neutron beam source} & (1 : \eta : 0)~,
\ea 
\right. 
\]
where the parameterized initial flavor-ratios are 
$1 : 2 \, (1 - \zeta) : 0$, $\eta : 1 : 0$ and $1 : \eta : 0$, respectively. 
For pion sources, the zeroth order expression is $T = \frac 13$, and with 
$|\Delta, \, \overline{\Delta}^2, \, \zeta, \, \eta | \ls 0.1$, 
deviations can be up to 
$ 15\% $ for oscillation-induced effects and of order $1\%$ for impure 
flavor mixes. 
For muon-damped sources, on the other hand, the effect 
of uncertain oscillation parameters is up to $30\%$, 
and the effect of nonzero $\eta$ is more than $5\%$. 
The effect on neutron beam sources may be even more dramatic: 
the observable $T$ can change by more than $50\%$ due to deviations 
from tribimaximality, and by order $10\%$ due to impurities. 
We have also considered the ratio $R$ of electron to tau neutrinos, which 
in the same $\sin^2 \theta_{12}=1/3$ limit reads 
\[ 
R = \frac{\Phi_e}{\Phi_\tau}\simeq \left\{ 
\bad \nonumber 
1 + 3 \, \Delta + \overline{\Delta}^2 + \frac{\zeta}{3} \,,
& \mbox{pion source} & (1 : 2 \, (1 - \zeta) : 0)~, \\[0.2cm]
\frac{4}{7} \left( 
1 + 18 \, \Delta + \frac{36}{7} \,  \overline{\Delta}^2 
+ \frac{54}{7} \,  \eta 
\right)\,,
& \mbox{muon-damped source}& (\eta : 1 : 0)~, \\[0.2cm]
\frac 52 \left( 
1 + \frac 92 \, \Delta - \frac{27}{20} \, \eta 
\right) \,,
& \mbox{neutron beam source}& (1 : \eta : 0)~.
\ea
\right.
\]
The magnitude of coefficients reveal that the effects 
of nonzero $\Delta$, $\overline{\Delta}^2$, $\zeta$ or $\eta$
is in general stronger on $R$ than on $T$. To be more quantitative 
(see Tables \ref{tab:pion}, \ref{tab:muon}, \ref{tab:neutron}), 
oscillation effects are up to 50\% for pion sources and almost a factor of two 
for muon-damped and neutron sources. Impurities in the initial 
flux composition of 0.1 lead to deviations in the flux ratios 
of 4\%, 20\% and 15\% for pion, muon-damped and neutron sources, respectively.

We gave several illustrative examples 
where the assumption of 
an idealized, pure initial flux ratio may easily (mis)lead to 
incorrect inferences. Wrong inferences may include 
the octant of $\theta_{23}$, the magnitude of $|U_{e3}|$ and the 
existence of leptonic CP-violation. 
We stress that in future analyses, the intrinsic flux 
uncertainty should be taken into account before inferences are drawn.

\vspace{0.3cm}
\begin{center}
{\bf Acknowledgments}
\end{center}
We would like to thank John Learned, 
Paolo Lipari, Halsie Reno and Todor Stanev for their generous help
in clarifying several issues for us. 
W.R.~was supported by the Deutsche Forschungsgemeinschaft 
in the Sonderforschungsbereich 
Transregio 27 ``Neutrinos and beyond -- Weakly interacting particles in 
Physics, Astrophysics and Cosmology'' and under project 
number RO--2516/3--2, as well as by the EU program ILIAS N6 ENTApP WP1. 
S.P.~and T.J.W.~thank M. Lindner and 
the Max-Planck-Institut f\"ur Kernphysik, Heidelberg 
for support and hospitality, 
and acknowledge support from U.S.~DoE  
grants DE--FG03--91ER40833 and DE--FG05--85ER40226.

%\newpage
\renewcommand{\theequation}{A\arabic{equation}}
  % redefine the command that creates the equation no.
\setcounter{equation}{0}
\begin{appendix}

\section{\label{sec:app}Appendix: Mixing Probabilities}
For the sake of completeness, we give here the explicit forms of the 
oscillation probabilities. 
The lepton mixing, or PMNS, matrix is parameterized in Particle 
Data Group format as 
\bea \label{eq:Upara}
U = 
\left( \bad 
c_{12} \, c_{13} & s_{12}   \, c_{13} & s_{13}  \, e^{-i \delta} \\[0.2cm] 
-s_{12}   \, c_{23} - c_{12}   \, s_{23}   \, s_{13}   \, e^{i \delta}  
& c_{12}   \,  c_{23} - s_{12}  \,   s_{23}  \,   s_{13}  \,  e^{i \delta}
& s_{23}   \,  c_{13}  \\[0.2cm] 
s_{12}  \,   s_{23} - c_{12}  \,   c_{23}  \,   s_{13}  \, e^{i \delta}& 
- c_{12}  \,   s_{23} - s_{12}  \,   c_{23}   \,  s_{13} \,    e^{i \delta}
& c_{23}   \,  c_{13}  \\ 
               \ea   \right)  ~, 
\eea
where $c_{ij} = \cos\theta_{ij}$, 
$s_{ij} = \sin\theta_{ij}$, and 
we have omitted the Majorana phases, which are irrelevant 
for neutrino oscillations. Using Eq.~(\ref{eq:Pab}), one obtains 
\bea \D 
P_{e \mu} =   2 \, c_{13}^2 \, 
\left\{c_{12}^2 \, s_{12}^2 \, c_{23}^2 
+ \left(c_{12}^4+ s_{12}^2 \right) \, s_{13}^2 \, s_{23}^2 
\right.
\\[0.2cm]
\left.
\D + c_{12} \, s_{12} \, 
    c_{23} \, s_{23} \, c_\delta \, (c_{12}^2 - s_{12}^2)  
    \, s_{13} 
 \right\}~,
\eea
where $c_\delta = \cos \delta$, and  
\bea \D 
P_{\mu \mu} = 1 - 2 \, c_{12}^4 \, c_{23}^2 \, s_{23}^2 \, s_{13}^2  \\[0.2cm]
 \D +2 \left\{\left(s_{12}^2 \left[ \left(s_{13}^4+\left[ 4
    \, c_\delta^2 - 1\right] \, s_{13}^2 +1 \right)
    \, s_{23}^2 - 1 \right] - c_{13}^2 \, s_{23}^2\right)
    \, c_{23}^2
\right.
\\[0.2cm]
\left. \D 
+ s_{13}^2 \, s_{23}^2 \left(c_{13}^2
    \, s_{12}^2-\left[ c_{13}^2+ s_{12}^2\right]
    \, s_{23}^2\right)\right] \, c_{12}^2 \\[0.2cm]
 \D + s_{23} \left[ -2
    \left( c_{23}^2 \, c_{13}^4+\left[ c_{13}^2+ c_{23}^2
    \, s_{12}^2\right] \, s_{13}^2\right) \, s_{23} \, s_{12}^2
\right.
\\[0.2cm]
\left.
 \D -2 \, c_{12} \, s_{12} \, c_{23} \, c_\delta \, (c^2_{12}- s^2_{12})
    \left(c_{13}^2+\left[ s_{13}^2+1\right] [ c^2_{23}- s^2_{23}]
    \right) \, s_{13}\right\}~.
\eea
With the help of the identities \cite{chef,WR} 
\bea \label{eq:t23repl}
P_{e \tau} = P_{e \mu}(\theta_{23} \ra \theta_{23} + \pi/2 \mbox{ or } 
\theta_{23} \ra \theta_{23} + 3\pi/2)  ~,\\[0.2cm]
P_{\tau \tau} = P_{\mu \mu}(\theta_{23} \ra \theta_{23} + \pi/2\mbox{ or } 
\theta_{23} \ra \theta_{23} + 3\pi/2)~,
\eea
and the unitary relations 
\bea \label{eq:sumrules}
P_{ee} = 1 - P_{e\mu} - P_{e\tau} ~,\\[0.2cm]
P_{\mu \tau} = 1 - P_{e \mu} - P_{\mu \mu} ~,\\[0.2cm]
P_{\tau \tau} = 1 -  P_{e \tau} - P_{\mu \tau} 
= P_{ee} + 2 \,  P_{e \mu} + P_{\mu\mu} - 1 = 
P_{e\mu} - P_{e\tau} + P_{\mu\mu}~,
\eea
all other probabilities can be readily obtained.

\end{appendix}

\vfill\eject

\pagestyle{empty}

%\newpage

\begin{table}[ht]
\begin{center}
\begin{tabular}{|c|c|c||c|c|} \hline  
& \multicolumn{2}{c||}{general case} & \multicolumn{2}{c|}{TBM} \\ \hline
composition & $T = \Phi_\mu/\Phi_{\rm tot}$ & $R = \Phi_e/\Phi_\tau$ 
& $T = \Phi_\mu/\Phi_{\rm tot}$ & $R = \Phi_e/\Phi_\tau$  \\ \hline \hline 
$1 : 2 : 0$ & $0.323 \div 0.389$ & $0.818 \div 1.476$ 
& $ 0.333 $ & $1.000 $ \\ \hline 
$1 : 1.90 : 0.001$ & $0.32 1\div 0.386$ & $0.834 \div 1.493$ 
& $ 0.331$ & $1.017 $ \\ \hline
$1 : 1.85 : 0.001$ & $0.321 \div 0.384$ & $0.842 \div 1.503$ 
& $ 0.330$ & $1.026 $ \\ \hline
$1 : 1.80 : 0.001$ & $0.320 \div 0.382 $ & $0.850 \div 1.513$ 
& $ 0.329$ & $1.036 $ \\ \hline
\end{tabular}
\caption{\label{tab:pion}Pion sources: ranges of the ratios  
$T = \Phi_\mu/\Phi_{\rm tot}$ and $R = \Phi_e/\Phi_\tau$ 
for the current $3\sigma$ ranges of the oscillation 
parameters. We used different flux compositions, pure $1 : 2 : 0$ and 
different impure cases. The values for tribimaximal mixing 
are also given.} \vspace{.4cm}
 
\begin{tabular}{|c|c|c||c|c|} \hline  
 & \multicolumn{2}{c||}{general case} & \multicolumn{2}{c|}{TBM} \\ \hline
composition & $T = \Phi_\mu/\Phi_{\rm tot}$ & $R = \Phi_e/\Phi_\tau$ 
& $T = \Phi_\mu/\Phi_{\rm tot}$ & $R = \Phi_e/\Phi_\tau$  \\ \hline \hline
$0 : 1 : 0$ & $0.33 \div 0.51$ & $0.34 \div 1.13$ 
& $ 0.39$ & $0.57 $ \\ \hline 
$0.05 : 1 : 0$  & $0.33 \div 0.49$ & $0.40 \div 1.17$ 
& $ 0.38$ & $0.63 $ \\ \hline
$0.1 :  1 : 0$ & $0.33 \div 0.48$ & $0.46 \div 1.21$ 
& $ 0.37$ & $0.68 $ \\ \hline 
\end{tabular}
\caption{\label{tab:muon}Muon damped sources: ranges of the ratios  
$T = \Phi_\mu/\Phi_{\rm tot}$ and $R = \Phi_e/\Phi_\tau$ 
for the current $3\sigma$ ranges of the oscillation 
parameters. We used different flux compositions, pure $0 : 1 : 0$ and 
different impure cases. The values for tribimaximal mixing 
are also given.}\vspace{.4cm}

\begin{tabular}{|c|c|c||c|c|} \hline   
& \multicolumn{2}{c||}{general case} & \multicolumn{2}{c|}{TBM} \\ \hline
composition & $T = \Phi_\mu/\Phi_{\rm tot}$ & $R = \Phi_e/\Phi_\tau$ 
& $T = \Phi_\mu/\Phi_{\rm tot}$ & $R = \Phi_e/\Phi_\tau$  \\ \hline \hline 
$1 : 0 : 0$ & $0.12 \div 0.35$ & $1.39 \div 5.35$ 
& $ 0.22$ & $2.50 $ \\ \hline 
$1 : 0.05 : 0$ & $0.14 \div 0.35$ & $1.35 \div 4.73$ 
& $0.23 $ & $2.34 $ \\ \hline
$1 : 0.1 : 0$ & $0.16 \div 0.35$ & $1.32 \div 4.26$ 
& $0.24 $ & $2.21 $ \\ \hline
\end{tabular}
\caption{\label{tab:neutron}Neutron beam sources: ranges of the ratios  
$T = \Phi_\mu/\Phi_{\rm tot}$ and $R = \Phi_e/\Phi_\tau$ 
for the current $3\sigma$ ranges of the oscillation 
parameters. We used different flux compositions, pure $1 : 0 : 0$ and 
different impure cases.  The values for tribimaximal mixing 
are also given.}
\end{center}
\end{table}

\end{document}